\documentclass{aa}
\usepackage{CJK}
\usepackage{enumitem}
\usepackage{verbatim}
\usepackage{amsmath}
\usepackage{overpic}
\usepackage{xcolor}
\usepackage{float}
\usepackage{multirow}
\newcommand{\XMM}{ XMM-{\em Newton}}
\newcommand{\Chandra}{{\em Chandra}}
\newcommand{\egs}{erg~cm$^{-2}$~s$^{-1}$}
\newcommand{\N}[1]{$N_\textrm{#1}$}
\newcommand{\arcdeg}{\mbox{$^\circ$}}
\newcommand{\ARF}{\mathrm{ARF}}
\newcommand{\RMF}{\mathrm{RMF}}

\newcommand{\A}[1]{\AA{}}

\usepackage[colorlinks=true,
            linkcolor=red,
            urlcolor=blue,
            citecolor=blue]{hyperref}

\begin{document}

\title{The eROSITA Final Equatorial-Depth Survey (eFEDS):}
   \subtitle{The AGN Catalogue and its X-ray Spectral Properties}
\author{Teng~Liu\inst{\ref{in:mpe}}\thanks{email: liu@mpe.mpg.de}\and
Johannes~Buchner\inst{\ref{in:mpe}}\and
Kirpal~Nandra\inst{\ref{in:mpe}}\and
Andrea~Merloni\inst{\ref{in:mpe}}\and
Tom~Dwelly\inst{\ref{in:mpe}}\and
Jeremy~S.~Sanders\inst{\ref{in:mpe}}\and
Mara~Salvato\inst{\ref{in:mpe}}\and
Riccardo~Arcodia\inst{\ref{in:mpe}}\and
Marcella~Brusa\inst{\ref{in:bologna1},\ref{in:bologna2}}\and
Julien~Wolf\inst{\ref{in:mpe}}\and
Antonis~Georgakakis\inst{\ref{in:athens}}\and
Thomas~Boller\inst{\ref{in:mpe}}\and
Mirko~Krumpe\inst{\ref{in:Potsdam}}\and
Georg~Lamer\inst{\ref{in:Potsdam}}\and
Sophia~Waddell\inst{\ref{in:mpe}}\and
Tanya~Urrutia\inst{\ref{in:Potsdam}}\and
Axel~Schwope\inst{\ref{in:Potsdam}}\and
Jan~Robrade\inst{\ref{in:hamburg}}\and
J\"orn~Wilms\inst{\ref{in:remeis}}\and
Thomas~Dauser\inst{\ref{in:remeis}}\and
Johan~Comparat\inst{\ref{in:mpe}}\and
Yoshiki~Toba\inst{\ref{in:Kyoto},\ref{in:Taiwan},\ref{in:Ehime}}\and
Kohei~Ichikawa\inst{\ref{in:Tohoku1},\ref{in:Tohoku2},\ref{in:mpe}}\and
Kazushi~Iwasawa\inst{\ref{in:icc}}\and
Yue~Shen\inst{\ref{in:illi1},\ref{in:illi2}}\and
Hector~Ibarra~Medel\inst{\ref{in:illi1},\ref{in:illi2}}
}
\institute{Max-Planck-Institut f\"ur extraterrestrische Physik, Giessenbachstra{\ss}e 1, D-85748 Garching bei M\"unchen, Germany \label{in:mpe}
\and Dipartimento di Fisica e Astronomia, Universit\`a di Bologna, Via Piero Gobetti 93/2, 40129 Bologna, Italy\label{in:bologna1}
\and INAF - Osservatorio di Astrofisica e Scienza dello Spazio di Bologna, Via Piero Gobetti 93/3, 40129 Bologna, Italy\label{in:bologna2}
\and Institute for Astronomy and Astrophysics, National Observatory of Athens, V. Paulou and I. Metaxa 11532, Greece\label{in:athens}
 \and Leibniz-Institut für Astrophysik Potsdam, An der Sternwarte 16, 14482 Potsdam, Germany\label{in:Potsdam}
\and Hamburger Sternwarte, University of Hamburg, Gojenbergsweg 112, 21029 Hamburg, Germany\label{in:hamburg}
\and Dr.~Karl Remeis-Sternwarte \& Erlangen Centre for Astroparticle Physics, Sternwartstr.~7, 96049 Bamberg, Germany\label{in:remeis}
\and Department of Astronomy, Kyoto University, Kitashirakawa-Oiwake-cho, Sakyo-ku, Kyoto 606-8502, Japan\label{in:Kyoto}
\and Academia Sinica Institute of Astronomy and Astrophysics, 11F of Astronomy-Mathematics Building, AS/NTU, No. 1, Section 4, Roosevelt Road, Taipei 10617, Taiwan\label{in:Taiwan}
\and Research Center for Space and Cosmic Evolution, Ehime University, 2-5 Bunkyo-cho, Matsuyama, Ehime 790-8577, Japan\label{in:Ehime}
  \and Frontier Research Institute for Interdisciplinary Sciences, Tohoku University, Sendai 980-8578, Japan\label{in:Tohoku1}
  \and Astronomical Institute, Tohoku University, Aramaki, Aoba-ku, Sendai, Miyagi 980-8578, Japan\label{in:Tohoku2}
  \and  ICREA and Institut de Ciències del Cosmos (ICC), Universitat de Barcelona (IEEC-UB), Martí i Franquès 1, 08028 Barcelona, Spain\label{in:icc}
\and Department of Astronomy, University of Illinois at Urbana-Champaign, Urbana, IL 61801, USA\label{in:illi1}
  \and National Center for Supercomputing Applications, University of Illinois at Urbana-Champaign, Urbana, IL 61801, USA\label{in:illi2}
}

  \date{} 
  \abstract
   {After the successful launch of the Spectrum-Roentgen-Gamma (SRG) mission in July 2019, eROSITA, the soft X-ray instrument aboard SRG, performed scanning observations of a large contiguous field, namely the eROSITA Final Equatorial Depth Survey (eFEDS), ahead of the planned four-year all-sky survey. eFEDS yielded a large sample of X-ray sources with very-rich multi-band photometric and spectroscopic coverage.}
   {We present here the eFEDS Active Galactic Nuclei (AGN) catalog and the eROSITA X-ray spectral properties of the eFEDS sources.}
   {Using a Bayesian method, we perform a systematic X-ray spectral analysis for all eFEDS sources.
     The appropriate model is chosen based on the source classification and the spectral quality, and, in the case of AGN, including the possibility of intrinsic (rest-frame) absorption and/or soft excess emission. Hierarchical Bayesian modelling (HBM) is used to estimate the spectral parameter distribution of the sample.}
   {X-ray spectral properties are presented for all eFEDS X-ray sources. There are $21952$ candidate AGN, which comprise 79\% of the eFEDS sample. 
     Despite a large number of faint sources with low photon counts, our spectral fitting provides meaningful measurements of fluxes, luminosities, and spectral shapes for a majority of the sources.
     This AGN catalog is dominated by X-ray unobscured sources, with an obscured ($\log$\N{H}$>$21.5) fraction of 10\% derived by HBM. The power-law slope of the catalog can be described by a Gaussian distribution of 1.94$\pm$0.22.
     Above a photon counts threshold of $500$, nine out of $50$ AGN have soft excess detected.
     For the sources with blue UV to optical color (type-I AGN), the X-ray emission is well correlated with the UV emission with the usual anti-correlation between the X-ray to UV spectral slope $\alpha_{OX}$ and the UV luminosity.
     }
   {}

\titlerunning{eFEDS AGN \& X-ray spectra}
\authorrunning{Liu et al.}
   \maketitle

\section{Introduction}
Among current imaging X-ray telescopes, eROSITA, launched on July 13, 2019 aboard the Spectrum-Roentgen-Gamma (SRG) mission has the largest grasp in the 0.3-3.5 keV band (Predehl  et  al.  2021).
Working in a continuous scanning mode, it is currently surveying the X-ray sky with high efficiency and is expected to detect millions of Active Galactic Nuclei (AGN) in the planned 8-pass, four-year eROSITA all-sky survey (eRASS:8; Predehl et al. 2021).
Meanwhile, it simultaneously provides X-ray spectroscopy with CCD energy-resolution over the 0.2-8 keV band. These spectra can be used to investigate the physical properties of large samples of AGN, as well as of other classes of X-ray sources.

During the SRG Performance Verification (PV) phase, four days of observations were dedicated to the eROSITA Final Equatorial Depth Survey (eFEDS; Brunner et al. submitted), reaching about 50\% deeper than the nominal exposure depth of the four-year eRASS:8.
The eFEDS field is a large extragalactic field (total area 142 deg$^2$) centered at RA=136\arcdeg, Dec=1.5\arcdeg (Galactic $b$=30\arcdeg) with extremely rich multi-wavelength coverage (Salvato et al. 2021, submitted).
eFEDS was designed to verify the survey capabilities of eROSITA in a number of different ways, and to test the science workflow in anticipation of the all-sky survey. This work describes the current status of the eROSITA X-ray spectral analysis pipeline and presents a catalog of the X-ray spectral properties of the eFEDS sources.

In the X-ray sky, AGN largely outshine and outnumber other types of astronomical objects \citep[e.g.,][]{Brandt2015}.
In the past two decades, \XMM{} and \Chandra{} have surveyed a number of particularly selected contiguous fields with various area and depth (see a summary in Brunner et al. submitted), from the deepest and smallest 7Ms \Chandra{} deep field south survey \citep[CDFS;][]{Luo2017} to the widest XMM-XXL surveys \citep[e.g.,][]{Pierre2016}.
AGN are always the dominant population in these extragalactic X-ray surveys, with non-active galaxies outnumbering AGN only at extremely low fluxes (0.5--2 keV flux below $10^{-17}$ \egs; to date only reached in the CDFS).
The eFEDS survey is relatively shallow, but covers a much larger area than these previous surveys (Brunner et al. submitted). It provides a larger X-ray catalog than any previous contiguous X-ray field  and a better observational coverage of bright AGN, which have a small number density.
Moreover, because of the relatively-high X-ray flux limit, the rich multi-wavelength imaging and spectroscopic data in this field have allowed Salvato et al. (submitted) to identify the optical counterparts for a large majority of the X-ray sources, and to derive their spectroscopic or photometric redshifts.

In this work, we present the AGN catalog selected from the eFEDS X-ray sources, which can be considered a prototype of the future eRASS:8 multi-million AGN catalog, and study their properties based on eROSITA X-ray spectral analysis.
It is a common choice to perform spectral analysis only for bright X-ray sources with a reasonable photon counts \citep[e.g.,][]{Liu2017}, because maximum-likelihood-based spectral fitting techniques do not work in the low-counts regime.
Instead, we analyze the spectra of {\it all} the eFEDS sources in this work using a Bayesian method.
In so doing, we can explore the lower limit of spectral constraining capability of eROSITA.
For the faintest sources, spectral analysis is expected to provide only a measurement of flux.
For the majority of the sources we can adopt simple, single-component spectral models. For the brightest sources, on the other hand, we can test if additional spectral components are detected.
In the spectral analysis, we adopt the WMAP cosmology with $\rm \Omega_{\Lambda}$ = 0.7 and $H_{0}$ = 70 km $\rm s^{-1}$ $\rm Mpc^{-1}$, and adopt the \citet{Verner1996} photoionization cross sections and the \citet{Wilms2000} abundances for absorption.

As the first systematical analysis of eROSITA spectra, in this work we test and demonstrate the performance of eROSITA X-ray spectroscopy and introduce the relevant software.
\S~\ref{sec:data} introduces the eFEDS AGN catalog and the eROSITA spectra extraction and stacking.
\S~\ref{sec:specfit} describes our spectral analysis methods.
\S~\ref{sec:results} presents X-ray spectral properties and the UV/optical luminosities of the eFEDS AGN.

\section{Catalog and X-ray spectra}
\label{sec:data}

\subsection{The eFEDS AGN catalog}
\begin{figure}[hp]
\begin{center}
\includegraphics[width=0.7\columnwidth]{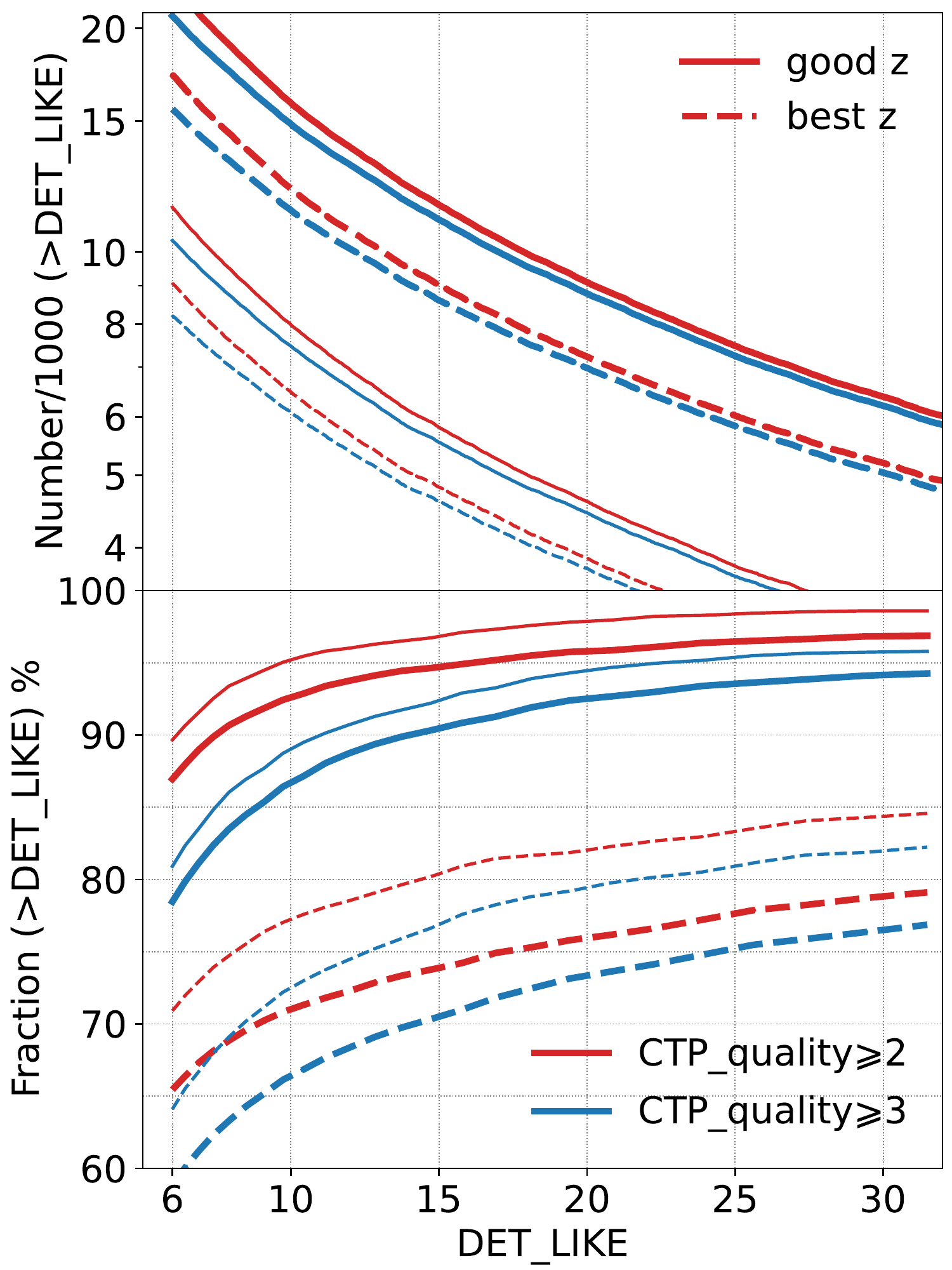}
\caption{Number of X-ray sources (upper panel) and fractions of sources (lower) resulted from counterpart quality and redshift quality selections.
  The red and blue lines indicate counterpart quality $\geqslant$2 and $\geqslant$3, respectively.
  The solid and dashed lines indicate good ($zG\geqslant$3) and best ($zG\geqslant$4) redshift measurements, respectively.
  The thick and thin lines indicate sources in the eFEDS 90\%-area region and in the KiDS region, respectively.
}
\label{fig:ctp_frac}
\end{center}
\end{figure}

\begin{figure}[h]
\begin{center}
\includegraphics[width=0.8\columnwidth]{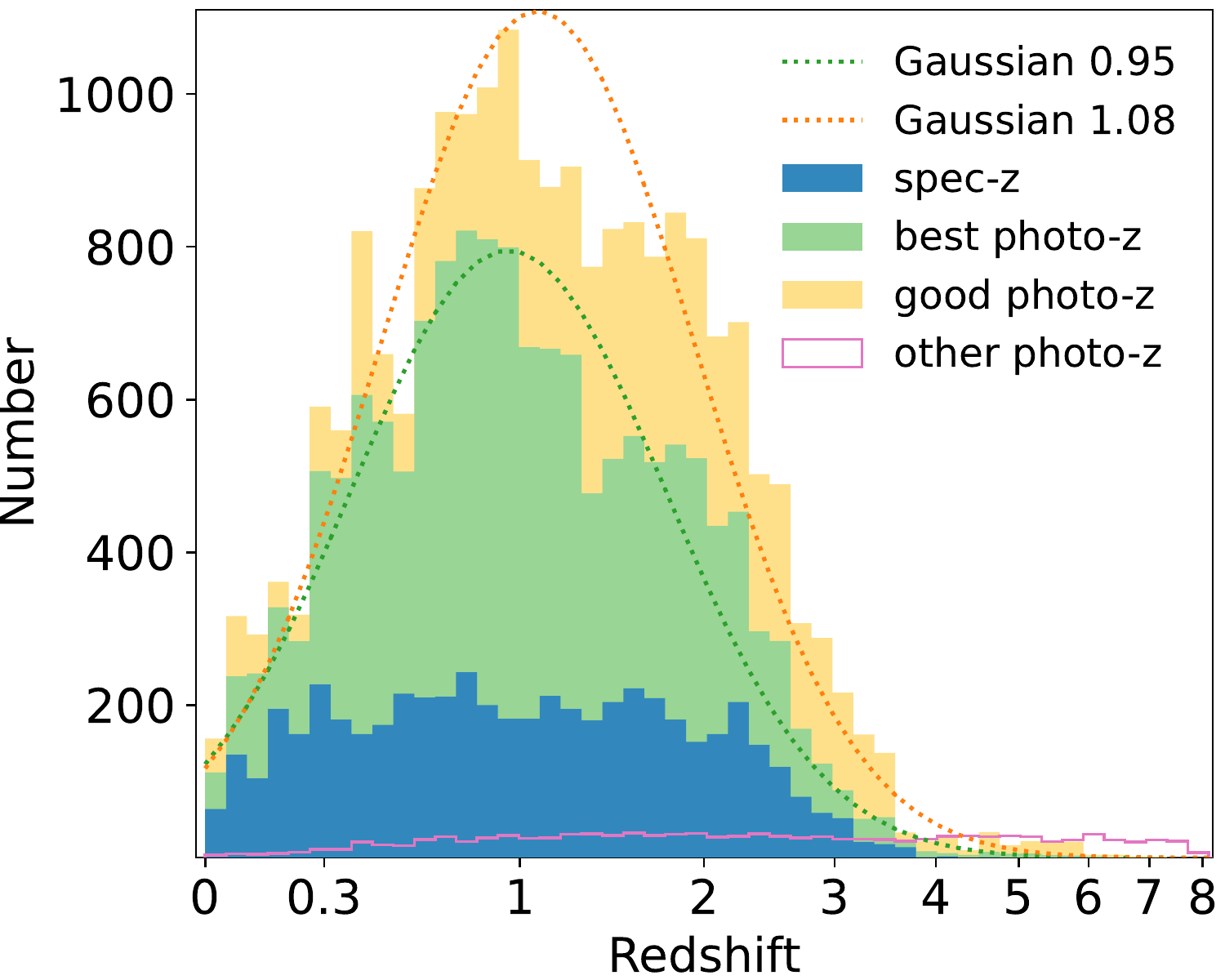}
\caption{
  The stacked filled histograms display the redshift distributions of the eFEDS AGN with spec-z ($5284$; $zG=$5; in blue), with best photo-z ($9668$; $zG=$4; in green), and with good photo-z ($5898$; $zG=$3; in yellow), respectively.
  The other AGN ($1102$; $zG<$3) are displayed in magenta empty histogram.
  For sources with photo-z, the redshift PDF is used in plotting.
  For comparison, we plot the Gaussian distributions centered at $z=0.95$ (green dotted line) and $z=1.08$ (orange dotted line), which are normalized to the number of AGN with best redshift measurements ($zG\geqslant$4) and good measurements ($zG\geqslant$3), respectively. Both of them have $\sigma=0.15$ in the space of $\log(1+z)$.
  }
\label{fig:hist_photz}
\end{center}
\end{figure}
  Brunner et al. (submitted; Paper I) presented the eFEDS main X-ray catalog.
  It contains $27910$ sources detected in the 0.2--2.3 keV band form the whole eFEDS region, most ($>$98\%) of which are point sources (unresolved, extent likelihood $=$0).
About $\sim3\%$ of the X-ray sources are located at the field border, where the data suffers from shorter exposure, stronger vignetting, and higher background.
With such sources excluded, the inner region of eFEDS, which comprise 90\% of the total area and has a relatively-flat sensitivity distribution, is recommended for AGN demography studies (Paper I).
  Salvato et al. (submitted; Paper II) identified the optical counterparts of the point sources from the DESI Legacy Imaging Survey DR8 (LS8; Dey et al. 2019) catalog, which comes with Gaia (Gaia collaboration et al. 2020) and WISE \citep{Lang2014} photometry.
  Each counterpart has a quality flag \texttt{CTP\_quality}.
The counterparts with \texttt{CTP\_quality} $\geqslant$3 are considered highly reliable, in the sense that two independent methods of counterpart identifications, a Bayesian method and a maximum-likelihood method, agree on the results.
The counterparts with \texttt{CTP\_quality}$=$2 are also relatively reliable, except that secondary counterpart candidates exist for such sources.

The eFEDS field has been observed by several spectroscopic surveys (Paper II). The SDSS I-IV \citep{Ahumada2020} survey provides the largest number of spectra over the eFEDs area (more than 60 thousand). Observations were carried out at the Apache Point Observatory \citep{Gunn2006} with the BOSS spectrograph \citep{Smee2013}. In addition to the public data from SDSS phases I-IV \citep{Ahumada2020}, in the SPIDERS program \citep{Dwelly2017,Comparat2020}, part of SDSS-IV \citep{Dawson2016,Blanton2017}, a dedicated campaign was performed in Spring 2020 to observe eFEDS X-ray sources. This data set will be part of the upcoming SDSS DR17, and the observations described in detail in Merloni et al. (in prep.). Paper II has collected the spectroscopic redshift (spec-z) measurements from all the available surveys, and carefully selected the high-quality ones.
The eFEDS field also has rich multi-band photometry coverage. In addition to LS8, Gaia, and WISE, it is also partly covered by the Galex survey, the Kilo-degree Survey (KiDS), the Viking survey, the VISTA/VHS survey, and the UKIDSS survey (see Paper II for more details).
Particularly, high-resolution photometry was also obtained with the Hyper Suprime-Cam \citep[HSC;][]{Miyazaki2018} Program \citep[HSC–SSP;][]{Aihara2018}; its S19A photometry data \citep[][Toba et al. (submitted)]{Aihara2019} was used in Paper II to construct the SED.
According to optical spectra or SED, Paper II classified each counterpart as galactic or extragalactic sources.

In this paper, we present the eFEDS AGN catalog ($21952$ sources), which is selected from the eFEDS main X-ray catalog as the point sources with \texttt{CTP\_quality}$\geqslant$2 and having the counterpart classified as either ``Secure'' or ``Likely'' extragalactic in Paper II.
This catalog contains $687$ sources located outside the inner 90\%-area region, which can be excluded when necessary with the \texttt{inArea90} flag.
It also includes a small number of normal galaxies at the lowest redshifts, which can be excluded based on their low X-ray luminosities (\S~\ref{sec:AGNLum}).

Fig.~\ref{fig:ctp_frac} displays the sample size and optical-classification completeness as a function of X-ray source detection likelihood, considering only the $26488$ point sources in the inner 90\%-area region.
The \texttt{CTP\_quality}$\geqslant$2 threshold corresponds to a completeness of $87\%$.
If selecting a subsample with X-ray detection likelihood $>$10 or $>$15, this counterpart completeness can be increased to $93\%$ and $95\%$, respectively.

Instead of limiting spectral analysis to the AGN catalog, which comprises $79\%$ of the whole X-ray catalog, we analyze and present the spectral fitting results for the whole X-ray catalog.
This is because all the sources must be considered during the spectra extraction for cleaning reason, and considering the incompleteness of AGN selection and the potential cases of mis-classifications, the spectral properties for the sources outside the current AGN catalog could be useful in the future when more multi-wavelength observations are available.
Among the sources with reliable counterparts (\texttt{CTP\_quality}$\geqslant$2), $2822$ have the counterparts classified as either ``Secure'' or ``Likely'' galactic. A very small number of galactic compact objects may also be in this category.
In the spectral analysis, we treat these galactic sources as stars and treat all the other sources as AGN.
Finally, the eFEDS main catalog includes also $541$ extended sources, which are candidates galaxy clusters.
We refer to Liu, A. et al. (submitted) for their spectral properties, where they are properly analyzed as galaxy clusters. 
The spectral properties of these extended sources presented in this work are only valid if the source is in fact an AGN mis-classified as extended source.

Based on the rich multi-band photometry data, Paper II measured the photometric redshift (photo-z) of all the sources through SED fitting.
High-quality spec-z is adopted when available with the highest priority, and the photo-z is adopted when spec-z is not available.
A redshift grade ($zG$) is provided for each source in Paper II.
The highest $zG$ of 5 corresponds to spec-z.
The grade $zG$=4 indicates photo-z measurement with the highest reliability, in the sense that an independent deep-learning based method results in a consistent photo-z measurement with the SED fitting.
The photo-z measurements with $zG=$3 are also considered highly reliable, even though not confirmed by another independent method. Hereafter, we consider the sources with $zG\geqslant$3 as having ``good'' redshift measurements.
Among the $21952$ eFEDS AGN, $5284$ have high-quality spec-z and $20850$ have ``good'' redshifts.
The completeness of redshift measurement is also displayed in Fig.~\ref{fig:ctp_frac}.
If a higher completeness of ``good'' redshift measurements is needed, one could select a subsample inside the region of the KiDS survey, where the photometry data from the KiDS and Viking surveys improve the photo-z measurements significantly.

Fig.~\ref{fig:hist_photz} displays redshift distribution of the AGN catalog, in which the probability distribution function (PDF) is considered in the cases of photo-z.
The redshift distribution peaks around redshift 1. High-z sources with $z>$4 are rare.
For sources relying on photo-z, we adopt the photo-z redshift estimate without propagating its uncertainty in the spectral fitting. This is because the relatively-simple spectral models cannot inform the redshift better than the the multi-wavelength photo-z, and error propagation on important parameters can be also performed post-hoc, if needed.

\subsection{Extraction of Spectra} \label{sec:extract}
The observation mode of eROSITA is continuous scanning of the sky.
This is different to previous large X-ray surveys, which were carried out with pointed observations in raster patterns. In scanning mode, a source moves across the Field Of View (FOV) multiple times rather than staying at a particular position in the FOV, and at the same position on the detector. Thus, during a scanning observation, each source is exposed to a local, effective exposure time corresponding to the duration of its passage in the FOV; during this effective exposure time, the source's signal is subject to varying Point Spread Function (PSF) and vignetting. The treatments of the data required by the eROSITA scanning mode are implemented in the eROSITA Science Analysis Software System (eSASS; version eSASS\_users201009; Paper I) task \texttt{srctool} (V1.63 used in this work), which creates spectral files of the OGIP format (OGIP/92-007, Arnaud et al. 1992; OGIP/92-002, George et al. 2007).
We introduce here the eROSITA scanning-mode spectral extraction, taking eFEDS as an example.

\subsubsection{Source and background regions} \label{sec:regions}
Source and background extraction regions are automatically defined by the versatile \texttt{srctool} task. 
The algorithm for building the extraction regions is described below. 
The extraction regions vary for each source depending on the source counts, background counts, and source extent model radius from the detection catalog. The local eROSITA PSF at 1 keV (\texttt{PSF\_ENERGY\_KEV}) is used throughout as the reference scale.

The source extraction region is chosen as a circle with a radius that maximises the nominal signal to noise ratio (S/N) given the local background surface brightness, clipped to a minimum radius of 10\arcsec (\texttt{MINIMUM\_SOURCE\_RADIUS} parameter) and a maximum radius of the 99\% energy enclosed fraction (EEF) radius of the PSF.
To remove contamination of a nearby source from the source extraction region, we compare the surface brightness along the line joining the two sources, and exclude a circular region around the contaminating source out to a radius where the PSF surface brightness of the confusing source is greater than 20\% (\texttt{MAX\_CONF\_MAP\_TO\_SRC\_MAP\_RATIO}) of the target source surface brightness. It is clipped to a minimum of $5\arcsec$ and a maximum of 99\% EEF radius. No exclusion zones are allowed to be centred within 10$\arcsec$ (\texttt{MINIMUM\_EXCLUDE\_DIST}) from the source.

\begin{figure}[hptb]
\begin{center}
  \includegraphics[width=0.8\columnwidth]{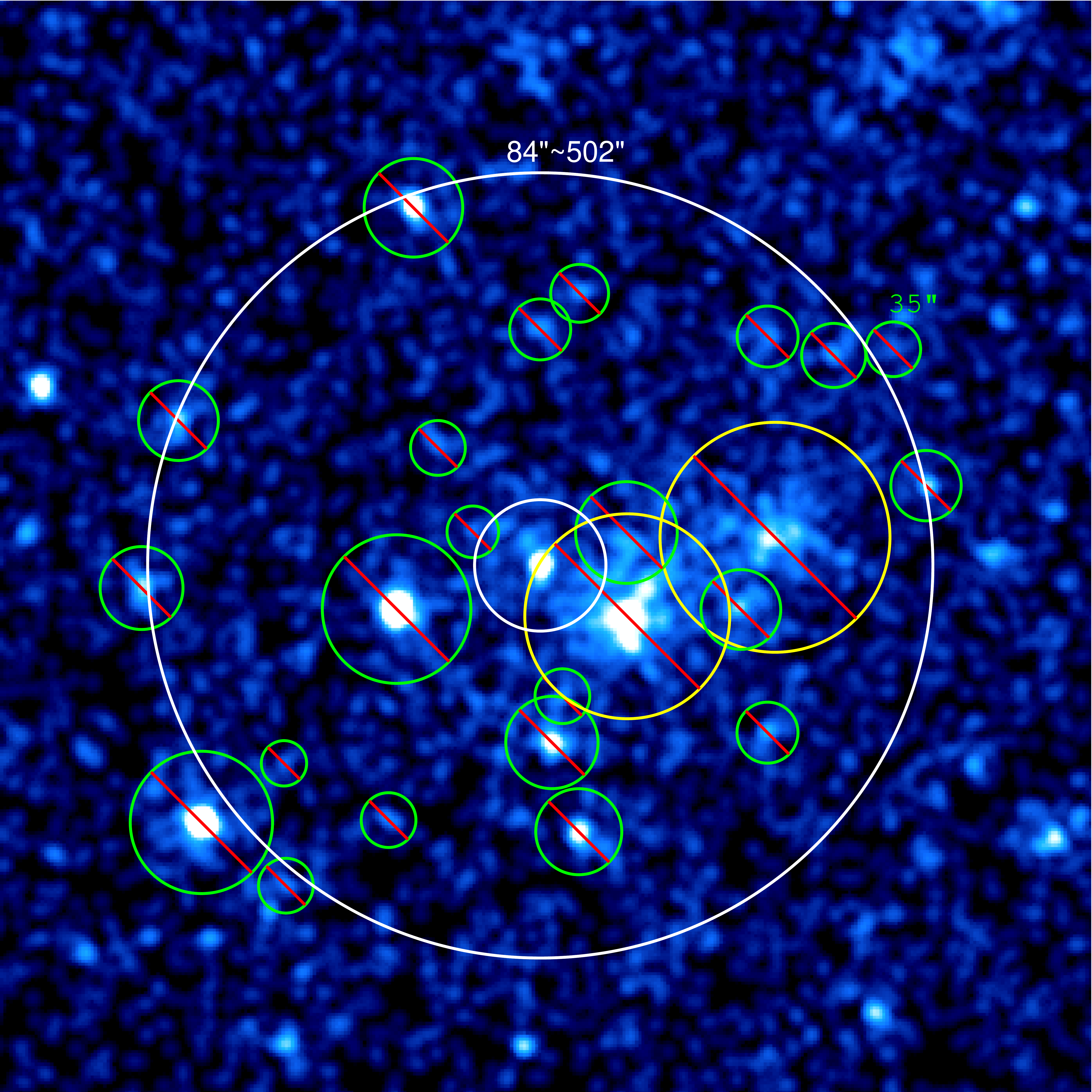}
\caption{An example of a background extraction region overlaid on the 0.2--2.3 keV image: for source 1702 (at the center), the background extraction region is defined by the area contained between the white annuli, after excluding nearby sources (circles with red stripes). The green and yellow circles indicate point- and extended sources, respectively.
  Physical scales are printed for the annuli and a source in terms of radii.
  }
\label{fig:extreg}
\end{center}
\end{figure}

We adopt an annular background extraction region. An example is shown in Fig.~\ref{fig:extreg}.
The inner radius is determined by increasing the radius step by step starting from twice (\texttt{INITIAL\_SRC\_R\_TO\_BACK\_R1}) the source extraction radius until the target source's surface brightness is less than 5\% (\texttt{MAX\_SRC\_MAP\_TO\_BG\_MAP\_RATIO}) of the local background surface brightness, adopting a maximum of three times (\texttt{MAX\_RATIO\_BACK\_R1\_TO\_RADIUS\_99PC}) the 99\% EEF radius of the PSF.
With the inner radius determined, the outer radius determines the geometric area of the background extraction region. 
In order to to sample the background spectrum with a good S/N in a sufficiently large area, we set the outer radius to a value corresponding to a background area that is $200$ times \footnote{The default value of \texttt{srctool} V1.63 is 150. In a field with deeper exposure, a smaller area is needed to obtain a well-sampled background spectrum.} (\texttt{BACK\_TO\_SRC\_AREA\_RATIO}) the source extraction area after excluding nearby sources from the background region.
Similarly, to remove contamination of a nearby source from the background extraction region we calculate an exclusion radius where the contaminating source surface brightness is 10\% (\texttt{MAX\_CONF\_MAP\_TO\_BACK\_MAP\_RATIO}) of the local background surface brightness.
The exclusion radius is clipped to a minimum of $5\arcsec$ and a maximum of 99\% EEF radius.
Meanwhile, it is restricted to be smaller than the distance from the target source, so that in the case of a target point source inside a big cluster, a part of the diffuse emission of the cluster near the target source position will be included in the background (see Fig.~\ref{fig:extreg} for an example).
We further apply a maximum outer radius of $15\arcmin$ to the \texttt{srctool} V1.63 output, so that background signal is always extracted near the source position.

\begin{figure}[htbp]
\begin{center}
  \includegraphics[width=\columnwidth]{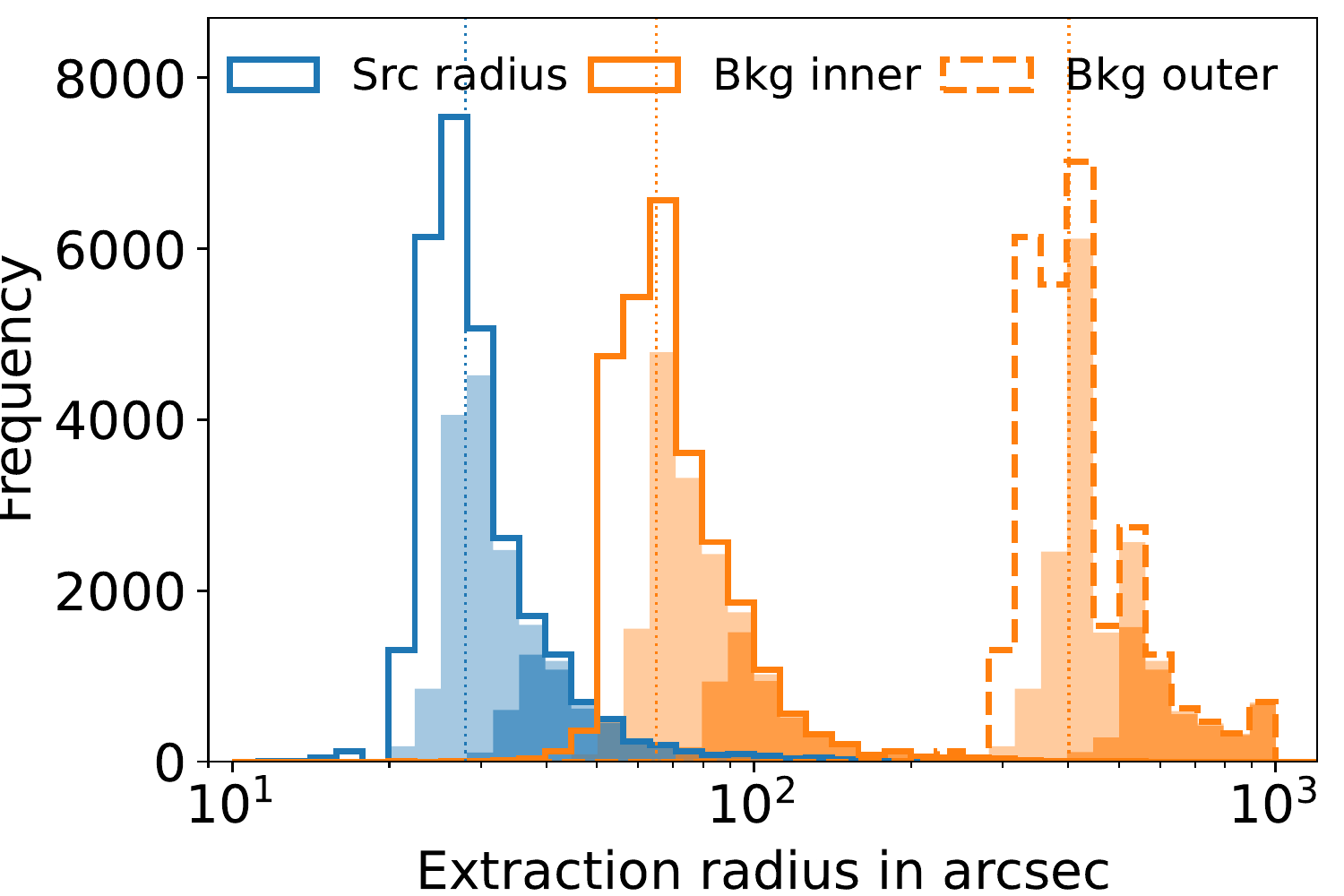}
  \includegraphics[width=\columnwidth]{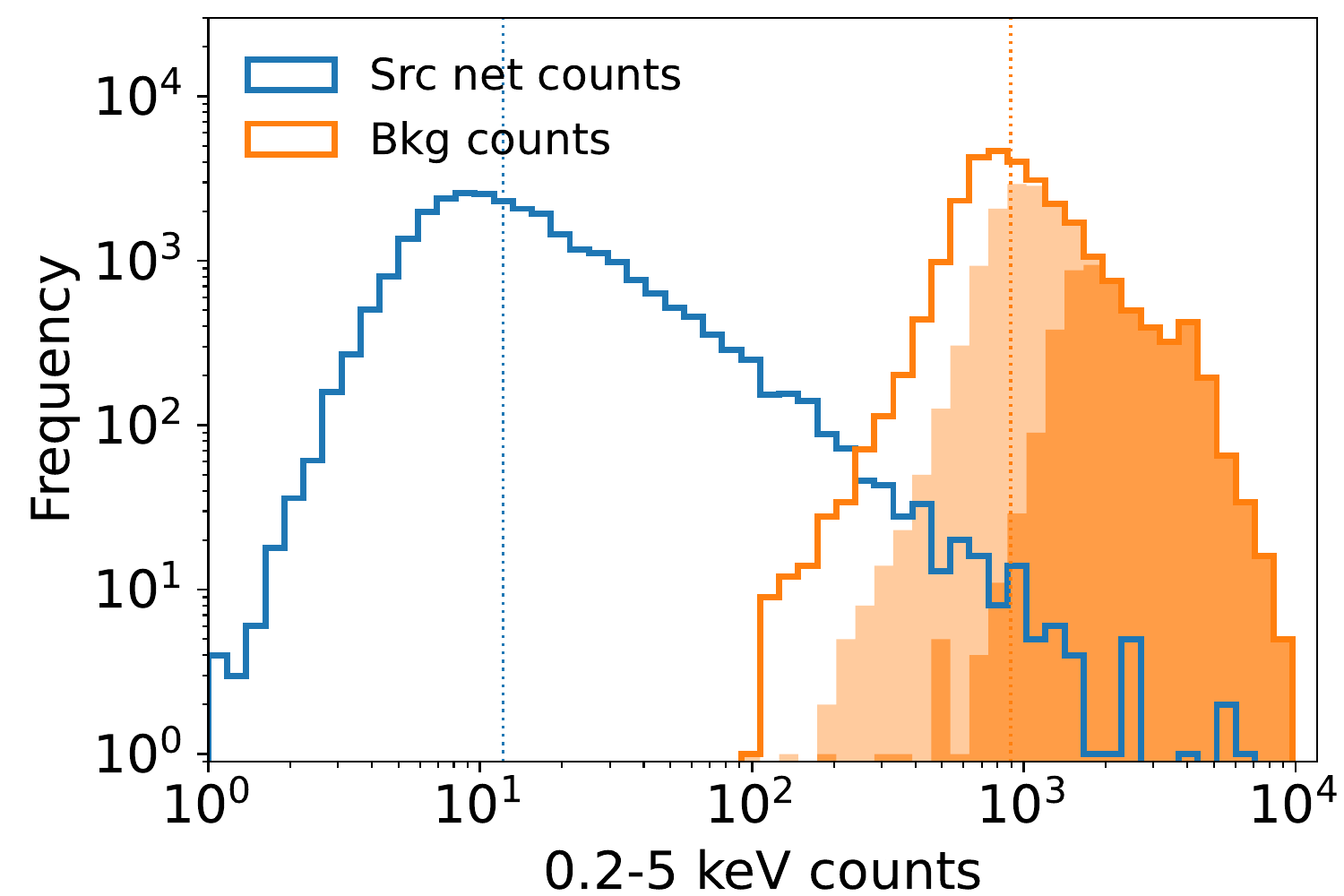}
\caption{
  The distributions of source (blue) and background (orange) extraction radii (upper panel) and the distributions of source net counts and background counts in the 0.2--5 keV band (lower panel). The median values of the whole sample are marked with dotted vertical lines.
  The two-level shaded regions indicate the subsamples of the sources with source net counts $\geqslant 10$ and $\geqslant30$, respectively.
}
\label{fig:radii_cts}
\end{center}
\end{figure}

Fig.~\ref{fig:radii_cts} displays the distributions of source (blue) and background (orange) extraction radii and spectra counts.
We measure the spectral source counts in the 0.2--5 keV band instead of the full 0.2-8 keV band because most of the sources have no signal above 5 keV.
Fainter sources have smaller source extraction regions to gain a better S/N, and consequently smaller background extraction regions.
The median of source radius, background inner radius, and background outer radius of the whole sample are ($28\arcsec$, $65\arcsec$, $401\arcsec$), respectively.
If selecting only the sources with at least 10 or 30 spectral net counts in the 0.2--5 keV band, these median values are increased to ($31\arcsec$, $74\arcsec$, $444\arcsec$), and ($41\arcsec$, $100\arcsec$, $588\arcsec$), respectively.
We adopted a large ratio ($200$) of background to source area in order to guarantee a well-sampled background spectrum with a large number of counts.
For the full sample, the median number of 0.2--5 keV background counts is $896$; selecting sources with source net counts $>10$ and $>30$ lead to median background counts of $1140$ and $2006$, respectively.
For the full sample, $90\%$ sources have at most $43\%$ of the useful background spectral channels ($20\sim900$) empty.
Selecting sources with source net counts $>10$ and $>30$, this maximum fraction of empty channel in $90\%$ sources is reduced to $34\%$ and $15\%$, respectively.
The median 0.2--5 keV source net counts of the whole sample is $12$.
A total of $5005$ (18\%) sources have at least $30$ 0.2--5 keV source net counts.

\subsubsection{Creation of spectral products}
With the regions defined, the counts can be extracted from the event files. However, each spectrum must be associated with the instrument sensitivity at the relevant observing time and location. In a scanning observation like eFEDS, different sky positions correspond to different observing time, different exposure length, and different instrument responses.

For the most rigorous spectral analysis, responses for the source and background regions should be computed separately to account for differences in the response. Indeed, in scanning mode, the instrument sensitivity is time and location dependent, as sources change their angle relative to the instrument. However, some simplifications can be made.
Firstly, the point sources considered here are spatially compact, and thus the same response can be used for the source and the nearby background spectra. 
The response between the source and background regions also become similar because as a source moves through the entire FOV multiple times, the PSF and vignetting effects are averaged in a similar way across the FOV.
A significant fraction of the sources with \texttt{EXT\_LIKE} between 6 (the lower boundary used in eFEDS source detection) and 14 are in fact point sources (Liu et al. 2021). Therefore, we treat all sources with \texttt{EXT\_LIKE}$<14$ as point sources by setting their source extent to zero before extracting spectral products. The original source extents were however used to define the extraction regions in \S~\ref{sec:regions}.
For the spectral response of the detector (response matrix file, RMF), the same ground calibration is used for all Telescope Modules (TMs) \citep{Dennerl2020}. 
The effective area as a function of energy (area response file, ARF) quantifies the exposed mirror area and vignetting effects. First, ARFs are computed in grids of time and space: The good time intervals (GTIs) of each source are sampled in time steps of 5\,ms. The effective source region (inside the FOV) is sampled in steps of $9.6\arcsec$, which corresponds to the the detector pixel scale.
Then, the energy-dependent area-loss correction (\texttt{CORRPSF}) is calculated by \texttt{srctool} as the fraction of source light falling inside the source extraction region and inside the FOV, as expected by the PSF-convolved source extent model (a $\delta$ function for point sources). Using the ``2D\_PSF'' mode for unresolved sources, this correction accounts for both the PSF enclosed energy fraction and the inside-FOV fraction of the source extraction region at each position of the source track in the FOV. The vignetting is computed at the source center, and assumed to be constant over the source extent. Finally, \texttt{srctool} calculates the total correction (\texttt{CORRCOMB}) which combines the contributions. The total correction is averaged separately for outside-region-loss and vignetting and thus not necessarily the product of \texttt{CORRPSF} and \texttt{CORRCOMB}.
 
In general, the X-ray background varies also over time and space. However, the eROSITA background is very stable, with barely any background flare as commonly seen in the \XMM{} observations \citep{Predehl2021}. A short background flare is found in the eFEDS observation and removed using the eSASS task \texttt{flaregti} (Paper I). Therefore, background variability between the source and the background regions is negligible.
To make sure the source and background regions share the same background level, \texttt{srctool} extracts background signals only when the source is in the FOV, so that the source and background spectra are always exposed at the same time (in the GTI of the source).
We remark that the statements above are valid only because we extract background in an annulus region near the source; in the case of large extended sources, when background has to be extracted from a region far away, the background responses should be extracted separately.

\subsubsection{Spectra stacking}
All seven roughly-identical TMs of eROSITA were activated for almost the entire eFEDS observation.
For each source, \texttt{srctool} extracts a source spectrum, corresponding response files, and a background spectrum for each TM. These are then summed. The combined spectrum is the equally weighted sum of the spectra from the individual telescopes. 
For each TM, the exposure time $T_i$ (\texttt{EXPOSURE} keyword) is calculated as the deadtime-corrected total exposure time inside the GTIs of the source. The exposure time of the combined spectrum is the mean of the exposure times over the seven TMs, with inactive TMs having zero for their exposure time.

The \texttt{BACKSCAL} keyword is the area (in square degrees) of the intersection of the extraction region with the FOV during the GTIs, which is thus smaller than the geometric area (the \texttt{REGAREA} keyword) of the extraction region in the scanning mode. The \texttt{BACKSCAL} of the combined spectrum is averaged over the activated telescopes.

The combined ARF is seven times the exposure-time weighted mean ARF of the activated telescopes
$$\ARF = 7\times\frac{\sum{\ARF_i\times T_i}}{\sum{T_i}} ,$$ where $i$ indexes the telescope units.
The summed RMF is the exposure and ARF weighted sum over the RMFs from the individual telescopes. The weighting is computed individually for each energy bin.
$$\RMF = \frac{\sum{\RMF_i(e)\times \ARF_i(e)\times T_i }}{\sum{\ARF_i(e)\times T_i}} .$$
Inactive telescopes (with $T_i =0$) contribute to the combined ARF but not RMF.

\section{Spectral analysis} \label{sec:specfit}

With the data products and instrument response in hand, we can analyse the spectra with astrophysical models.
Our automated spectral analysis procedure first characterizes the total background emission empirically (§\ref{subsec:backgroundmodel}), which is then jointly analysed with a astrophysical source spectral model (\S~\ref{sec:AGNmodel}). The fitting procedure is described in (\S~\ref{sec:pow_apec}).

\subsection{Background model}\label{subsec:backgroundmodel}
Background photons are present in both the source and background regions.
The source spectrum is composed of a source component and a background component.
We model the background spectrum (extracted from the background region), and use the best-fit model together with the area scaling factor to account for the background component in the source spectrum.

The detected X-ray background partially corresponds to celestial X-ray photons focused by the mirrors and partially corresponds to secondary emission caused by soft or hard particles hitting the detector directly.
To analyze the background spectrum, we have to model the vignetted celestial component and the unvignetted particle component separately \citep[see more discussions in][Liu et al. submitted]{Freyberg2020,Predehl2021}.
However, this is not necessary for the present work, where we are interested in analyzing the source properties, rather than the backgorund.
With the aim of measuring the background in the source region, all we need is to rescale both the background components properly from the background region to the source region.
We have chosen small source and background regions, which are nearby to each other, and extracted both the source and background spectra in the source GTI.
Thus they have exactly the same exposure time and approximately the same response.
Therefore, both the vignetted and unvignetted background components can be rescaled using the ratio of \texttt{BACKSCAL} between the source and the background regions.

We use the automatic background fitting method described in the appendix of \citet{Simmonds2018} and implemented in BXA \citep{Buchner2014}. In this method, the background spectrum is modelled phenomenologically as a function of detector channels (instead of energy). Briefly, principal component analysis (PCA) is run on the unbinned background spectra of all the eFEDS sources, after a $\log(1+\mathrm{counts})$ transformation. The first six principal components (PCs) are then linearly combined to fit any particular background spectrum of interest.
An individual background spectrum does not necessarily show all the features of the six PCs. Starting from the mean spectrum, PCs are iteratively added as long as the Akaike information criterion \citep[AIC;][]{Akaike1974} of the fit is significantly improved.
After finding the linear combination of PCs that describe the spectrum best, Gaussian lines are added (in count-space) as long as they improve the fit further. These added Gaussians can model features that might appear in some individual spectra and were missed by the PCA.
Finally, the best-fit background model spectrum is converted into \texttt{XSPEC} table model, which is dedicated only to this particular source. The model has a scale parameter, which should be set to the ratio between \texttt{BACKSCAL}
of source and background when fitting the source spectrum. It also has a normalization parameter, which is expected to be unity in the simultaneous source and background fitting, in the sense that it is only determined by the background spectrum.

\begin{figure}[htbp]
\begin{center}
  \includegraphics[width=\columnwidth]{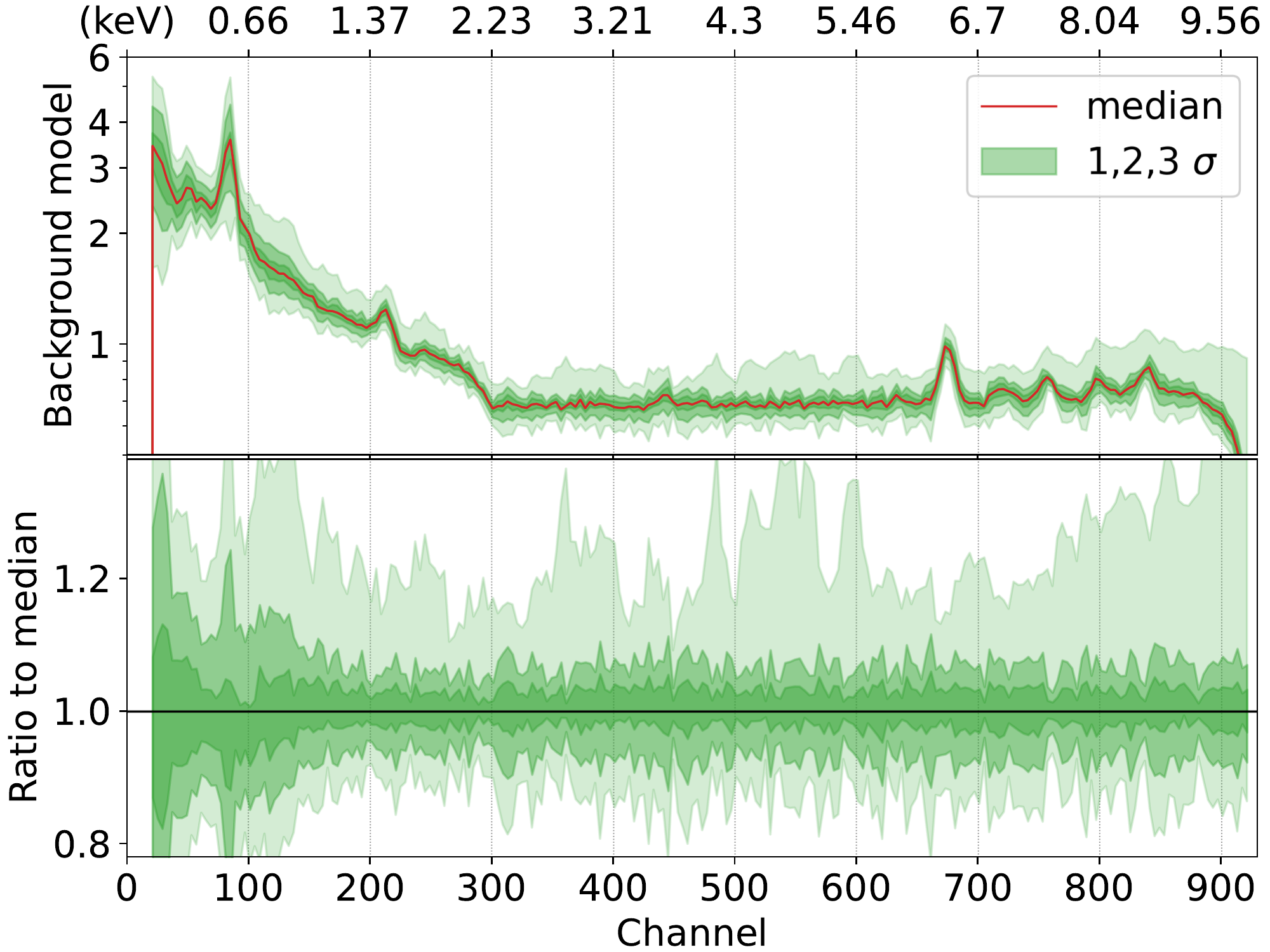}
  \caption{
    The background model normalized to a mean value of 1. The red line displays the median of the models of all the sources, and the three-level green shaded regions indicate the 1,2,and 3 $\sigma$ percentiles. The lower panel displays the ratio to the median. 
  }
\label{fig:bkgmodels}
\end{center}
\end{figure}

Fig.~\ref{fig:bkgmodels} displays the best-fit background spectral shapes, which are functions of channel.
These spectral models are normalized to the mean value over the full range (between channel 20 and 900) and thus only show the variety of background spectral shapes among the eFEDS sources.
The background spectral shape is relatively stable across the four-days observation of eFEDS. The most variable part is at the softest energies below 1 keV ($\sim$ channel 150).
For a small fraction of sources, one or more Gaussian component are added in addition to the PCA components. They appear as the features in the 3-$\sigma$ upper percentiles of all the sources.
The mean background values over the full range, to which the models in Fig.~\ref{fig:bkgmodels} are normalized, has a mean and standard deviation of $25.2\pm1.5\times 10^{-3}$ counts~s$^{-1}$~channel$^{-1}$~deg$^{-2}$. Therefore, the background flux is also highly stable across the eFEDS observations.

\subsection{Source models}

The sources are classified as galactic or extragalactic based on their SED or optical spectroscopy in Paper II. However, the classifications of a small fraction of sources are uncertain and may need to be revised as more data is obtained in the future. To allow for this, all sources are analysed with a variety of physically motivated models appropriate for different source classes, and the derived properties are released as catalogs.

\subsubsection{Stellar models}
\begin{figure*}[!hp]
  \centering
  \vspace{1cm}
  \begin{overpic}[scale=0.25,tics=40,abs,unit=1mm]{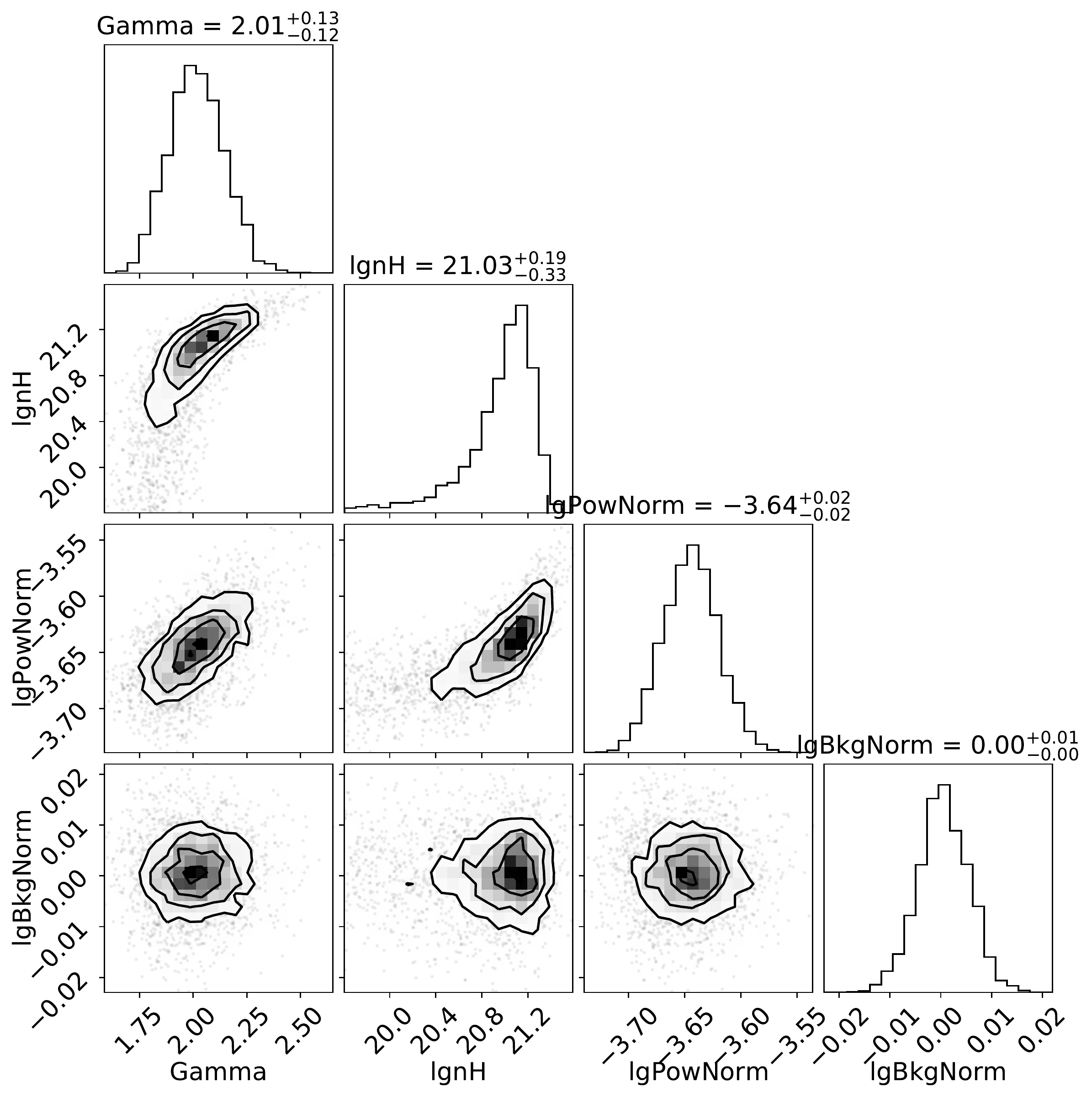}
    \put(20,50){\includegraphics[scale=0.27]{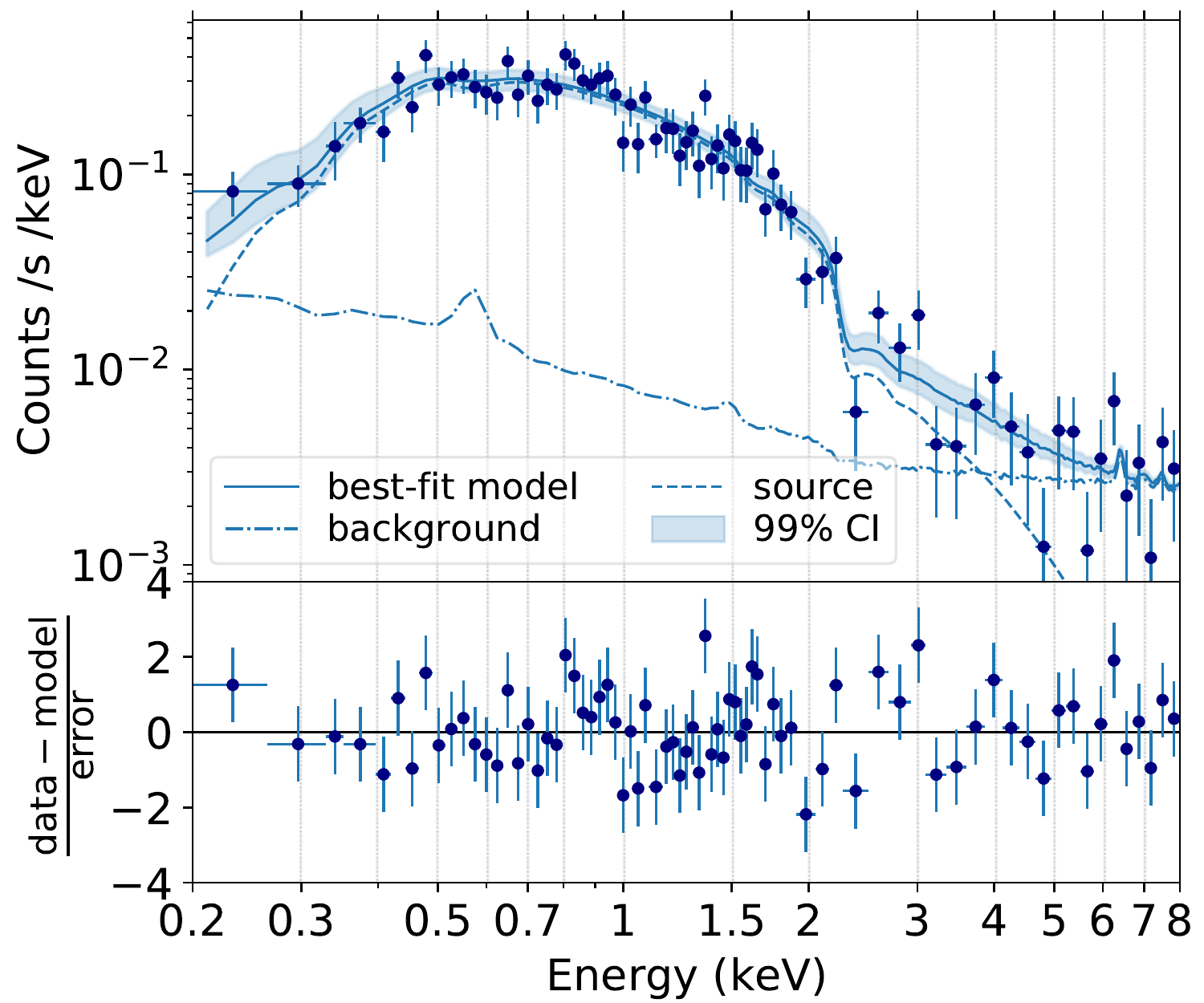}}
  \end{overpic}
  \hspace{2cm}
  \begin{overpic}[scale=0.25,tics=40,abs,unit=1mm]{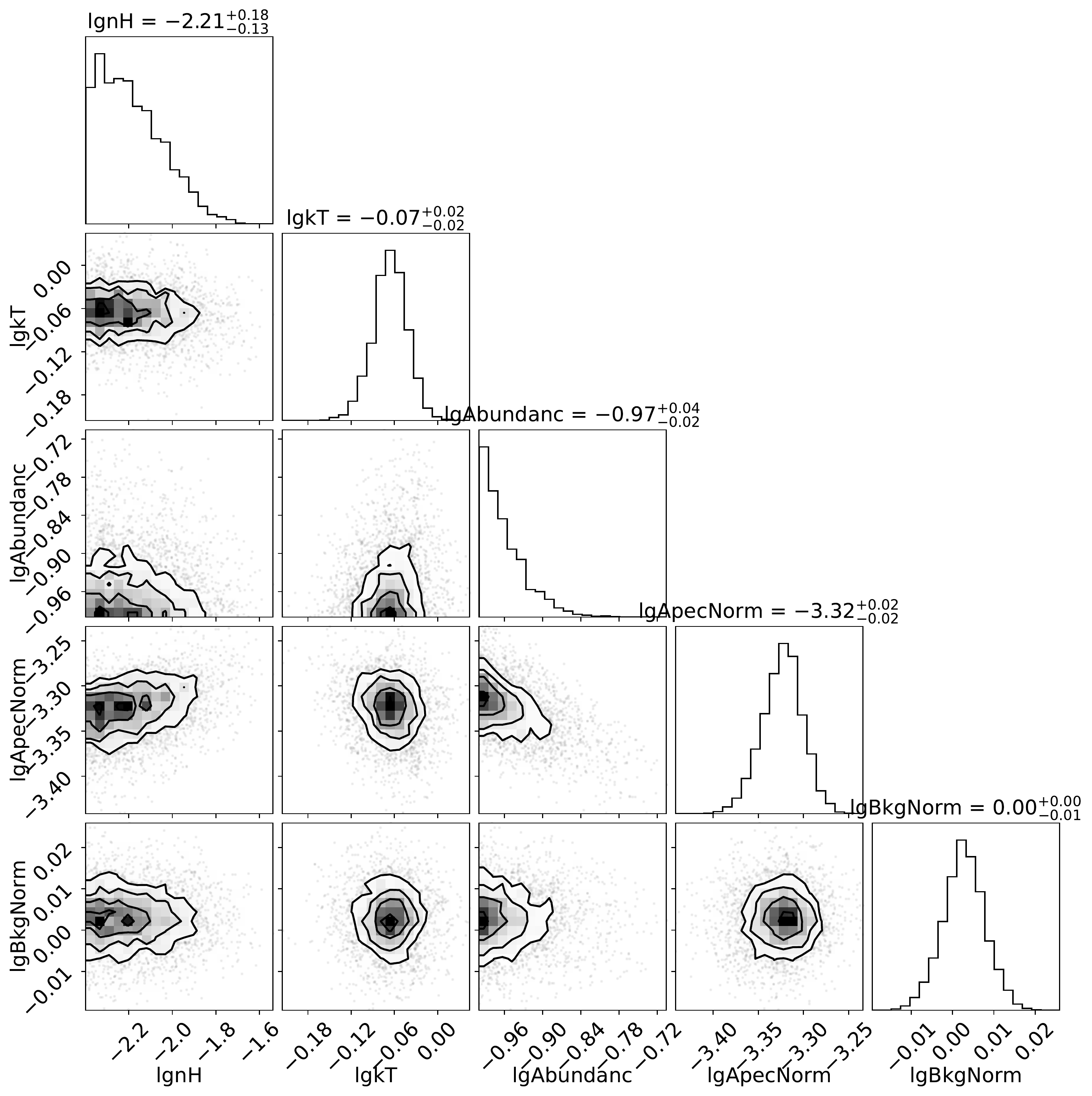}
    \put(33,50){\includegraphics[scale=.27]{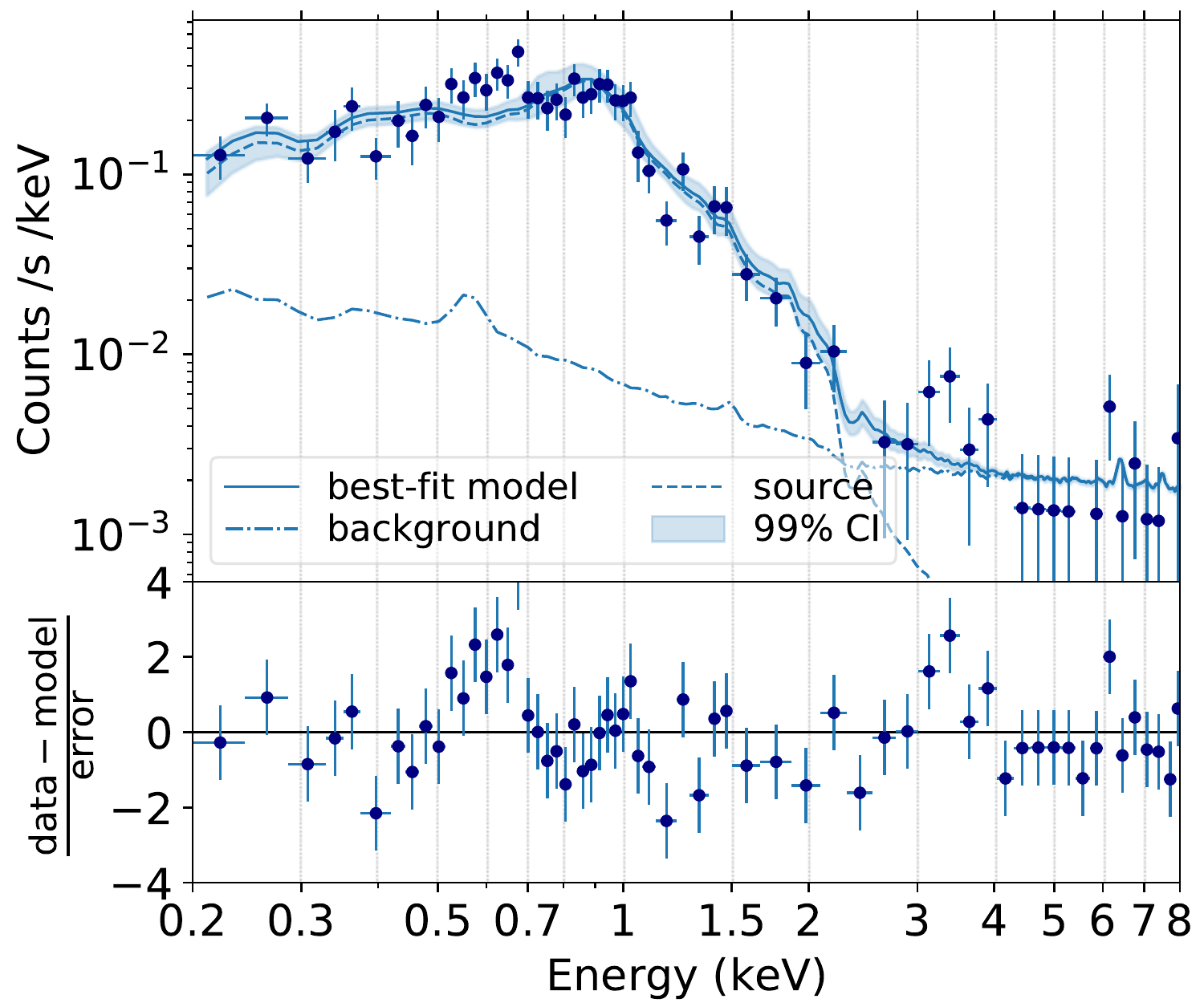}}
  \end{overpic}
  \caption{Examples: posterior parameter distributions and spectra (inlaid at upper right) for an AGN (ID$=$34) fitted with an absorbed power-law (left) and for a star (ID$=$50) fitted with an apec model (right). The posterior distribution of each individual parameter and each pair of parameters are plotted using the ``corner'' package \citep{corner}, with the median and 68\% percentile interval around the median printed on top of each individual distribution. The spectral shape parameters for the absorbed power-law model include the slope ($\Gamma$) of the power-law and the $\log$\N{H} of the intrinsic absorber (in cm$^{-2}$).
    For the apec model, the spectral shape parameters include the Galactic absorption column density $\log$\N{H}, the temperature $\log kT$ (in keV), and the abundance.
    All the normalization parameters are measured in logarithm space. The background normalization parameter (lgBkgNorm) is set free in the fitting and always equals one.
    In the spectral plot, the data (dark blue points) are rebinned just for representation to reach a S/N of 3 but allowing at most 6 adjacent bins (each bin includes 4 channels) to be grouped.
    The blue solid, dashed, and dot-dashed lines display the best-fit model folded with the instrument responses, the folded best-fit source model, and the best-fit background model rescaled to the source extraction region. The shaded region indicate the 99\% percentile confidence interval of the model around the median.
    The lower panel compares the data with the best-fit model in terms of (data-model)/error, where error is calculated as the square root of the model predicted number of counts.
\label{fig:pow_apec}
\\
\\
    }

  \begin{overpic}[scale=0.25,tics=0,percent]{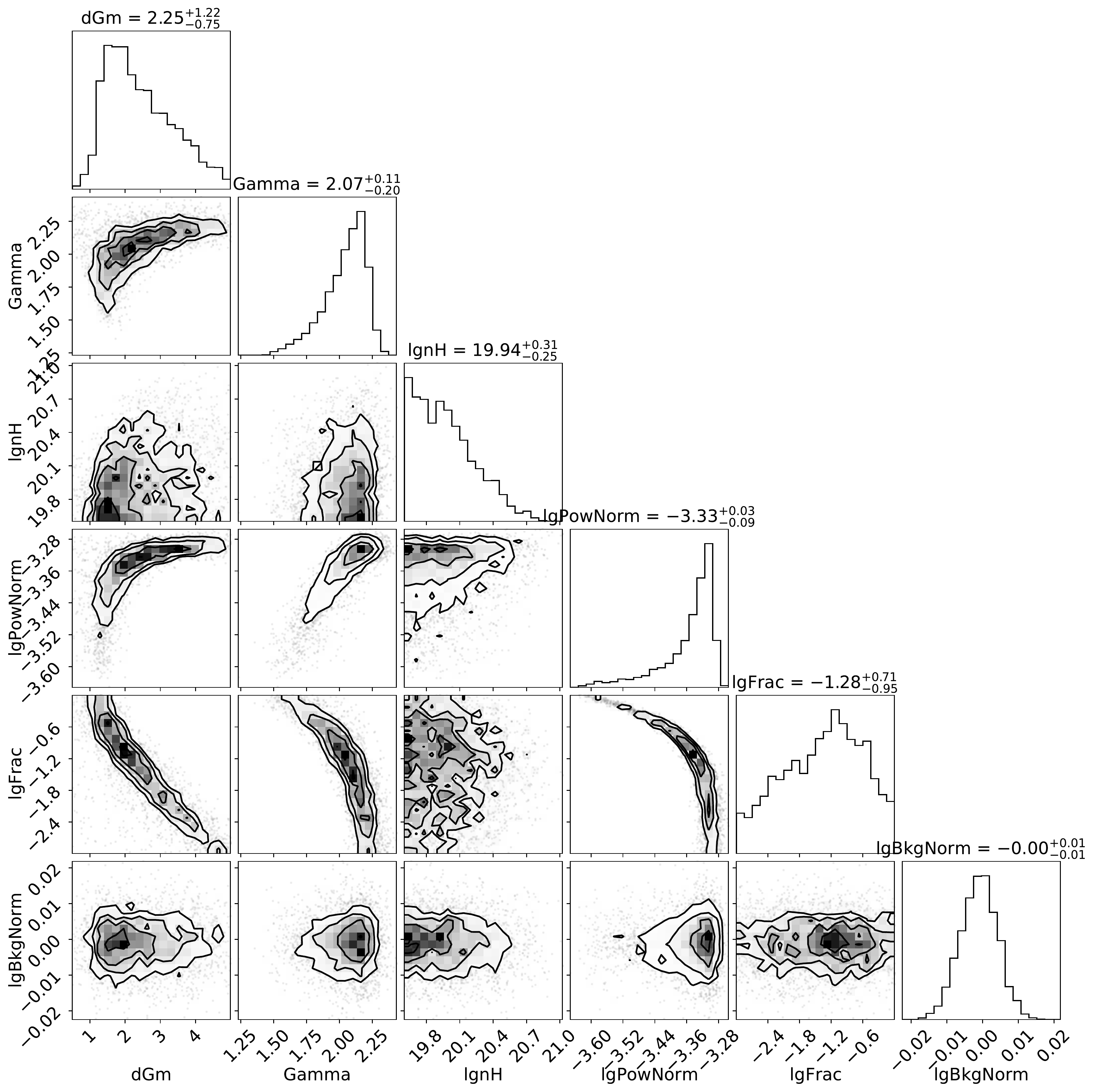}
    \put(52,60){\includegraphics[scale=0.27]{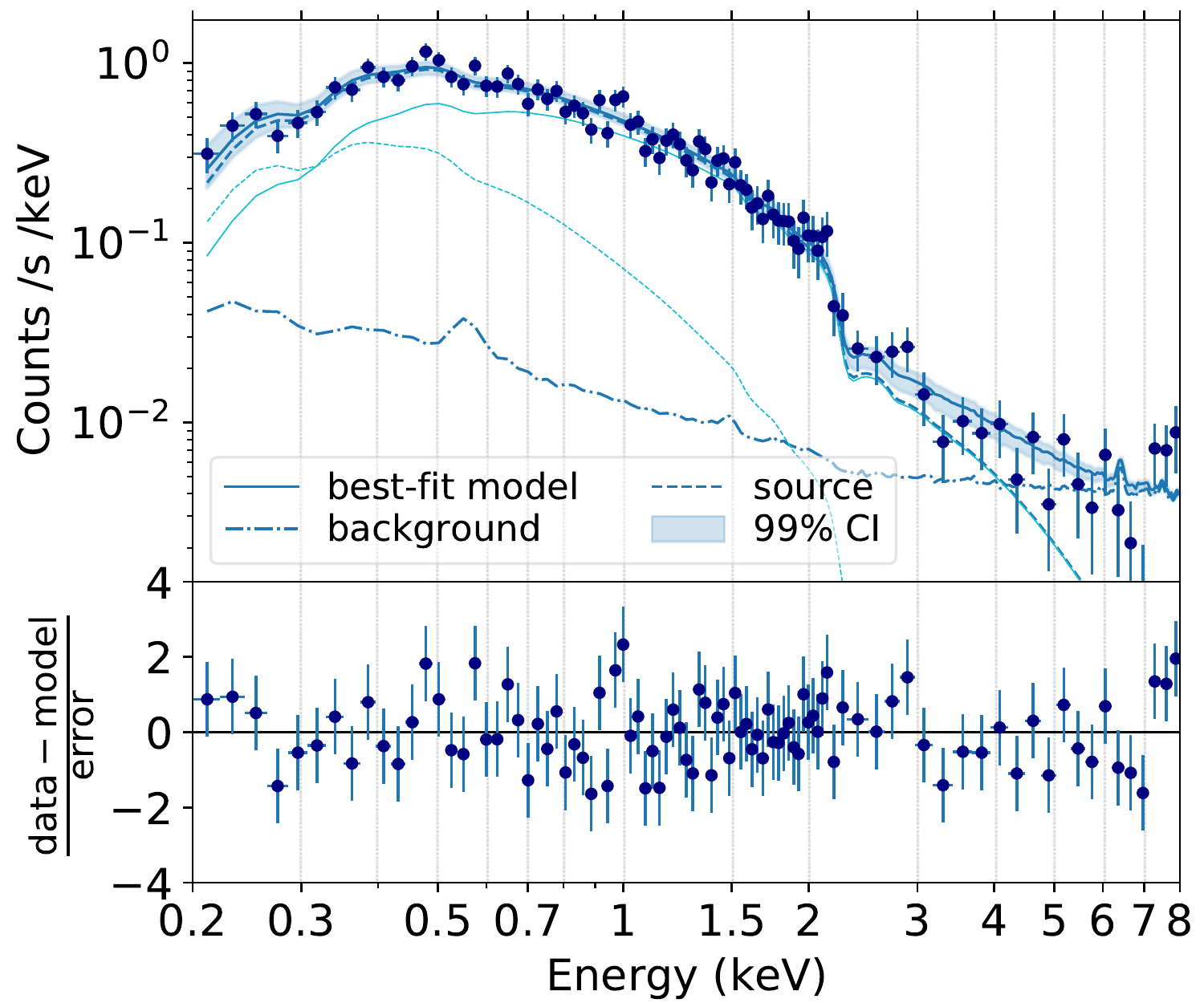}}
  \end{overpic}
  \begin{overpic}[scale=0.25,tics=0,percent]{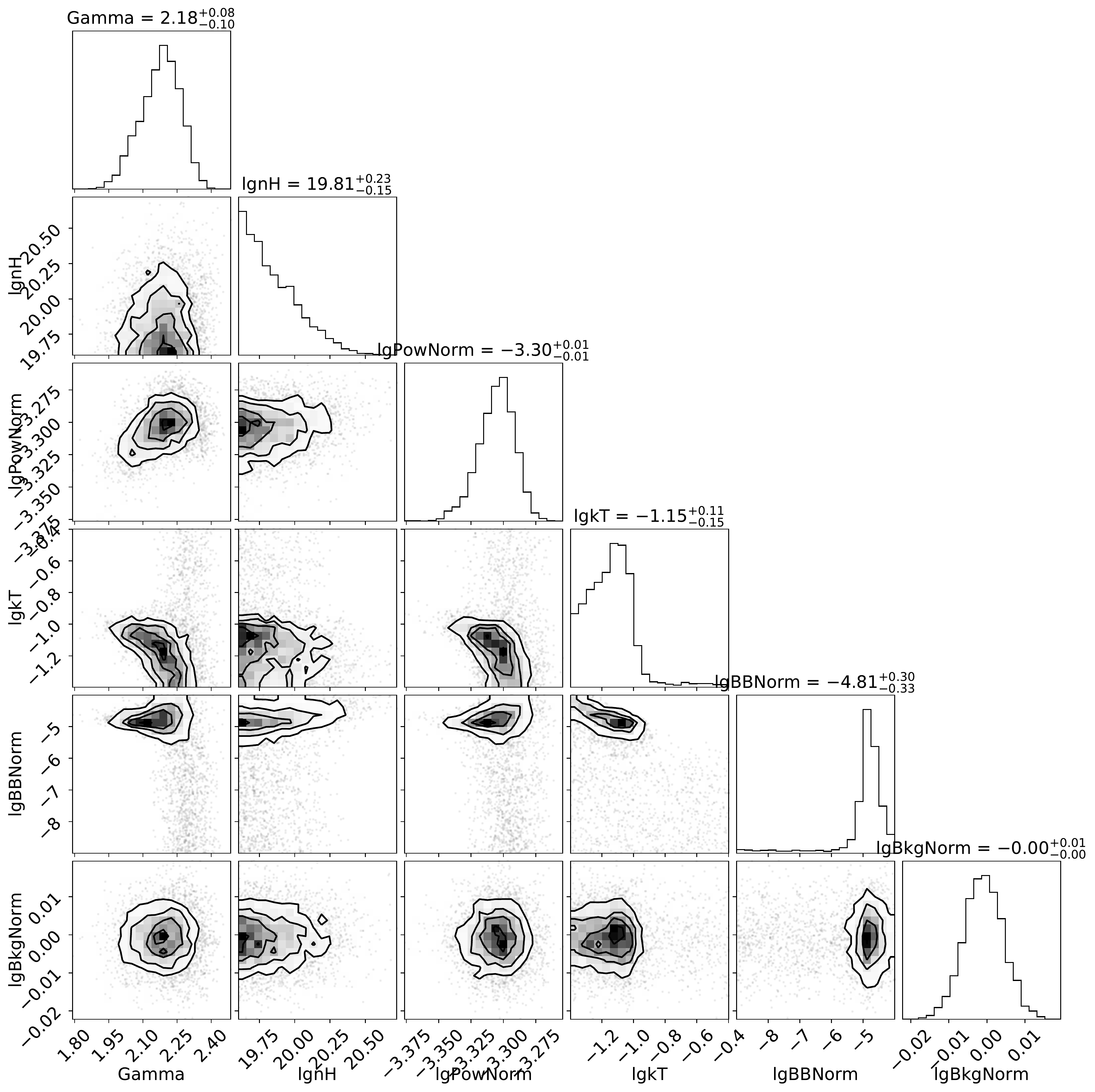}
    \put(54,60){\includegraphics[scale=.27]{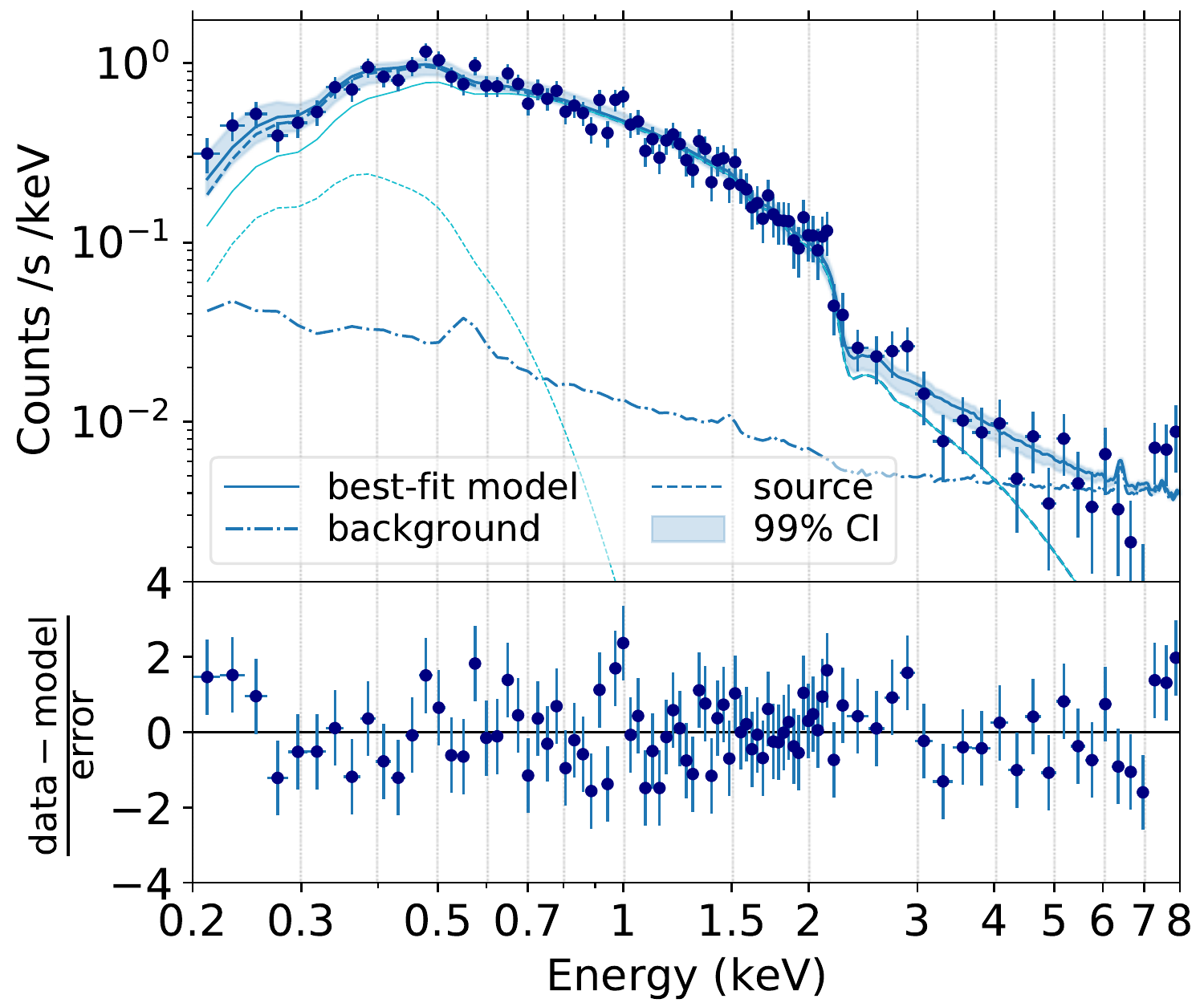}}
  \end{overpic}
  \caption{Same as Fig.~\ref{fig:pow_apec}, but for an AGN (ID$=$7) fitted with two different models: an absorbed power-law with an additional soft power-law (left) and an additional black-body component (right), respectively.
    For the first model, the spectral shape parameters include the $\Delta\Gamma$ for the additional power-law, the $\Gamma$ of the primary power-law, the $\log$\N{H}, and the flux ratio of the secondary to the primary power-law at 1 keV ($\log$Frac), plotted from left to right.
    For the second model, in addition to the power-law normalization, the parameters determining the spectral shape include the $\Gamma$ of the power-law, the $\log$\N{H}, the $\log kT$ of the blackbody, and the blackbody normalization (lgBBNorm).
\label{fig:po_bb}
}
\end{figure*}

A total of $2822$ X-ray point sources have reliable counterparts that are classified as ``Likely galactic'' or ``Secure galactic'' in Paper II.
To account for them, we use a model of collisionally-ionized gas emission \citep[APEC;][]{Smith2001} at a redshift of $0$ to fit all the spectra in the 0.2-8 keV band.
An example of spectral fitting results with this model is shown in the right panel of Fig.~\ref{fig:pow_apec}.
We use a log-uniform prior between 0.05 and 5 keV for the temperature and a log-uniform prior between 0.1 and 1 for the abundance.
We also add a Galactic absorption with ``TBabs'' \citep{Wilms2000}.
Instead of adopting the Galactic \N{H} at the source position, which is usually done for AGN, the Galactic column densities of these stars are allowed to vary in a narrow range between $4\times10^{19}$ and $4\times10^{20}$ cm$^{-2}$ with a log-uniform prior.
This model is called model 0 hereafter.
It is the only model applied to galactic sources, as this work mainly focuses on the AGN, which comprise the majority of the catalog. 

\subsubsection{AGN spectral models}
\label{sec:AGNmodel}
Our baseline spectral model for AGN is an absorbed power-law, which is expressed as ``powerlaw*zTBabs*TBabs'' in the \texttt{XSPEC} terminology (see the right panel of Fig.~\ref{fig:pow_apec} for an example).
AGN spectra have a typical power-law slope $\Gamma$ between 1.7 and 2.0, which varies depending on the sample selection \citep[e.g.,][]{Nandra1994,Buchner2014,Liu2017,Ricci2017}.
To cope with the potential variety of spectral shapes, we adopt for $\Gamma$ a Gaussian prior with a $\sigma$ of 0.5, centered at $2.0$ and truncated at $-2$ and $6$, which can be expressed as Gaussian(2.0,0.5).
Even though not being non-informative, this is a weak prior in the sense that it is much broader than the intrinsic $\Gamma$ scatter of AGN \citep[e.g.,][]{Nandra1994,Liu2017}. 
The prior center 2.0 is chosen based on the spectral fitting results of this sample (\S~\ref{sec:specproperties}).
The AGN intrinsic (rest-frame) absorption is modeled with ``zTBabs'' \citep{Wilms2000}.
For its column density \N{H}, a log-uniform prior is adopted between $4\times10^{19}$ and $4\times10^{24}$ cm$^{-2}$. The range is sufficiently wide, extending to an unmeasurable low \N{H} below the Galactic \N{H} and an unmeasurable high \N{H} in the Compton-thick regime.
In all the models, we always apply a Galactic absorption (``TBabs'') using the total \N{H} measured by HI4PI \citep{HI4PI2016} in the direction of the eFEDS field. 

The baseline model described above is called the ``single-powerlaw'' model (model 1).
In order to fit potential soft excess, we add additional soft components to the ``single-powerlaw'' model (model 2 and 3) in the following section.
In order to cope with low-count sources, a modified model of ``single-powerlaw'', with the powerlaw slope fixed at $\Gamma=2.0$ is also adopted, called ``$\Gamma$-fixed-powerlaw'' (model 4).
A further modified variation, ``shape-fixed-powerlaw'' (model 5), is a simple, unabsorbed power-law (\N{H}=0) with $\Gamma=2.0$ fixed.
The models 1-4 are used to fit the broad-band (0.2-8 keV) spectra, but model 5 is used to fit the source detection band (0.2--2.3 keV), where even the faintest sources are still detectable.
For each source, the most appropriate model will be chosen according to its counts and properties.

\subsubsection{Soft excess models}

It is common to see deviations from a power-law in the X-ray spectra of AGN as observed by \XMM{} and \Chandra{} when the spectral S/N is sufficiently high. The most prominent feature is the soft excess in type-I AGN \citep[e.g.,][]{Walter1993,Bianchi2009}.
Such a soft component could be detected by eROSITA, which has a larger effective area, higher energy resolution, and a wider energy range (down to 0.2 keV) in the soft band than \XMM{}.
To model the spectra with soft excess, we use two multi-components models to fit the broad-band (0.2-8 keV) spectra of the sources with at least 10 net counts.
The first one adds an additional soft power-law to the power-law model, that is, ``TBabs*zTBabs*(powerlaw+constant*powerlaw)'' (model 2). 
The second one adds an additional blackbody component, that is, ``TBabs*zTBabs*(powerlaw+bbody)'' (model 3).
In both cases, we adopt the same prior as above for the primary power-law.
In the first model, we define the additional power-law with a $\Delta\Gamma$ parameter, which is the deviation of its slope to the slope of the primary power-law and has a uniform prior between $0.5$ and $5$.
Setting a positive lower boundary of $0.5$ for $\Delta\Gamma$, the additional power-law is always significantly softer than the primary one.
For the constant factor regulating the relative strength of the additional power-law, we adopt a log-uniform prior between $0.001$ and $1$. This ensures that its 1 keV monochromatic flux is always lower than the primary one.
In the second model, we use a log-uniform prior between $0.04$ and $0.4$ keV for the temperature of the blackbody component.
Examples of these two double-components models are shown in Fig.~\ref{fig:po_bb}.

\subsection{Spectral fitting procedure}
\label{sec:pow_apec}
The \texttt{XSPEC} software \citep{Arnaud1996} is used to load the spectra files and calculate the Poisson likelihood ({\sl C} statistic, \citealt{Cash1979}) for each set of parameters.
We fit the source and background spectra simultaneously, modelling the background spectrum with the background model (see \S~\ref{subsec:backgroundmodel}), and the source spectrum with a source model convolved with the X-ray responses plus the background model convolved with a diagonal matrix response.
In addition to the parameters of the source model, the background model adds the additional normalisation parameter, which is expected to be unity.

A Bayesian spectral fit is performed with \texttt{BXA}\footnote{\url{https://github.com/JohannesBuchner/BXA}} \citep{Buchner2014,Buchner2021BXA}, which connects \texttt{XSPEC}  with the \texttt{UltraNest}\footnote{\url{https://github.com/JohannesBuchner/UltraNest/}} nested sampling package \citep{Buchner2021}.
Given a prior distribution for each parameter of the model, the robust MLFriends algorithm \citep{Buchner2016,Buchner2019} implemented in \texttt{UltraNest} explores the whole parameter space and samples equal-weighted (same probability) points. These represent the posterior distributions of the model parameters (as illustrated in Fig.~\ref{fig:pow_apec} and Fig.~\ref{fig:po_bb}).

Spectral fitting results in a best-fit model. The parameters of this model and the fluxes and luminosities predicted by this model are reported in our spectral property catalog (see Appendix). However, in this work we adopt the Bayesian interpretation of the models. From the posterior distribution of each parameter, we measure the median and the 1-$\sigma$ percentile confidence interval around the median, i.e., the 68\% equal-tailed interval. 
By calculating the flux and luminosity predicted by each set of parameters, we also obtain the posterior distribution of flux and luminosity. We also present the median and 1-$\sigma$ interval for fluxes and luminosities measured from the posterior distributions.

The Bayesian method, in combination with the background modelling, does not require rebinning of the data. However, in order to speed up the fitting of the large number of sources, the spectra are regrouped four-fold, i.e., grouping every four channels into one, since high energy-resolution analysis of narrow line or edge features is out of the scope of this work and has negligible impact on our results.

\subsection{Goodness of fit}
\label{sec:goodness}
In order to test the goodness of fit, we rebin the source spectrum to guarantee at least 25 counts in each bin, and then calculate the $\chi^2$ value for the rebinned spectrum against the best-fit model. The quality of the fit is judged by comparing the $\chi^2$ value with a $chi^2$ distribution with the degrees of freedom (DOF) of the rebinned spectrum.
This method only applies to the brightest sources, as the $\chi^2$ cannot be calculated in most cases, where the rebinned spectrum has no DOF at all.
For the ``double-powerlaw'' model or the power-law plus blackbody model, there are only $153$ AGN with at least $5$ DOF and $85$ AGN with at least $10$ DOF.
The brightest sources might exhibit complex spectral features that are not modeled by our simple models.
With the ``double-powerlaw'' model, three AGN (ID 28, 62, and 171) with at least $10$ DOF have reduced-$\chi^2>2$, because of strong narrow emission/absorption features in the soft band. 
Among them, source 62 is likely due to a galaxy cluster.
However, such bright sources comprise only a tiny fraction of our sample.
The goodness of fit is not relevant for most sources that are faint, because, with a poor S/N, such sources are already over-fitted by the model.
The goal of this work is to analyze the whole sample in a systematical way.
Therefore, rather than goodness of fit, it is more important for this work to investigate the constraints on the spectral parameters.

\subsection{Constraints on the Spectral Shape}
Many sources have too few counts to constrain the parameters regulating the shape of the spectral model, e.g., \N{H} and power-law slope $\Gamma$.
To explore the spectral-shape constraint power of eROSITA spectra, we develop and present our criteria to objectively measure the constraints on the spectral shape following two approaches.

In the first approach, we calculate the 1-$\sigma$ (68\%) uncertainty intervals defined in two ways for each parameter from the posterior samples.
The first one is the 68\% (equal-tail) percentile interval. It defines a lower limit and an upper limit in a classic way, as long as the lower and upper limits exist in the parameter range, i.e., the parameter domain should be wide enough to allow the PDF to drop towards zero at both the boundaries of the range.
This requirement can be easily met for a well-constrained parameter like the normalization, but not for \N{H}.
For an unobscured AGN, the lower limit of \N{H} is unmeasurable and the PDF does not drop towards lower \N{H}.
To cope with such cases, we adopt the highest-density-intervals (HDI) as the second definition.
If the \N{H} PDF monotonically increases towards lower \N{H}, the HDI lower limit will be pegged at the lower boundary. This is an indicator of no absorption.
In order to calculate HDI with a better accuracy, rather than using the posterior sample of a parameter directly, we first smooth the distribution using Gaussian kernel density estimation (KDE) with a minimum bandwidth of 0.1. The kernel is renormalized to take into account only the part of the kernel within the domain. Then we extract 10,000 points following the smoothed distribution and use them to calculate the HDI limits.

In the second approach, we quantify how much the posterior distribution of a parameter differ from the prior using the Kullback-Leibler ($KL$) divergence for a parameter $X$, 
$$\mathrm{KL}(X)=\int{\mathrm{posterior}\times\ln\frac{\mathrm{posterior}}{\mathrm{prior}}dX}.$$

For uniform/log-uniform prior, the comparison is made in the full parameter range; in the case of a Gaussian prior (for power-law slope), the comparison is made within the 3-$\sigma$ range.
This $KL$ value is a measurement of information gain from the data in units of nats.
A small value of $KL$ means the posterior is the same as the prior and thus the posterior gains little information from the data. A large value of $KL$ means a significant difference between the posterior and the prior, which is attributed to the constraint provided by the data.

Based on the $KL$ divergence and the HDI confidence interval, we quantify the constraint for each AGN on \N{H} and $\Gamma$ of the ``single-powerlaw'' model in \S~\ref{sec:NH_Gm_KL}.

\subsection{Luminosity and flux measurement}
\label{sec:lum_flux}

For each AGN model, the rest-frame 0.5-2\,keV and 2-10\,keV intrinsic (absorption-corrected) luminosities are computed. In the cases of ``double-powerlaw'' fitting, the soft component is included in the luminosity measurement. The most appropriate model is chosen to present the X-ray luminosity of each source in \S~\ref{sec:AGNLum}.

We present a measurement of the observed fluxes in the 0.5-2 keV and 2.3-5 keV bands for all the X-ray sources. 
The 2-10 keV fluxes are not a direct measurement from the spectra because the eROSITA data are dominated by background at $>5$ keV. Assuming an unabsorbed power-law with a slope of 1.7, 1.8, 1.9, or 2.0, respectively, the 2.3-5 keV flux can be converted to a 2-10 keV flux by a factor of 2.27, 2.20, 2.13, or 2.07.
We choose the fluxes measured with the most appropriate model as follows and assign a \texttt{FSclass}/\texttt{FHclass} for the flux measurement of each source in the soft/hard band.

We choose the 0.5-2 keV flux measurements as follows.
\begin{enumerate}
\item When available, more flexible models are preferred in flux measurement. For the sources  with at least 10 net counts, we have fitted the spectra with the ``double-powerlaw'' model. We adopt the soft-band fluxes measured with this model and call it \texttt{FSclass} 1 ($14707$ sources).
\item For the sources with less than 10 counts, we perform a narrow-band fitting in the 0.4-2.2 keV with ``$\Gamma$-fixed-powerlaw'' model and call this fitting model 6. We adopt the soft-band fluxes measured with this model and call the flux measurement \texttt{FSclass} 2 ($7026$ sources).
  \item The above models are for AGN. For stars, we adopt the soft-band fluxes measured with APEC model (model 0). We call this class \texttt{FSclass} 3, which includes $2568$ sources.
  \item Regardless of the models adopted above, when the measured 1-$\sigma$ percentile lower limit of flux is lower than $20\%$ of the median value, such sources ($3609$) are too faint for spectral fitting. We call it \texttt{FSclass} 0 and adopt the fluxes measured with the ``shape-fixed-powerlaw'' model (model 5). Note that this model is applied to the source detection band (0.2--2.3 keV), which guarantees that the source signal is detectable.
\end{enumerate}

Since eROSITA is much more sensitive in the 0.2--2.3 keV band than in the band above 2.3 keV, the broad-band fitting is dominated by the soft band signal. Therefore, it provides an accurate flux measurement in the soft band but not necessarily in the hard band. 
To measure the 2.3-5 keV flux, we perform a narrow 2.3-8 keV band fitting for the sources with at least 3 2.3-5 keV spectral net counts using the ``single-powerlaw'' model. This fitting is called model 7.
Again, we adopt the flux measurement from the spectral fitting only if the 1$\sigma$ lower limit is at least 20\% of the median value. This selection results in a \texttt{FHclass} 1 of $621$ sources.
For the other $27289$ sources, we adopt the 2.5-5 keV fluxes measured using the ``shape-fixed-powerlaw'' model (model 5). Such a hard-band flux, classified as \texttt{FHclass}$=$0, corresponds to extrapolating of the source signal in the source-detection band (0.2--2.3 keV) but not any hard-band detected signal.

\subsection{Sample distributions}
\label{sec:hbm}
Most sources detected in eFEDS have only few counts, and thus poor spectral constraints.
This means that for the majority of sources the "best-fit" is not a well-defined point, and degeneracies between, e.g., the photon index and the column density can be substantial.
Nevertheless, we aim to produce basic parameter distributions, for example for the column
density distribution, or the photon index distribution.

Hierarchical Bayesian models (HBM) are a suitable method to propagate uncertainties and learn a sample distribution in this setting. In our approach \citep[see e.g.,][]{Baronchelli2020}, a parametric sample distribution, such as a Gaussian with a specific mean $\mu$ and standard deviation $\sigma$ is assumed. For each object, a fixed number of posterior samples (1000) is selected. It is necessary that the posterior samples are obtained under flat priors (e.g., for $\log$\N{H}). For parameters were this is not the case (the photon index $\Gamma$), the posterior samples are first resampled according to the inverse of the prior. 
For each object, the Gaussian probability density of the sample distribution is evaluated and averaged at the values of the posterior sample.
Then, the densities are multiplied for all the sources to obtain a likelihood, which is a function of the assumed $\mu$ and $\sigma$ parameters. This encodes that all objects are described by the sample distribution.
Assuming flat priors on $\mu$ and $\log\sigma$, the posterior sample distribution can be explored with Monte Carlo samplers.
We use the PosteriorStacker\footnote{\url{https://github.com/JohannesBuchner/PosteriorStacker}} python tool to compute the posterior probability distribution for various parameters. Internally, PosteriorStacker also uses UltraNest.

Beyond a Gaussian model, PosteriorStacker also implements a non-parametric histogram model. In this model, bins are allowed to vary their density. A flat Dirichlet prior on the bin densities assures that the sample distribution sums to unity. This histogram model allows investigating the sample distribution without assuming a specific (e.g., symmetric, mono-modal) model shape.

\section{Results}
\label{sec:results}

\subsection{The spectral property catalog}
\begin{table*}[!ht]
  \centering
  \begin{tabular}{p{0.03\textwidth}p{0.095\textwidth}p{0.38\textwidth}p{0.40\textwidth}}
    \hline
    \hline
    Index & Table name & Table description & Sample selection criteria\\
     & \& number &  & or Xspec model and prior\\
    \hline
 1 & AGN\ \ \ \ \ \ \ \ 21952            & The eFEDS AGN catalog & sources from the main X-ray catalog with extent likelihood $=$0, counterpart quality $\geqslant$2, and classified as ``Likely galactic'' or ``Secure galacitc''.\\
    \hline
 2 & Spec\ \ \ \ \  27910   & Basic spectral properties & all sources in the main X-ray catalog \\
    \hline
 3 & m0:apec0\ \ \ \ 27910  &Broad-band fitting results using a single-temperature APEC model; only useful for stars& \texttt{TBabs*apec}  $z=0$; log-uniform prior between 0.05 and 5 keV for $kT$; log-uniform prior between $4\times 10^{19}$ and $4\times 10^{20}$ cm$^{-2}$ for Galactic \N{H}\\
    \hline
 4 & m1:pow\ \ \ \ \ \ \ 27910    &Broad-band fitting results using the ``single-powerlaw'' model & \texttt{TBabs*zTBabs*powerlaw} Gaussian(2.0,0.5) prior between -2 and 6 for $\Gamma$; log-uniform prior between  $4\times 10^{19}$ and $4\times 10^{24}$ cm$^{-2}$ for AGN \N{H} \\
    \hline
 5 & m2:powpow\ \ \ \ \ 16957 &Broad-band fitting results using the ``double-powerlaw'' model for sources with at least 10 counts & \texttt{TBabs*zTBabs*(powerlaw+constant*spowerlaw)~\footnotemark} same priors as model 1 for $\Gamma$ and \N{H}; uniform prior between 0.5 and 5 for $\Delta\Gamma$; log-uniform prior between 0.001 and 1 for the constant factor \\
    \hline
 6 & m3:powbb\ \ \ \ \ 16957  &Broad-band fitting results using the powerlaw plus blackbody model for sources with at least 10 counts & \texttt{TBabs*zTBabs*(powerlaw+zbbody)} same priors as model 1 for $\Gamma$ and \N{H},  log-uniform prior between 0.04 and 4 keV for $kT$\\
    \hline
 7 & m4:pow2d0\ \ \ \ \ 27910    &Broad-band fitting results using the ``$\Gamma$-fixed-powerlaw'' model & same as model 1  but fixing $\Gamma=$2.0 \\
    \hline
 8 & m5:powfix\ \ \ \ \ 27910    &Detection-band (0.2--2.3 keV) fitting results using the ``shape-fixed-powerlaw'' model & same as model 1 but fixing \N{H}=0 and $\Gamma=2.0$ \\
    \hline
 9 & m6:soft\ \ \ \ \ 27910   &Soft-band (0.4-2.2 keV) fitting results using the ``single-powerlaw'' model; only used to measure the 0.5-2 keV fluxes. & same as model 1\\
    \hline
10 & m7:hard\ \ \ \ \ 2836   &Hard-band (2.3-6 keV) fitting results using the ``single-powerlaw'' model for sources with at least 3 counts in the 2.3-5 keV band; only used to measure the 2.3-5 keV fluxes. & same as model 1\\
    \hline
    \end{tabular}
    \caption{A list of catalogs and tables presented in this work. The index is the extension number in the merged FITS-format file. The spectral fitting result tables are named so as to include the model index $i$ in terms of m$i$ before the colon.
    }
    \label{table:tables}
\end{table*}
\footnotetext[5]{The \texttt{spowerlaw} is a power-law model whose slope equals the slope of the primary \texttt{powerlaw} plus $\Delta\Gamma$.}

We present the AGN catalog selected from the eFEDS X-ray sources.
In addition to the AGN catalog, we also present the basic spectral properties of all the eFEDS X-ray sources, e.g., spectra extraction information, source and background count rate or flux.
We performed spectral fitting with eight models/settings, which are named as model 0$\sim$7.
In this section, we select the most appropriate model for each source for particular purposes, e.g., \N{H} measurement or luminosity measurement.
Considering that future multi-band follow-up could enrich or correct the current optical-counterpart identifications of the eFEDS X-ray sources, we present the spectral fitting results of all the models.
Table.~\ref{table:tables} lists all the catalogs/tables that are available with this paper, including the AGN catalog, the basic spectral property catalog, and the spectral fitting results of the eight models.
These tables are available on the eROSITA website\footnote{\url{https://erosita.mpe.mpg.de/edr/eROSITAObservations/Catalogues/}} and the table columns are described in Appendix~\ref{sec:tabledescription}.

\subsection{Model comparison}
\label{sec:modelselection}

\texttt{BXA} calculates the logarithmic Bayesian evidence $\log Z$ for each fit. A relatively larger value of $\log Z$ can be used to as a model preference indicator \citep{Buchner2014}.

\begin{figure*}[!htp]
  \centering
\includegraphics[width=0.245\textwidth]{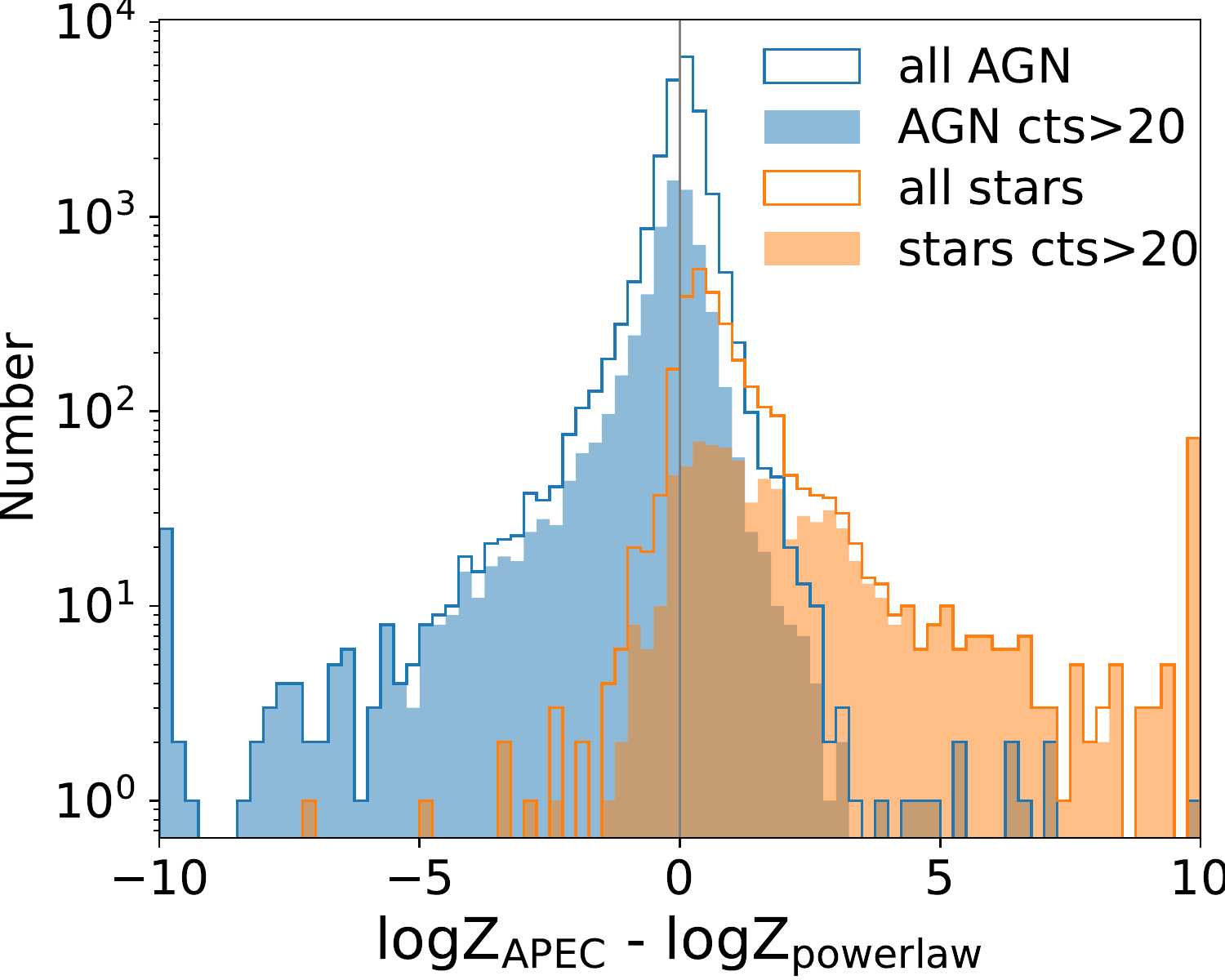}
\includegraphics[width=0.245\textwidth]{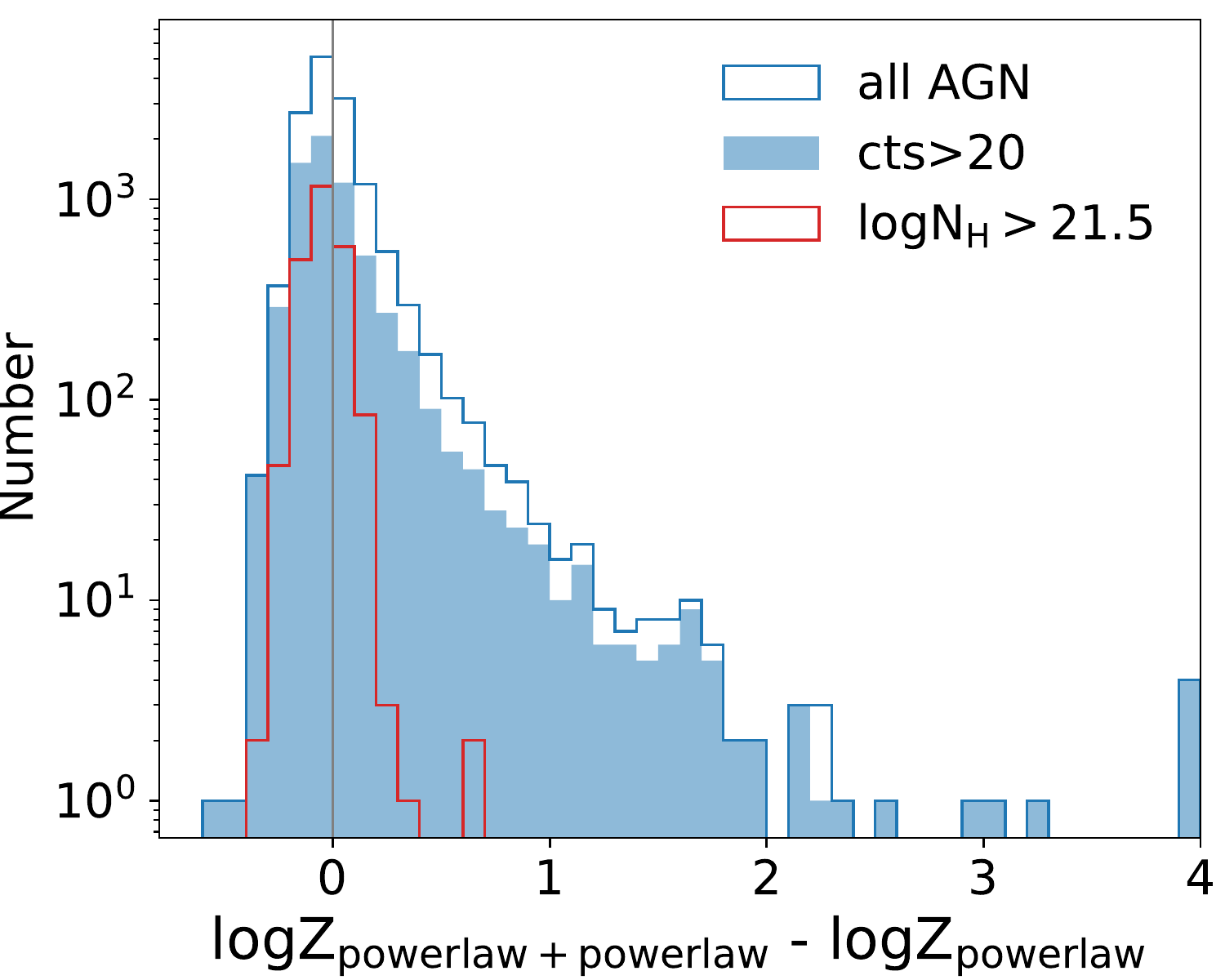}
\includegraphics[width=0.245\textwidth]{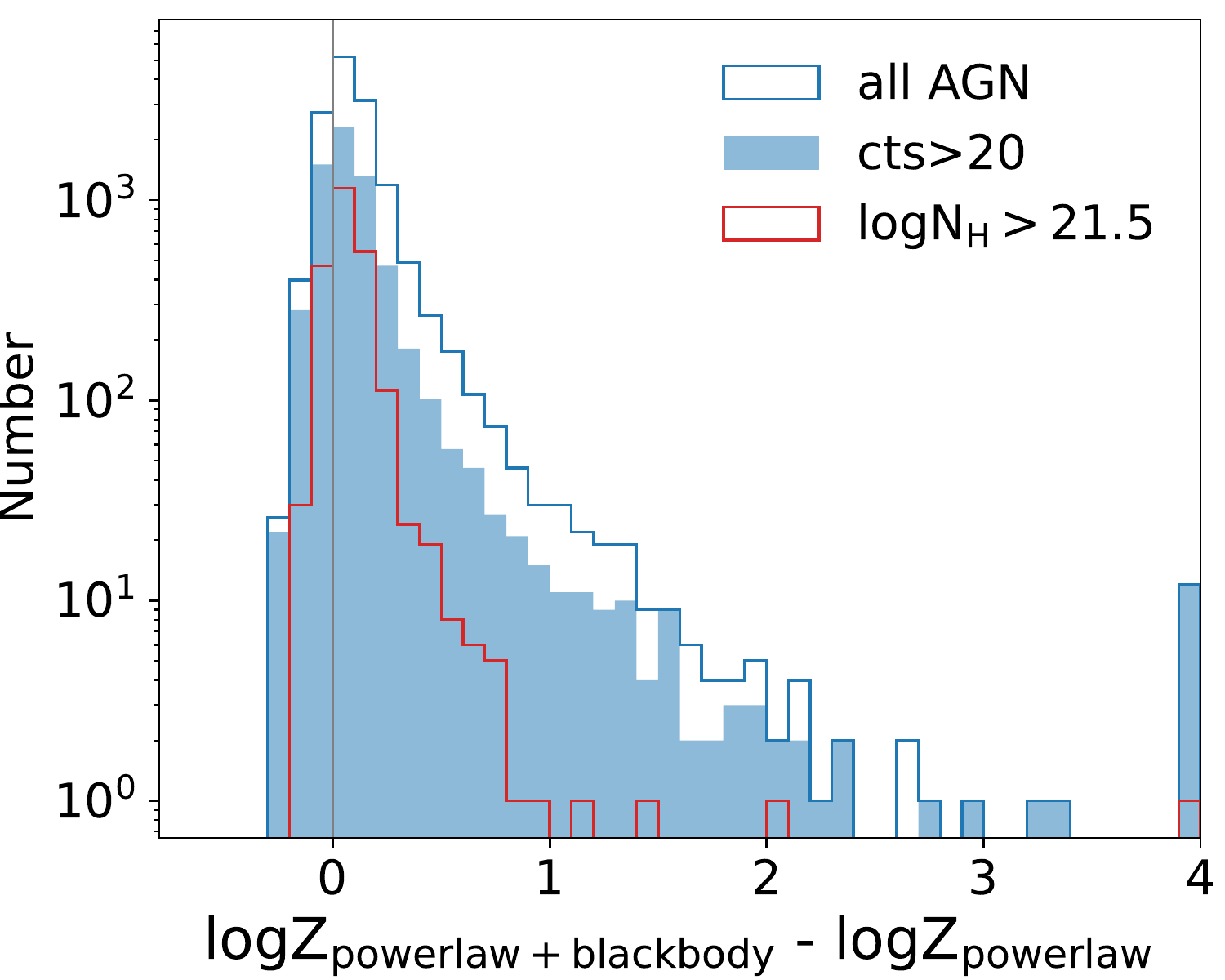}
\includegraphics[width=0.245\textwidth]{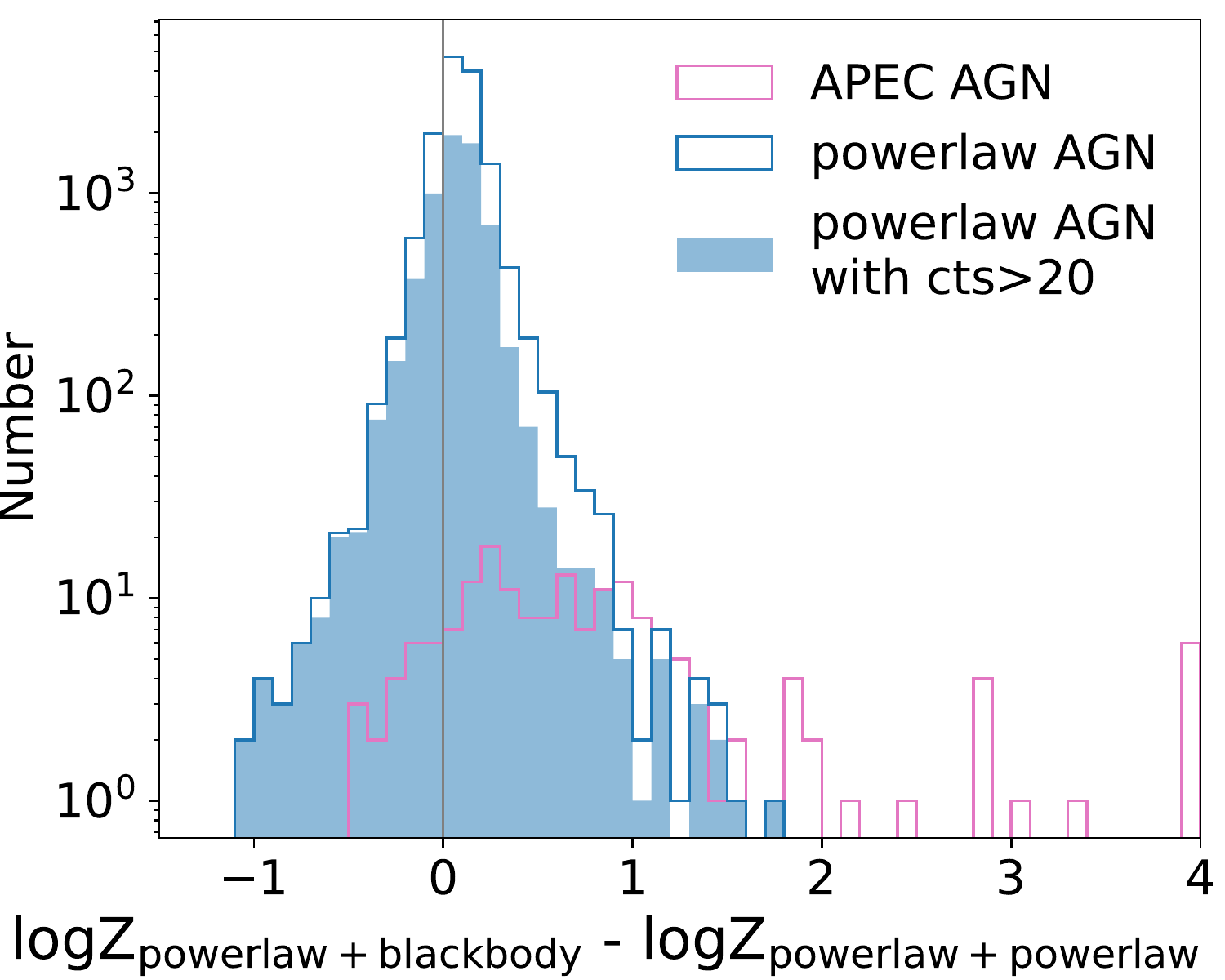}
\caption{Model comparison on the basis of $\Delta\log Z$ between pairs of models, i.e., 1) the ``single-powerlaw'' model vs the APEC model; 2) the ``double-powerlaw'' model vs the ``single-powerlaw'' model; 3) the power-law plus blackbody model vs the ``single-powerlaw'' model; 4) the power-law plus blackbody model vs the ``double-powerlaw'' model.
  Empty histogram indicates all the sources and filled histogram corresponds to the subsamples of bright sources with at least 20 net counts in 0.2--5 keV band.
  In the first panel, the blue and orange color indicate AGN and stars respectively.
  In the second and third panel, the red color indicates the subsamples of AGN with a median $\log$\N{H} above 21.5.
  In the last panel, the blue and magenta color indicate sources with $\log Z_\textrm{APEC}-\log Z_\textrm{powerlaw}$ below and above $1.3$ respectively.
  \\
  \\
\label{fig:modelcomp}
  }

\includegraphics[width=0.70\textwidth]{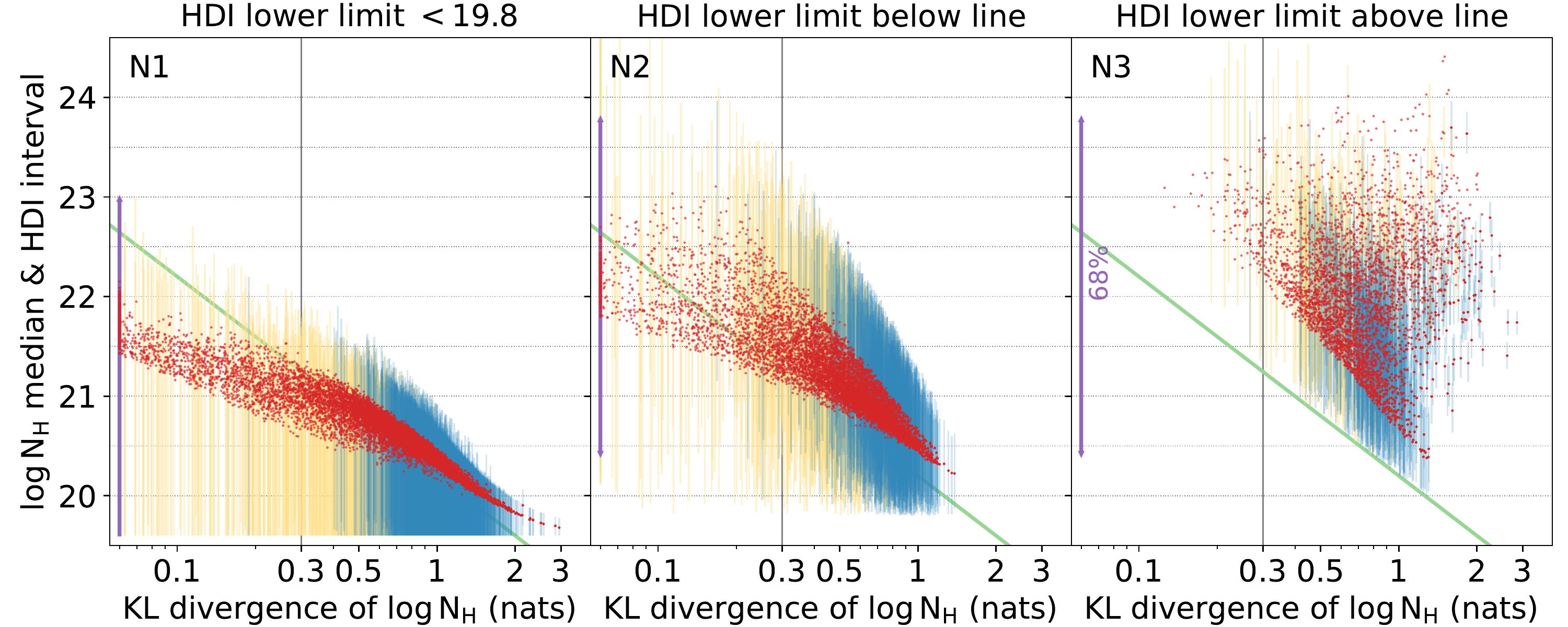}
\includegraphics[width=0.286\textwidth]{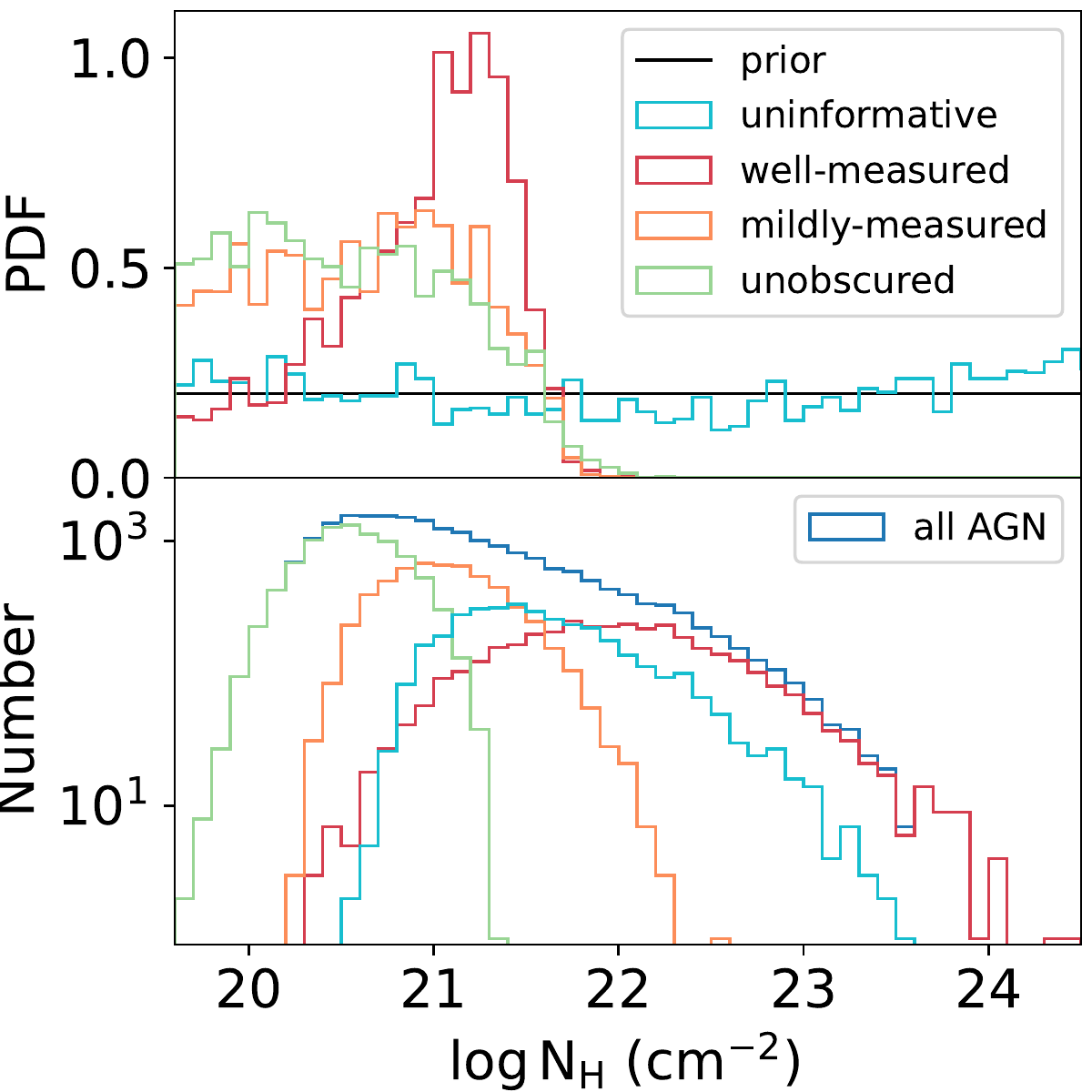}
\includegraphics[width=0.70\textwidth]{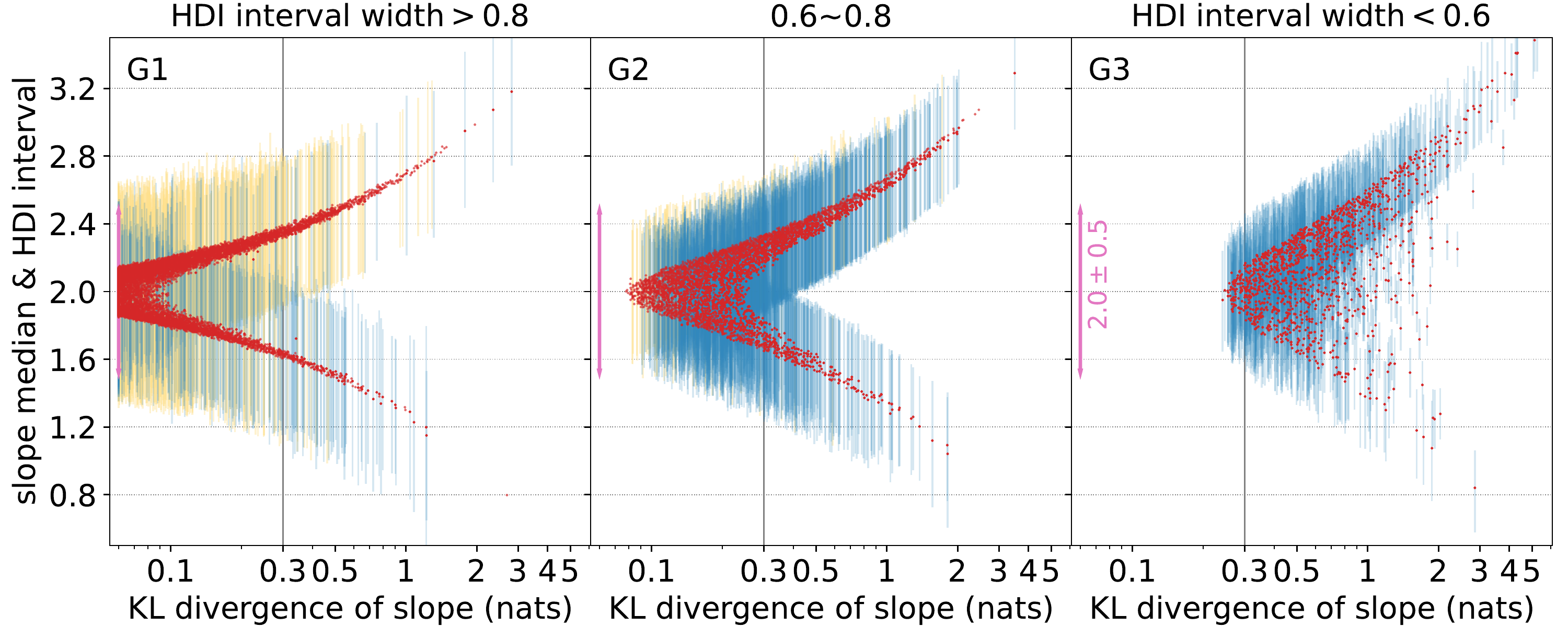}
\includegraphics[width=0.286\textwidth]{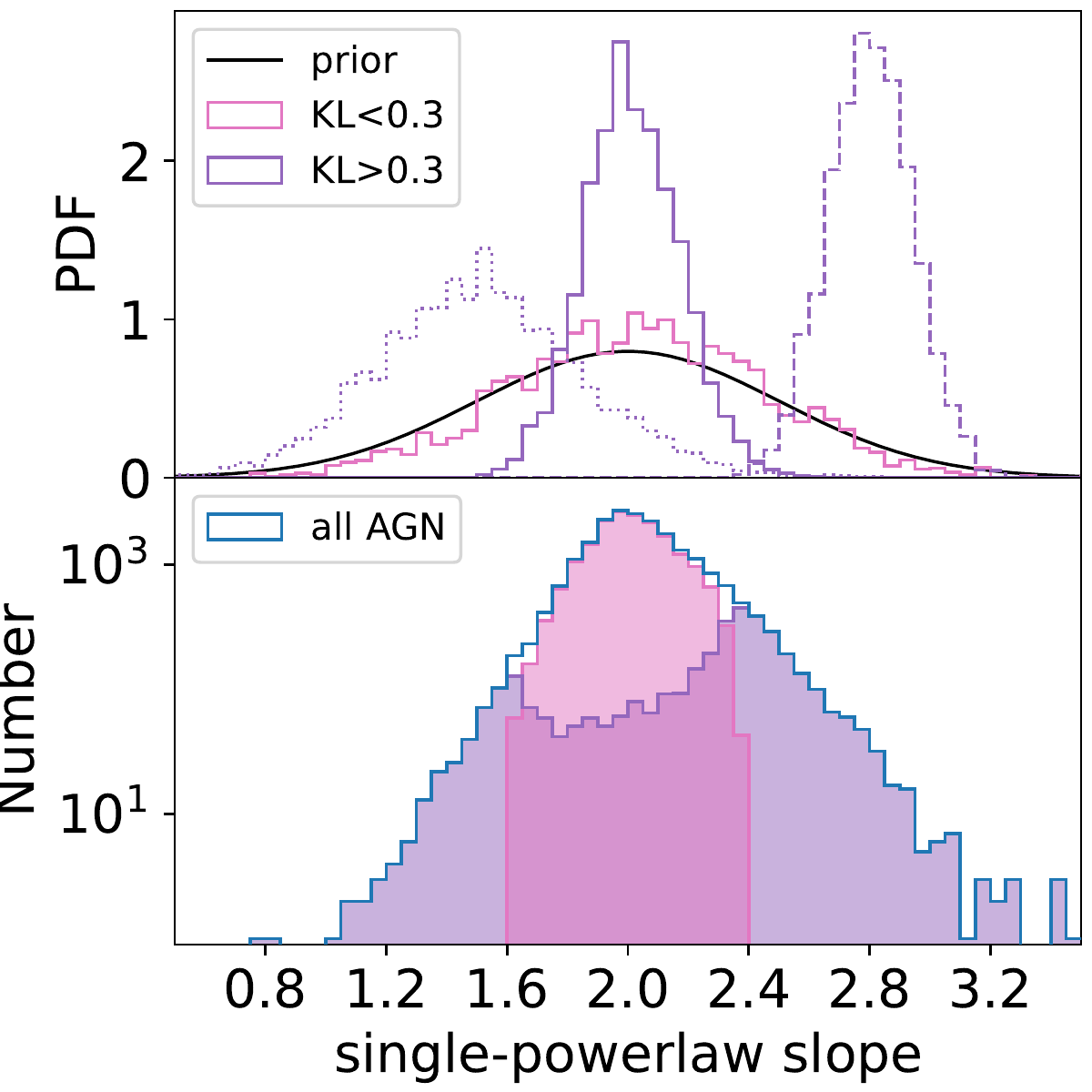}
\caption{The distributions of the posterior median and 1-$\sigma$ HDI interval as a function of $KL$ divergence for \N{H} (upper panels) and for power-law slope $\Gamma$ (lower), respectively, as measured using the ``single-powerlaw'' model.
  Examples of source posterior PDF and sample distributions of posterior median values are also displayed on the right for \N{H} and $\Gamma$, respectively.
  The HDI intervals are plotted as blue errorbars for the sources with at least 30 net counts and yellow errorbars for the other faint sources.
  For representation, we only plot the yellow errorbars for a random 10\% of these faint sources.
  In the \N{H}--$KL_\textrm{NH}$ plot, the green lines correspond to $\log$\N{H} $= 20.2-2\log KL$; the AGN catalog is plotted separately in three groups, i.e., N1: sources with the HDI lower limit truncated by the boundary ($<$19.8); N2 and N3: other sources with the HDI lower limit below and above the green line.
  In the $\Gamma$--$KL_\Gamma$ plot, the catalog is also divided into three, i.e.,  sources with HDI interval width $>$0.8 (G1), between 0.6 and 0.8 (G2), and $<$0.6 (G3).
  The purple arrowed lines indicate 68\% of the \N{H} parameter range at one side of the boundary or at the center of the range.
  The magenta arrowed lines indicate the 68\% range of the Gaussian(2.0,0.5) prior for $\Gamma$.
  In the median \N{H} and median $\Gamma$ distribution plots, the blue empty histograms indicate the whole AGN sample;
  the cyan, green, orange, and red colors indicate the four classes of \N{H} measurements, i.e., 1) uninformative, 2) unobscured, 3) possibly obscured, and 4) obscured AGN; the purple and magenta colors indicate sources with $KL_\Gamma$ above and below $0.3$.
  In the PDF plots, the black lines indicate the prior distributions of \N{H} and $\Gamma$; the ID of example sources for \N{H} PDF are 878 (cyan), 4526 (green), 2076 (orange), and 7274 (red); the ID of example sources for $\Gamma$ PDF are 864 (magenta), 326 (purple, dashed), 595 (purple, dotted), and 9 (purple, solid).
\label{fig:NH_Gm_KL}
  }
\end{figure*}

As displayed in Fig.~\ref{fig:modelcomp}, most stars prefer the apec model over the power-law model, i.e., $\log Z_\textrm{APEC} > \log Z_\textrm{powerlaw}$. On the contrary, AGN favor the power-law model.
There is a small number of AGN that favor the APEC model, which are possibly stars mis-classified as extragalactic sources. 
There are also a few galactic sources that favor the power-law model, which might be Galactic X-ray binary or mis-classified AGN.

Comparing the ``single-powerlaw'' model with the ``double-powerlaw'' model, the $\Delta\log Z$ show a significant tail in favor of the flexible double-component model, indicating the existence of soft excess in a small fraction of sources.
Since the sample is dominated by faint sources, whose spectral shapes are not well constrained, a majority of the sources reside in the peak around $\Delta\log Z=$0, especially for sources with less than 20 counts.
The AGN with a posterior median $\log$\N{H} above 21.5 have $\Delta\log Z\sim$0 because soft excess is irrelevant in such cases.

The power-law plus blackbody model shows similar behavior as the ``double-powerlaw'' model in the comparison with the ``single-powerlaw'' model.
Between the ``double-powerlaw'' model and the power-law plus blackbody model, it seems the latter is more preferred.
However, it is only because the power-law plus blackbody model is more flexible and does not indicate the soft excess component is better described by a blackbody model. The ``double-powerlaw'' model is restricted to the cases of AGN with a soft excess, which must be softer ($\Delta\Gamma$ always $>$0.5) than the primary power-law and weaker than the primary power-law at 1 keV.
However, the blackbody component is free to be stronger than the primary power-law in the model.
We select $224$ AGN with $\log Z_\textrm{APEC}-\log Z_\textrm{powerlaw}>1.3$ as a special class and call them APEC AGN.
As displayed in Fig.~\ref{fig:modelcomp}, the APEC AGN favor the blackbody model, because the spectral shape of hot plasma emission is more similar to blackbody than power-law.
For the other AGN (called power-law AGN in the figure), the ``double-powerlaw'' model and the power-law plus blackbody model could fit the data equally well.

\subsection{Constraints on \N{H} and power-law slope}
\label{sec:NH_Gm_KL}

In this section, we quantify the constraints on \N{H} and $\Gamma$ for each AGN based on the ``single-powerlaw'' model.
Compared with the power-law slope $\Gamma$, a varying \N{H} changes the overall wide-band spectral shape more prominently. We test whether the \N{H} of each source is constrained by the eROSITA data based on the HDI lower limit and the $KL$ divergence of \N{H}.
As displayed in the upper panels of Fig.~\ref{fig:NH_Gm_KL}, we divide the \N{H} measurement into four classes (\texttt{NHclass} in the catalog) as follows in sequence.
At a low $KL$ (e.g., $<0.3$), the 68\% interval widths are large and even close to 68\% of the parameter range (purple arrowed lines in Fig.~\ref{fig:NH_Gm_KL}).
We call them class (1) -- \texttt{uninformative} sources, as there is no information about \N{H} in the data, because of the low S/N. The posterior \N{H} distribution of an example source (ID$=$878) is shown in the upper right panel of Fig.~\ref{fig:NH_Gm_KL} in cyan.
As displayed in Fig.~\ref{fig:NH_Gm_KL}, fluctuation tends to bring the median \N{H} of an \texttt{uninformative} source to a large value in the middle of the parameter range, but such values should not be adopted as reliable measurements.
The class (2) comprise the sources with HDI lower limit $\log$\N{H,HDI,lower} pegged at the parameter lower boundary ($4\times10^{19}$ cm$^{-2}$), adopting $\log$\N{H,HDI,lower}$<$19.8 practically.
With the posterior \N{H} distribution monotonically increasing towards lower values, such sources are classified as \texttt{unobscured}. The posterior \N{H} distribution of an example source (ID$=$4526) is shown in Fig.~\ref{fig:NH_Gm_KL} in green.
The median \N{H} of such sources, which should not be used either, show a strong correlation with the $KL$, reflecting the effect of the lower boundary.
This boundary effect is the strongest in these HDI-pegged cases, but also exists in other cases (panel N2 of Fig.~\ref{fig:NH_Gm_KL}) where the \N{H} is near the lower boundary and the \N{H} uncertainty is not narrow enough to separate it from the lower boundary (see an example source, ID$=$2076, in orange color in Fig.~\ref{fig:NH_Gm_KL}).
To distinguish between such sources and significantly-obscured sources with well-constrained lower limits of \N{H} (see example source, ID$=$7274, in red color in Fig.~\ref{fig:NH_Gm_KL}), we adopt a criterion $\log$\N{H,crit} $= 20.2-2\log KL$ (green line in Fig.~\ref{fig:NH_Gm_KL}) and compare it to the 1-$\sigma$ HDI lower limit $\log$\N{H,HDI,lower}.
Sources with $\log$\N{H,HDI,lower} below and above $\log$\N{H,crit} are assigned to class (3) -- \texttt{mildly-measured} and class (4) -- \texttt{well-measured}, respectively.
Such a criterion represents a natural selection bias against measurement of low \N{H} when the constraining power ($KL$) of the data is low.
Note that, the \texttt{well-measured} class is defined in the sense that absorption is significantly detected, irrespective of the measured \N{H} value, which can be as low as $10^{20}$ cm$^{-2}$. 
When a sample of AGN that are obscured at a certain level is needed, we recommend selecting the \texttt{well-measured} and \texttt{mildly-measured} sources with $\log$\N{H} above a threshold of 21.5 or 22.

\begin{figure}[h]
\begin{center}
\includegraphics[width=0.8\columnwidth]{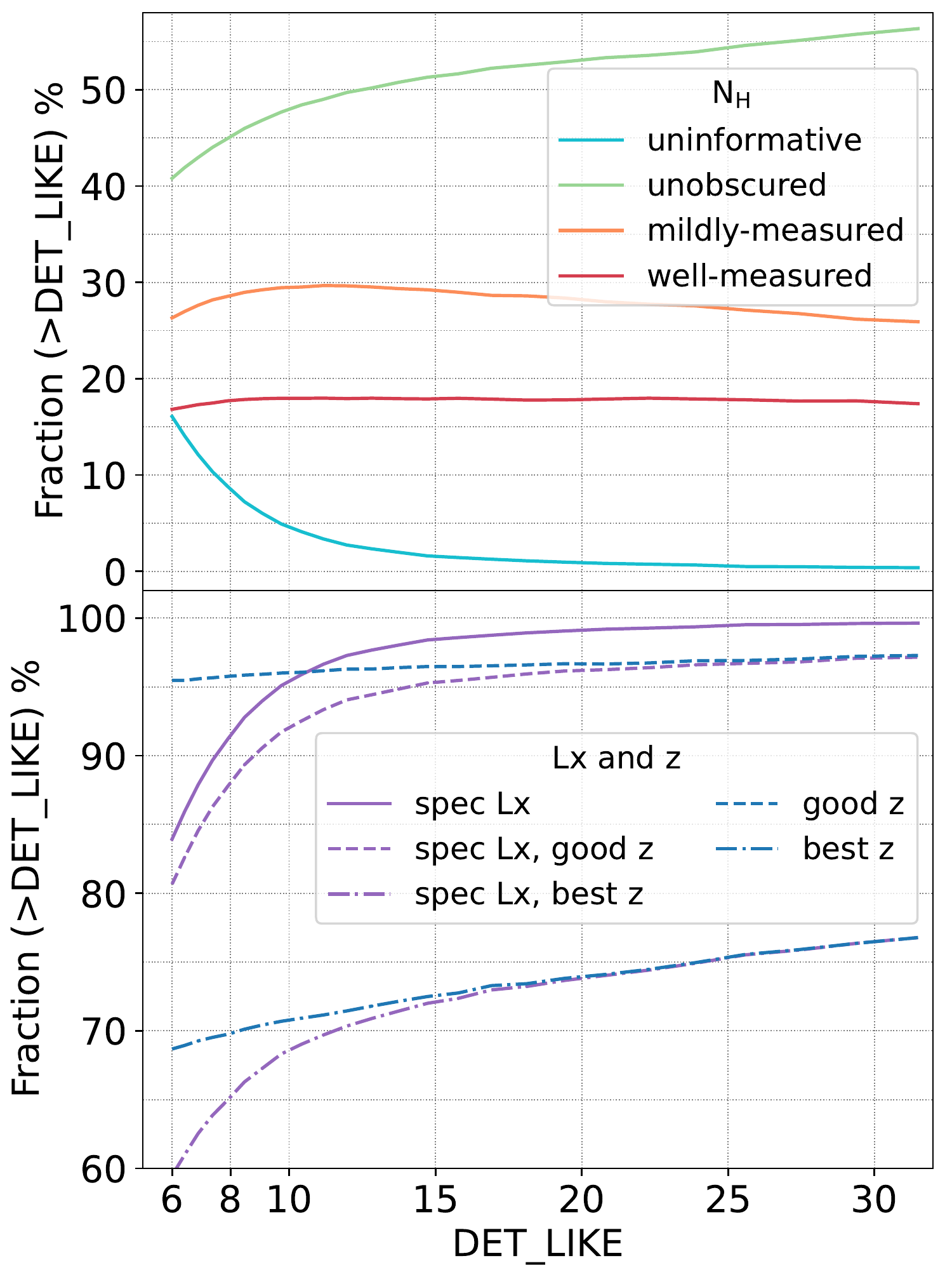}
\caption{Fractions of sources selected by X-ray spectral properties among the AGN detected in the eFEDS 90\%-area region as a function of the 0.2--2.3 keV source detection likelihood.
  In the upper panel, concerning the \N{H} measurement with the ``single-powerlaw'' model, the AGN are divided into four classes, i.e., 1) uninformative (cyan), 2) unobscured (green), 3) mildly-measured (orange), and 4) well-measured (red).
  In the lower panel, the blue and purple lines indicate all the AGN and the AGN with spectral measurements of $L_X$ (discussed in \S~\ref{sec:AGNLum}), respectively; and the solid, dashed, and dash-dotted lines indicated all the AGN, the AGN with good redshift measurements, and the AGN with the best redshift measurements, respectively.
}
\label{fig:AGN_frac}
\end{center}
\end{figure}

Since the sources with \texttt{uninformative} \N{H} are too faint for a reasonable spectral fitting, their luminosities cannot be robustly measured either.
Fig.~\ref{fig:AGN_frac} displays the fractions of each of the four classes as a function of the 0.2--2.3 keV source detection likelihood and the fractions of sources with robust luminosity measurements (discussed later in \S~\ref{sec:AGNLum}).
To suppress the fraction of \texttt{uninformative} \N{H} measurements to below $5\%$, a detection likelihood $>10$ is required. To suppress this fraction to $1\%$, a detection likelihood $>19$ is needed.

Similarly, the KL divergence for the power-law slope, $KL_\Gamma$, measures how significant the posterior $\Gamma$ distribution differs from the prior Gaussian(2.0,0.5) distribution.
The lower panels of Fig.~\ref{fig:NH_Gm_KL} displays the distributions of the posterior $\Gamma$ and the $KL_\Gamma$, dividing the sample into three according to the $\Gamma$ HDI interval width.
There are two causes that could lead to a high $KL_\Gamma$.
One cause is a significant offset of the measured $\Gamma$ from the center of the prior ($2.0$), as shown by the two branches in the $\Gamma$--$KL_\Gamma$ plots extending to very-large and very-small $\Gamma$ at high $KL_\Gamma$.
The upper branch corresponds to steep-slope sources. The posterior PDF of an example source (ID$=$326, purple dashed line) is displayed in Fig.~\ref{fig:NH_Gm_KL}.
A major reason for their steep slopes is the existence of soft excess, which is not modeled with the ``single-powerlaw'' model.
With the ``double-powerlaw'' model, these steep slopes will be flattened (\S~\ref{sec:specproperties}).
The lower branch corresponds to flat-slope sources (e.g., radio-loud AGN) or sources with complex absorption, which is not modeled either.
An example source (ID$=$595, in purple dotted line) is also displayed in the $\Gamma$ PDF panel of Fig.~\ref{fig:NH_Gm_KL}.
The second cause of a high $KL_\Gamma$ is a small uncertainty of $\Gamma$, in other words, the posterior PDF is narrower than the Gaussian prior with a scale of $0.5$. The PDF of an example source (ID$=$9, in purple solid line) is also displayed in Fig.~\ref{fig:NH_Gm_KL}. As shown by the G3 panel, most of the sources with small $\Gamma$ uncertainties have $KL_\Gamma>$0.3.
Whatever the cause, the information (center offset or small uncertainty) from the data should be adopted, and thus the $\Gamma$ parameter should be let free in the fitting.
When $KL_\Gamma<$0.3, there is barely any information imported from the data, as the posterior and prior are similar or even identical (see an example, ID$=$864, in Fig.~\ref{fig:NH_Gm_KL}).
As displayed in the panel G1 and G2 of Fig.~\ref{fig:NH_Gm_KL}, the $\Gamma$ of such sources have large uncertainties, which can be as large as the prior width (magenta arrowed lines) in the worst cases.
Such sources always have median $\Gamma$ within $2.0\pm0.3$ (magenta filled histogram in the lower right panel of Fig.~\ref{fig:NH_Gm_KL}), and the measured $\Gamma$ are always consistent with $2.0$ within uncertainties.
In order to avoid huge uncertainties in their measurements of \N{H} and luminosity, we can adopt a stronger prior in the fitting by fixing $\Gamma$ at $2.0$.
The sources with $KL_\Gamma>$0.3 have a bimodal distribution (purple filled histogram in the lower right panel of Fig.~\ref{fig:NH_Gm_KL}). It is because the $KL_\Gamma>$0.3 criterion selects all the sources with abnormal slopes (either significantly steeper or flatter than $2.0$); most of the normal AGN (consistent with $\Gamma=$2.0) are included in the $KL_\Gamma<$0.3 sample, where we can safely fix $\Gamma$ at $2.0$.

\subsection{AGN Spectral properties}
\label{sec:specproperties}
\begin{figure}[!h]
\begin{center}
\includegraphics[width=0.80\columnwidth]{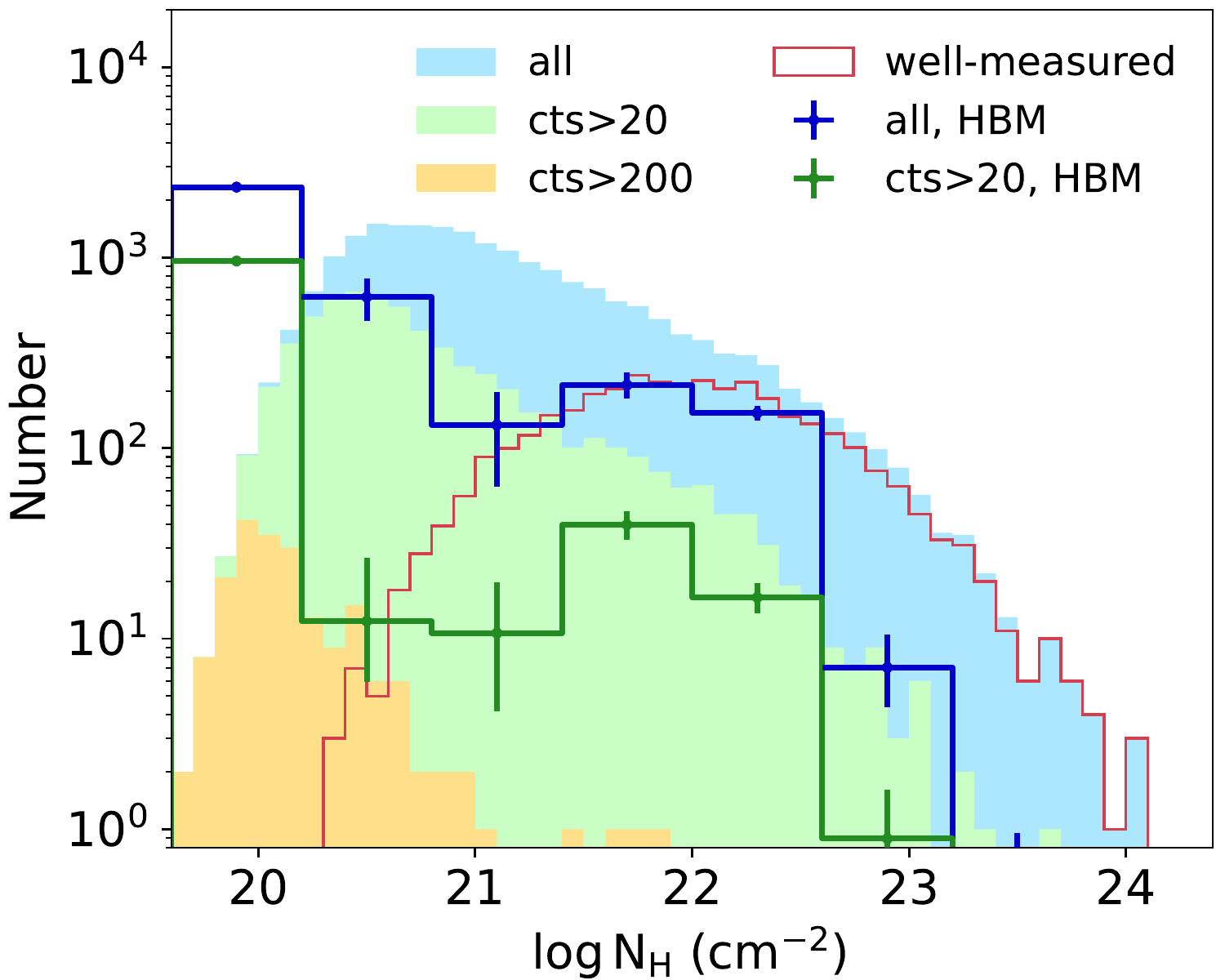}
\caption{The \N{H} distribution of the $20850$ eFEDS AGN with good redshifts ($zG\geqslant$3).
  The upper panel displays the distribution of median \N{H} of the ``single-powerlaw'' model for AGN.
  The stacked filled histograms display the median \N{H} distributions of the sources with 0.2--5 keV net counts $>$200 ($198$ sources; yellow), $>$20 ($6247$; yellow and green), and $>$0 ($20850$; yellow, green, and blue).
  The red empty histogram display the subsample of sources with \texttt{well-measured} \N{H} (\texttt{NHclass}$=$4).
  The blue and green empty histograms (with 1-$\sigma$ errorbars) present the inferred intrinsic \N{H} distribution using the HBM method for the whole sample ($20850$) and for the subsample with at least 20 counts ($6247$), respectively.
  }
\label{fig:hist_nh}
\end{center}
\end{figure}

\begin{figure}[!h]
\centering
\includegraphics[width=0.8\columnwidth]{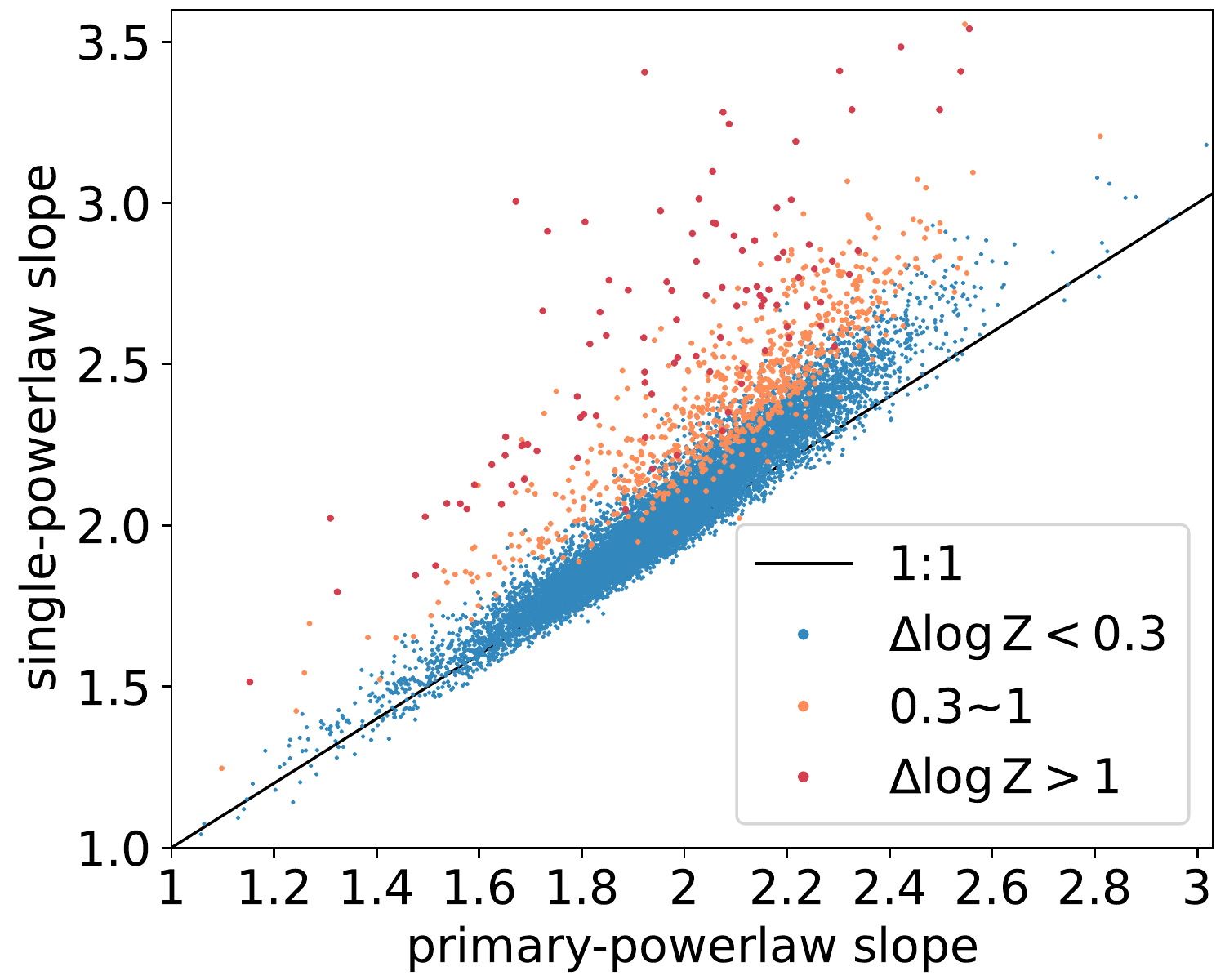}
\caption{
  Comparison between the posterior median $\Gamma$ measured using the ``single-powerlaw'' model and the median $\Gamma$ of the primary power-law component of the ``double-powerlaw'' models for AGN with at least 10 counts in the 0.2--5 keV band.
  The sources with $\log Z_\textrm{double-powerlaw}-\log Z_\textrm{single-powerlaw}$ below 0.3, between 0.3 and 1, and above 1 are plotted in blue, orange, and red, respectively.
}
\label{fig:comp_gamma}
\end{figure}

\begin{figure}[!h]
\begin{center}
\includegraphics[width=0.8\columnwidth]{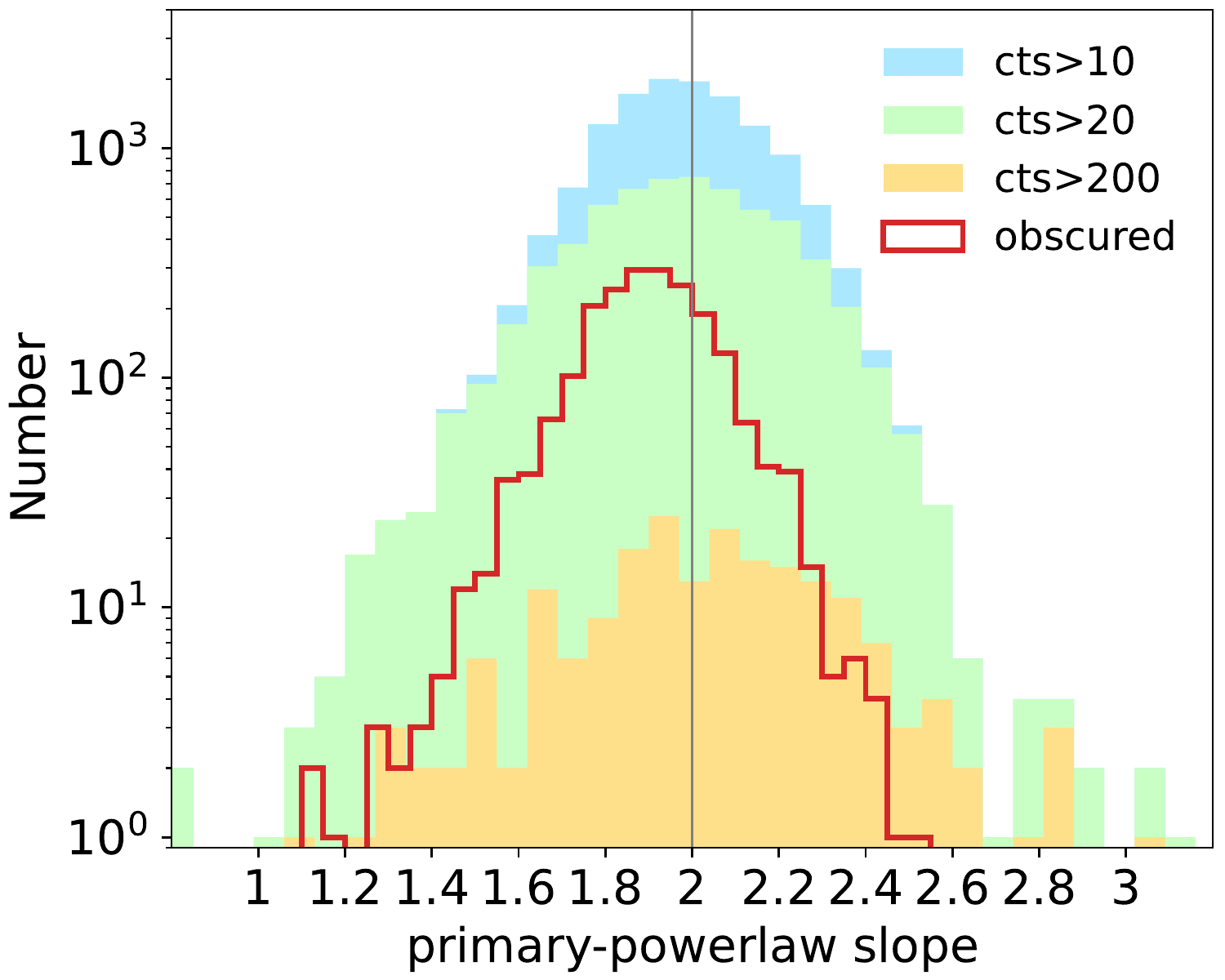}
\includegraphics[width=0.8\columnwidth]{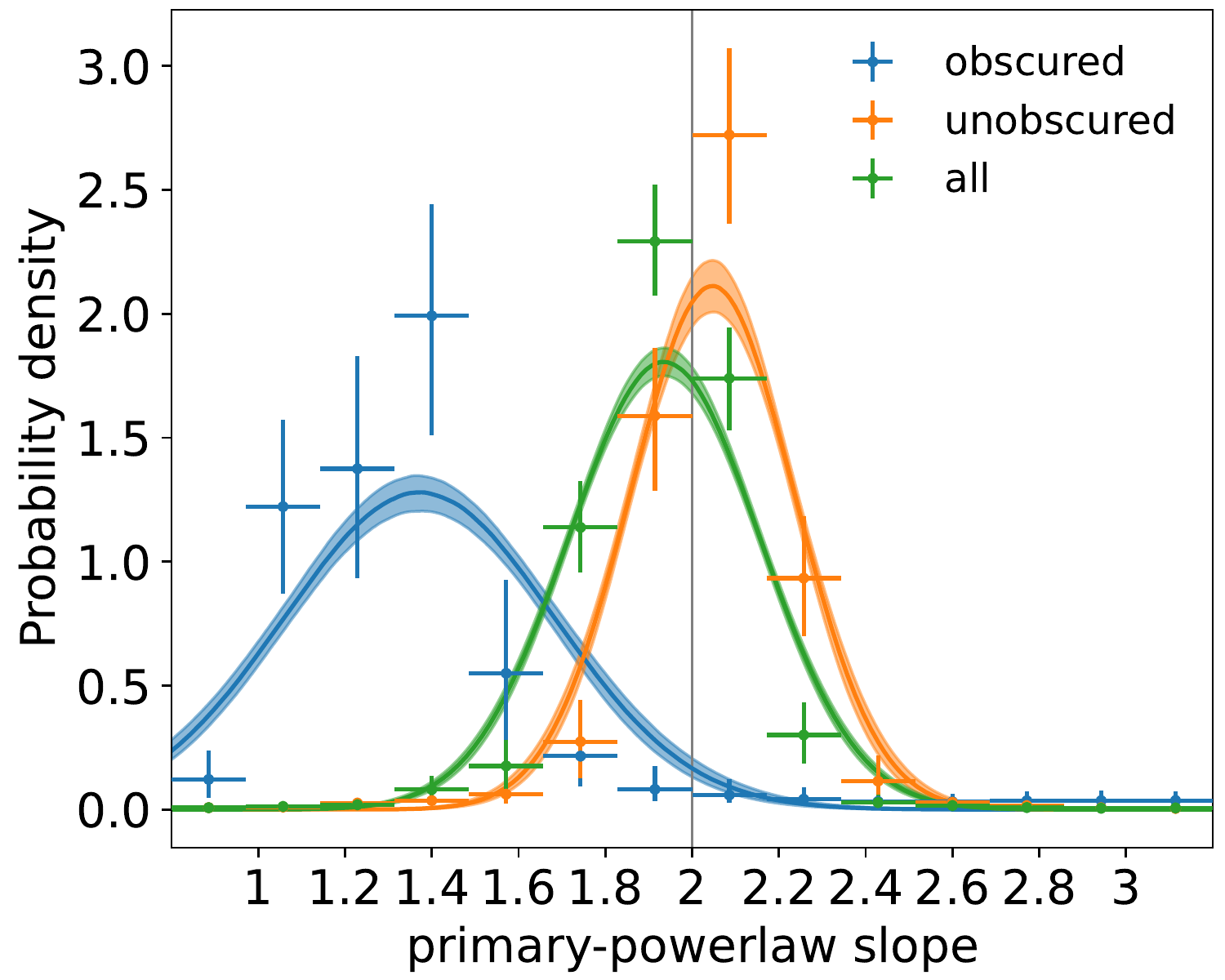}
\caption{
  The upper panel displays the distribution of the median primary power-law slope of the AGN with good redshifts ($zG\geqslant$3) measured with the ``double-powerlaw'' model.
  The stacked filled histograms correspond to the sources with 0.2--5 keV net counts $>$200 ($198$ sources; yellow), $>$20 ($6247$; yellow and green), and $>$10 ($13471$; yellow, green, and blue).
  The red empty histogram corresponds to the obscured subsample ($2070$) with \texttt{NHclass}$\geqslant$3 and median $\log$\N{H} above 21.5 (\S~\ref{sec:NH_Gm_KL}).
  The lower panel presents the inferred intrinsic $\Gamma$ distribution using the HBM method (\S~\ref{sec:hbm}).
  The error bars indicate the HBM histogram model.
  The lines with shaded regions (1-$\sigma$ uncertainty) indicate the HBM Gaussian model.
  The green, blue, and orange lines correspond to all the AGN with at least 10 counts, the obscured subsample, and the subsample with \texttt{NHclass}$=$2 (\texttt{unobscured}), respectively.
}
\label{fig:hist_gamma}
\end{center}
\end{figure}

Fig.~\ref{fig:hist_nh} displays the distribution of the median \N{H} measured with the ``single-powerlaw'' model.
The whole sample shows a wide \N{H} distribution. If selecting a subsample according to source brightness, a higher flux limit leads to relatively lower \N{H}.
This is because eROSITA has a much larger effective area in the soft band, which dominates the source counts/fluxes, a source brightness based selection is more of a soft-band selection and is biased against obscured sources.
As discussed in \S~\ref{sec:NH_Gm_KL}, these median values should not be used directly without checking the \texttt{NHclass} first.
In most cases, the median \N{H} is dominated by fluctuation.
Only in the \texttt{well-measured} cases (\texttt{NHclass}$=$4; red histogram in Fig.~\ref{fig:hist_nh}), where both the lower limit and the upper limit of \N{H} are well measurable, can the median value be considered as a good proxy of the expected \N{H}.
Thanks to the power of the Bayesian method, we can measure the \N{H} of obscured AGN in some cases with fewer than 20 counts (the overlapping between red and blue histograms in Fig.~\ref{fig:hist_nh}).
However, high \N{H} values measured in low counts cases should be treated with caution, because of the limited spectral model and potential, additional uncertainties induced outside the spectral fitting, e.g., in the spectra extraction or background estimation.

The apparent distribution of median \N{H} is largely broadened by the measurement uncertainties, which are often not only large but also asymmetric.
To measure the intrinsic \N{H} distribution of the sample, we run HBM to deconvolve the uncertainties.
We adopt the non-parametric histogram model for the \N{H} distribution, where there is no assumption on the shape of the distribution.
All the AGN with good redshift measurements are involved in the HBM calculation, including the \texttt{uninformative} ones, which practically have no impact to the results.
The inferred \N{H} histograms are normalized to the sample size in Fig.~\ref{fig:hist_nh} for comparison with the median \N{H} distributions.
The intrinsic distribution is largely dominated by unobscured sources.
Based on the HBM inferred histogram, the fractions of obscured AGN with $\log$\N{H}$>$21.5 and $\log$\N{H}$>$22 are 10\% and 5\%, respectively.
Selecting the sources with at least 20 counts, the obscured fractions are even lower, which are 5\% and 2\%, respectively.
Based on the median \N{H}, the obscured AGN with $\log$\N{H}$>$21.5 ($3311$ sources) and $\log$\N{H}$>22$ ($1673$ sources) comprise 16\% and 8\% of the sample, which are larger. This is because these high \N{H} sources have wide PDF, a substantial fraction of which resides in the less-obscured regime.

In Fig.~\ref{fig:comp_gamma}, we compare the slope of the primary power-law measured using the ``double-powerlaw'' model with that measured using the ``single-powerlaw'' model. Having the soft excess fitted with the additional power-law component, a much flatter slope of the primary power-law is obtained for the steep-slope sources as displayed by the upper $\Gamma$--$KL_\Gamma$ branch in Fig.~\ref{fig:NH_Gm_KL}.

The distributions of the median primary-power-law slope measured with the ``double-powerlaw'' model are displayed in Fig.~\ref{fig:hist_gamma}. Again, such median values must be considered with caution. In the case of faint sources where the parameter is not constrained, the measured values only reflect the assumed prior.
For the brightest sources with more than 200 counts, the distribution has a median value of $2.03$.
To measure the intrinsic slope distribution of the AGN with at least 10 counts, we use the HBM method to deconvolve the large uncertainties.
We adopt two models for the $\Gamma$ distribution, a histogram model and a Gaussian model.
As displayed in the bottom panel of Fig.~\ref{fig:hist_gamma}, the inferred histogram and Gaussian model are roughly consistent with each other.
With the Gaussian model, we derive a mean of $1.94\pm0.01$ and a standard deviation of $0.22\pm0.01$.

If considering only the obscured subsample ($2070$ sources), the posterior median $\Gamma$ distribution has a median of $1.90$ (red histogram in Fig.~\ref{fig:hist_gamma}), but through HBM, we derive a much flatter mean slope of $1.37\pm0.02$ and a standard deviation of $0.31\pm0.02$.
This is because the obscured subsample includes some extremely flat ($\Gamma<$1.4) sources, which might have ionized absorber.
As obscured sources have poor constraints on $\Gamma$, such extremely flat sources, which are negligible in the whole sample, become prominent in the small obscured subsample and bring down its HBM $\Gamma$ measurement.
If considering only the unobscured subsample with \texttt{NHclass}$=$2 ($6678$ sources), the inferred Gaussian model has a mean of $2.05\pm0.01$ and a standard deviation of $0.19\pm0.01$.
Based on the analysis above, in our spectral models (\S~\ref{sec:AGNmodel}), we adopt a Gaussian prior for the primary power-law slope, which is centered at $2.0$ but have a much larger scale (0.5) than the intrinsic scatter of the whole sample.
Catalog-wise, this is an unbiased, weak prior. Only in the cases with extremely steep or flat slopes, the prior causes a bias in the posterior $\Gamma$ towards the median slope of $2.0$.
Compared with hard-band selected AGN samples \citep[e.g.,][]{Liu2017}, the median slope of the eFEDS AGN is relatively steeper, because the soft-band-dominated sample selection favors steeper sources (see more discussions in Nandra et al. submitted).

\begin{figure}[h]
  \centering
\includegraphics[width=0.79\columnwidth]{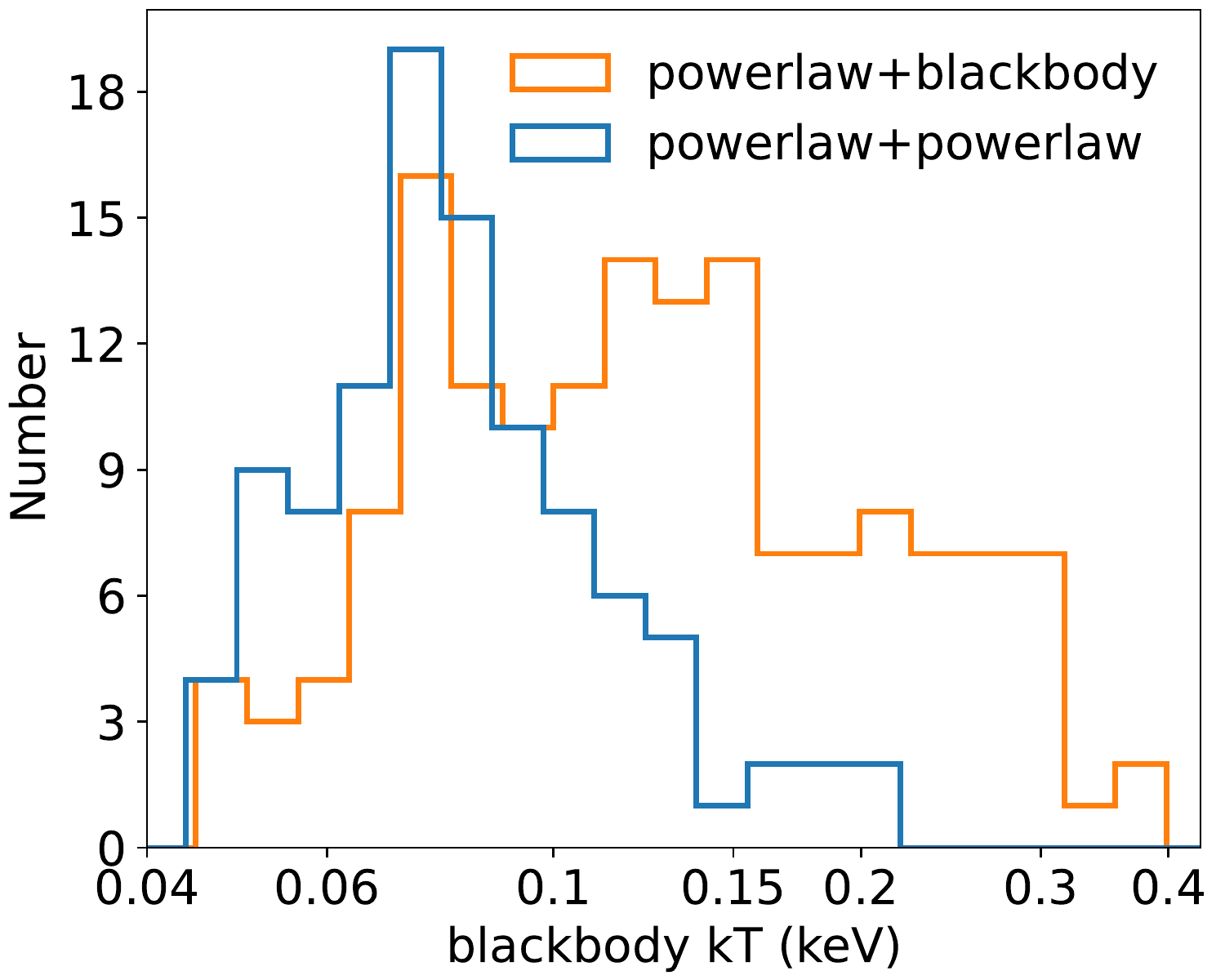}
\includegraphics[width=0.79\columnwidth]{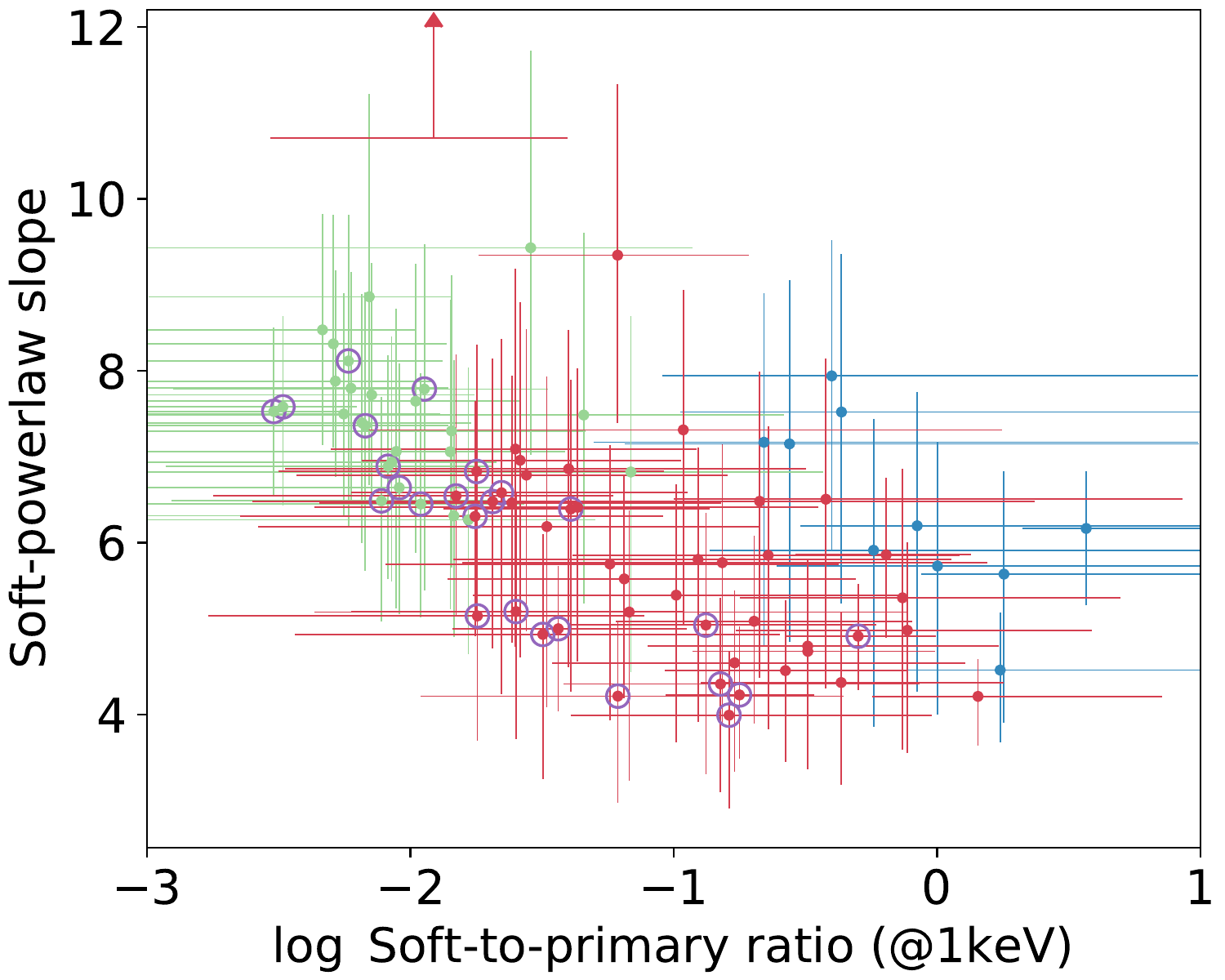}
\caption{
  The upper panel displays the blackbody temperature distributions of the $151$ AGN with $\log Z_\textrm{powerlaw+blackbody}-\log Z_\textrm{powerlaw}>1$ (orange) and the $102$ AGN with $\log Z_\textrm{powerlaw+powerlaw}-\log Z_\textrm{powerlaw}>1$ (blue).
  The lower panel displays the slope--strength distribution of the soft power-law for the $82$ AGN with well-detected soft excess. The strength is expressed as the ratio of the normalization at 1 keV between the soft power-law and the primary power-law.
  The ones with the relative-strength factor pegged at the lower boundary (practically, HDI lower limit $<$-2.8) and the ones with the factor pegged at the upper boundary (practically, HDI upper limit $>$0.99) are plotted in green and blue, respectively. The others are plotted in red.
  One source with the $\Delta\Gamma$ parameter pegged at the upper boundary (practically, HDI upper limit $>11.9$) is plotted with an arrow.
  The sources with well-constrained primary power-law slopes ($KL>$0.3) are marked with purple circles.
  }
\label{fig:softexcess}
\end{figure}

To select sources with soft excess detected, we can adopt a typical threshold of $\log Z_\textrm{powerlaw+powerlaw}-\log Z_\textrm{powerlaw}>1$ based on the ``double-powerlaw'' model, or adopt $\log Z_\textrm{powerlaw+blackbody}-\log Z_\textrm{powerlaw}>1$ based on the power-law plus blackbody model \citep{Buchner2014}.
The two selections result in a number of $102$ and $151$ AGN, respectively.
The blackbody temperature of these sources measured through the power-law plus blackbody model are displayed in Fig.~\ref{fig:softexcess}.
The temperature of the ``double-powerlaw'' selected soft-excess AGN tends to concentrate below 0.1 keV, as typically found in AGN \citep[e.g.,][]{Winter2012,Ricci2017}.
But the blackbody-model selected sources show a significant excess above 0.1 keV.
As discussed in \S~\ref{sec:modelselection}, the power-law plus blackbody model is too flexible, as it selects not only sources with soft excess but also sources whose spectra are dominated by blackbody-like (instead of powerlaw-like) emission.
Therefore, we stick with the ``double-powerlaw'' model in this work. More detailed discussions about soft excess detected by eROSITA are to be presented in Waddell et al. (in preparation).

We noticed that the parameter ranges we adopted in the ``double-powerlaw'' model is sufficiently wide for the majority of the sources but not for a few extreme cases. To measure the soft excess shape more accurately, we perform a special fitting for the $102$ $\Delta\log Z$ selected AGN with the ``double-powerlaw'' model, enlarging the $\Delta\Gamma$ range from $0.5\sim5$ to $0.5\sim12$ and enlarging the range of the soft-powerlaw relative-strength factor from $0.001\sim1$ to $0.001\sim10$.
Then we further select sources with good redshift measurements $(zG\geqslant$3), and select the ones whose KL divergence of the $\Delta\Gamma$ parameter is $>$0.3.
These criteria lead to $82$ AGN with soft excess detected, for which the spectra show curvature with respect to the ``single-powerlaw'' model and the shape of the soft component can be constrained by the data.
With the special ``double-powerlaw'' fitting, the measured soft-powerlaw parameters of these soft-excess AGN are displayed in Fig.~\ref{fig:softexcess}.
If considering only the bright AGN with more than 500 net counts and good redshift measurements, there are $50$ sources in total, out of which nine (18\%) have soft excess.
Even with the enlarged parameters ranges, some sources still have the HDI limits of the relative strength pegged at the boundaries.
When the factor is pegged at the upper boundary, the soft excess is very strong, or the primary power-law is in a very-low state, like the case of the eROSITA observation of 1H0707-495 \citep{Boller2021}.
When the factor is pegged at the lower boundary, the soft excess is extremely soft (slope $>$6) and appears mostly in the very-soft band ($<$0.5 keV).
Caution should be paid to such sources, since there might be additional calibration uncertainties caused by light leak in TM5 and TM7 \citep{Predehl2021}.
In Fig.~\ref{fig:softexcess}, we mark the $25$ sources with the KL divergence of the primary-powerlaw slope $>0.3$.
Strictly speaking, only in these sources with well-constrained primary power-law slopes are the soft excess well constrained by the data.
There is a strong correlation between the soft-powerlaw slopes and strength of these sources, just because of parameter denegeracy.

\subsection{X-ray luminosities}
\label{sec:AGNLum}

\begin{figure}[h]
\begin{center}
\includegraphics[width=0.9\columnwidth]{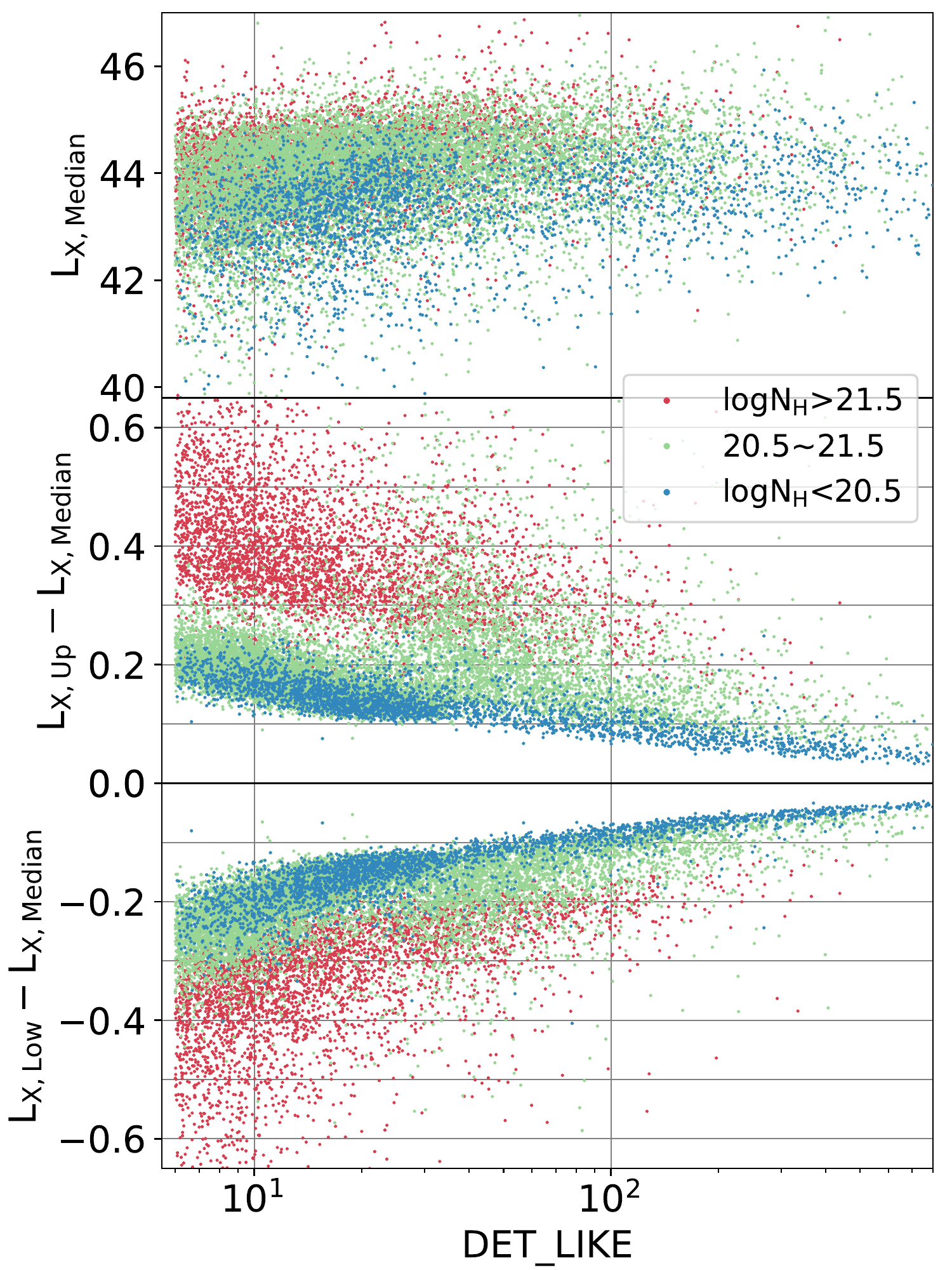}
\caption{The rest-frame 0.5-2 keV intrinsic luminosity $L_X$ of AGN as a function of the 0.2--2.3 keV source detection likelihood. Sources with posterior median $\log$\N{H} below 20.5, between 20.5 and 21.5, and above 21.5 are plotted in blue, green, and red, respectively.
  }
\label{fig:Lx_cts}
\end{center}
\end{figure}

We have measured the 0.5-2 and 2-10~keV luminosities using a few spectral models (\S~\ref{sec:lum_flux}).
Since most sources have few counts in the hard-band, the 2-10 keV luminosities have large uncertainties and must be treated with caution. We only discuss the 0.5-2~keV luminosity $L_X$ in this work. The 2-10 keV luminosity is discussed in more details for the hard-band selected sample (Nandra et al. submitted).
Among the used models, we choose the most appropriate model to measure $L_X$ for each AGN, based on the principle of adopting weaker priors for high-quality data and stronger priors for low-counts data.
When the \N{H} parameter of the ``single-powerlaw'' model is unconstrainable, i.e., \texttt{NHclass} is \texttt{uninformative}, we consider the data quality as too low for a reasonable spectral fitting.
For such sources ($3570$), we adopt the ``shape-fixed-powerlaw'' model (model 5) to calculate $L_X$.
For sources with high spectra quality, we give preference to the flexible ``double-powerlaw'' model, but avoid using it for obscured AGN.
Because of the high degeneracy between absorption and soft excess in the soft band, the soft excess model is too flexible to be used for absorption correction in the case of obscured AGN.
Based on the ``single-powerlaw'' fitting, we select obscured AGN as the ones with an \N{H} class of \texttt{well-measured} or \texttt{mildly-measured} and having a median $\log$\N{H} above 21.5. 
For these obscured AGN ($3472$ sources), we choose the ``single-powerlaw'' model (model 1).
For the other sources, we adopt the ``double-powerlaw'' model (model 2; $5703$ sources) as long as it has at least $20$ counts in the 0.2--5 keV band.
For the faint sources with less than $20$ counts, we adopt model 1 if it leads to a $KL_\Gamma>0.3$ ($888$ sources).
For the other sources with $KL_\Gamma<0.3$ ($8319$), we import a stronger spectral shape prior and thus adopt the ``$\Gamma$-fixed-powerlaw'' model (model 4).
In the cases of model 1,2, and 4, there is at least an \N{H} parameter constrained by the data; we call their $L_X$ as spectral measurements.
In the cases of model 5, there is only a normalization parameter and no spectral shape information; we call the $L_X$ as counts measurements.
The counts measurements of $L_X$ are less reliable in the sense that the spectral shape uncertainty is not considered; such $L_X$ are presented but excluded from further analysis in this paper.

The lower panel of Fig.~\ref{fig:AGN_frac} displays the fraction of AGN with a spectral $L_X$ measurements (model 1,2,or 4).
The sources with a counts-measured $L_X$ concentrate at detection likelihood $<10$.
Therefore, we recommend a sample selection threshold of detection likelihood $>10$ for the sake of a reasonable X-ray spectral analysis.

Fig.~\ref{fig:Lx_cts} displays the spectral-measured $L_X$ and their uncertainties as a function of the 0.2--2.3 keV detection likelihood. The $L_X$ uncertainty is largely affected by \N{H}. For sources with $\log$\N{H}$<$21, the $L_X$ uncertainty is mostly below $0.2$ dex.

\begin{figure}[h]
\begin{center}
\includegraphics[width=\columnwidth]{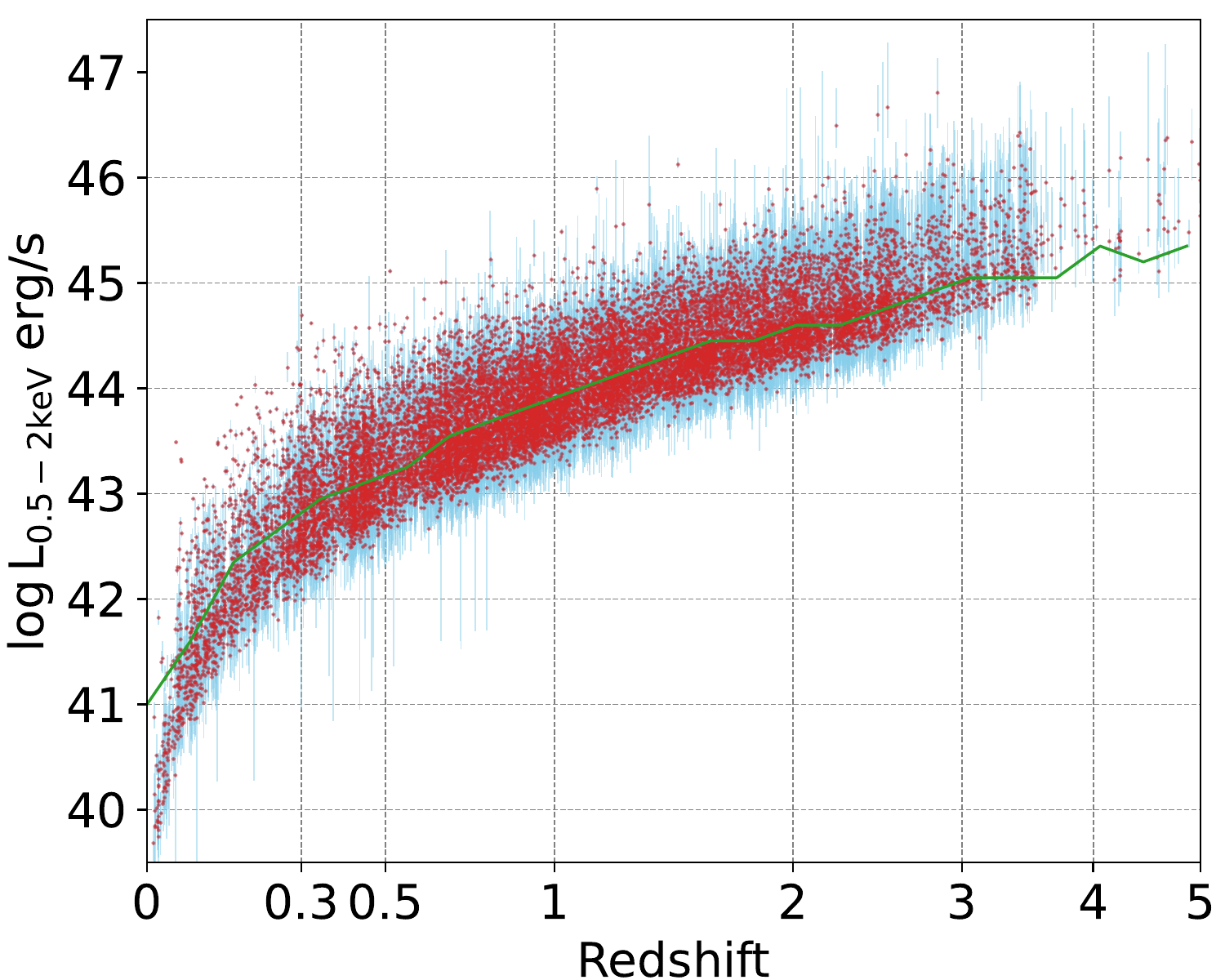}
\caption{
  The rest-frame 0.5-2 keV intrinsic luminosity $L_X$ and redshift (in the scale of $\log(1+z)$) distribution of $17826$ AGN with good redshift measurements ($zG\geqslant$3) and with spectral measurements of $L_X$ (model 1,2,or 4).
  The $L_X$ median and 68\% percentile interval measured from the posterior distributions are plotted with red points and blue error bars.
  The purple line is the 90\%-detection curve of AGN with $\log$\N{H}$<21$ measured through simulation (Liu et al. submitted).
}
\label{fig:Lz}
\end{center}
\end{figure}

Fig.~\ref{fig:Lz} displays the $L_X$ and redshift distribution of the AGN with both a spectral measurement of $L_X$ and a good redshift measurement ($zG\geqslant$3).

\subsection{X-ray -- UV correlations}
\begin{figure*}[h]
\begin{center}
\includegraphics[width=0.32\textwidth]{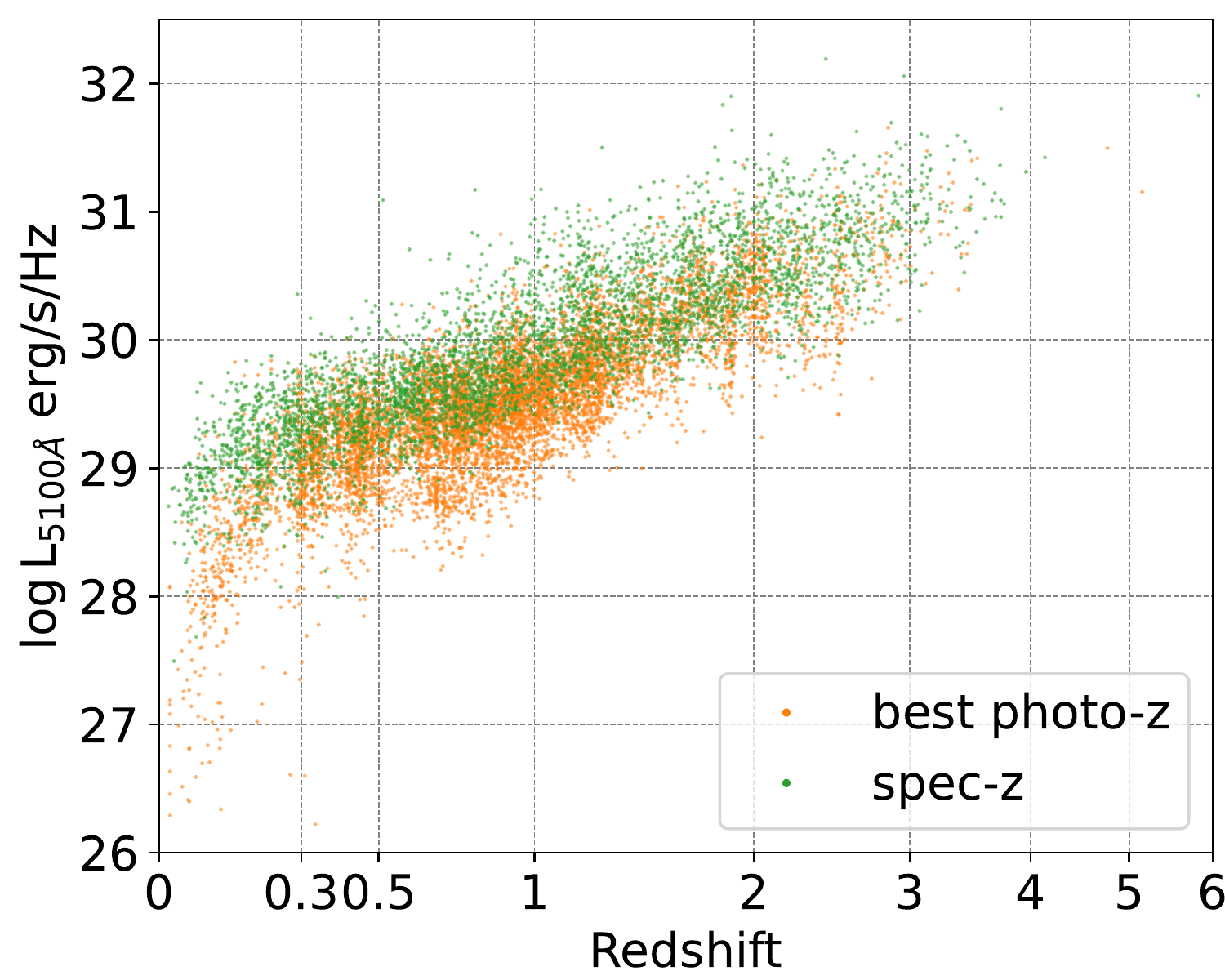}
\includegraphics[width=0.32\textwidth]{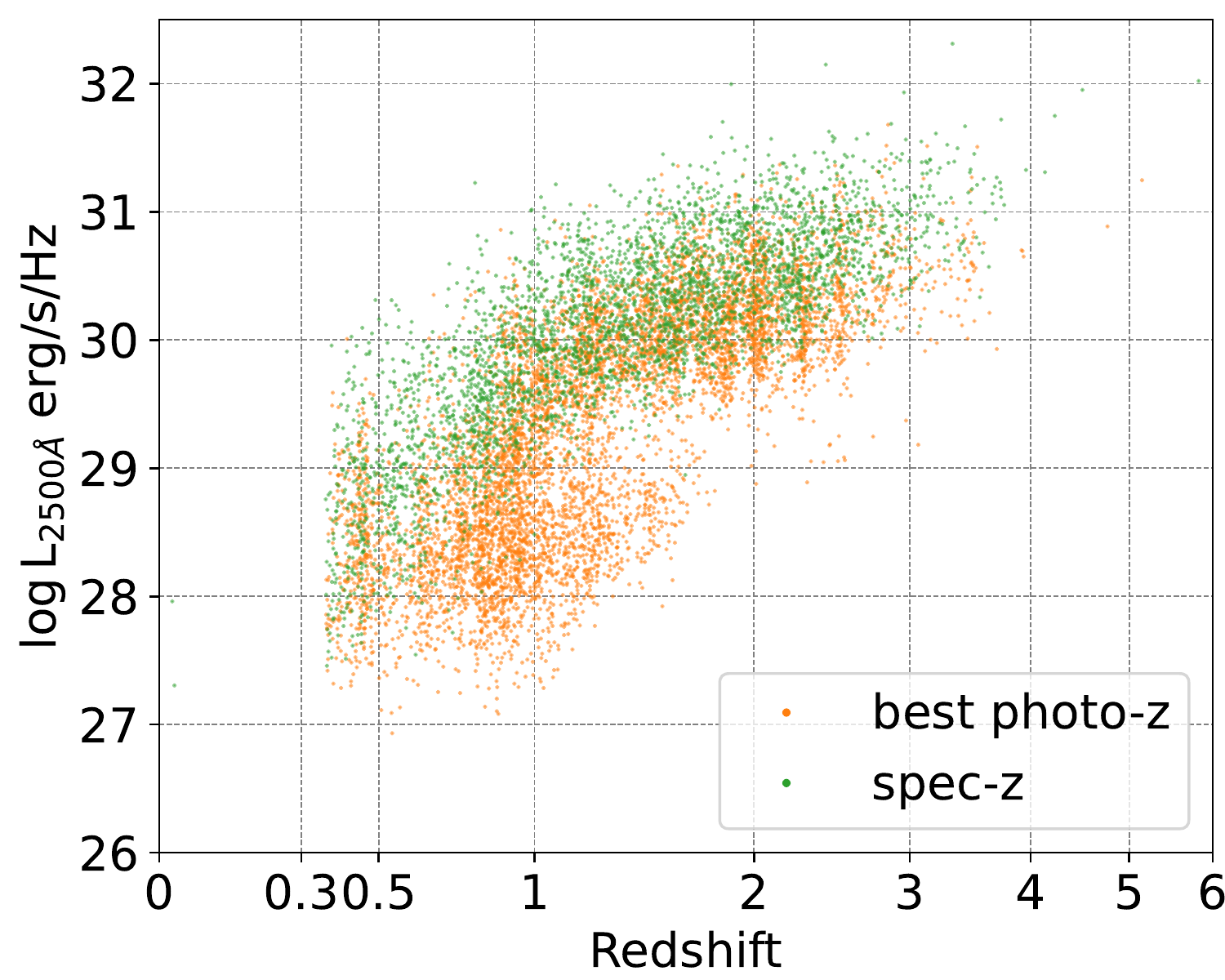}
\includegraphics[width=0.32\textwidth]{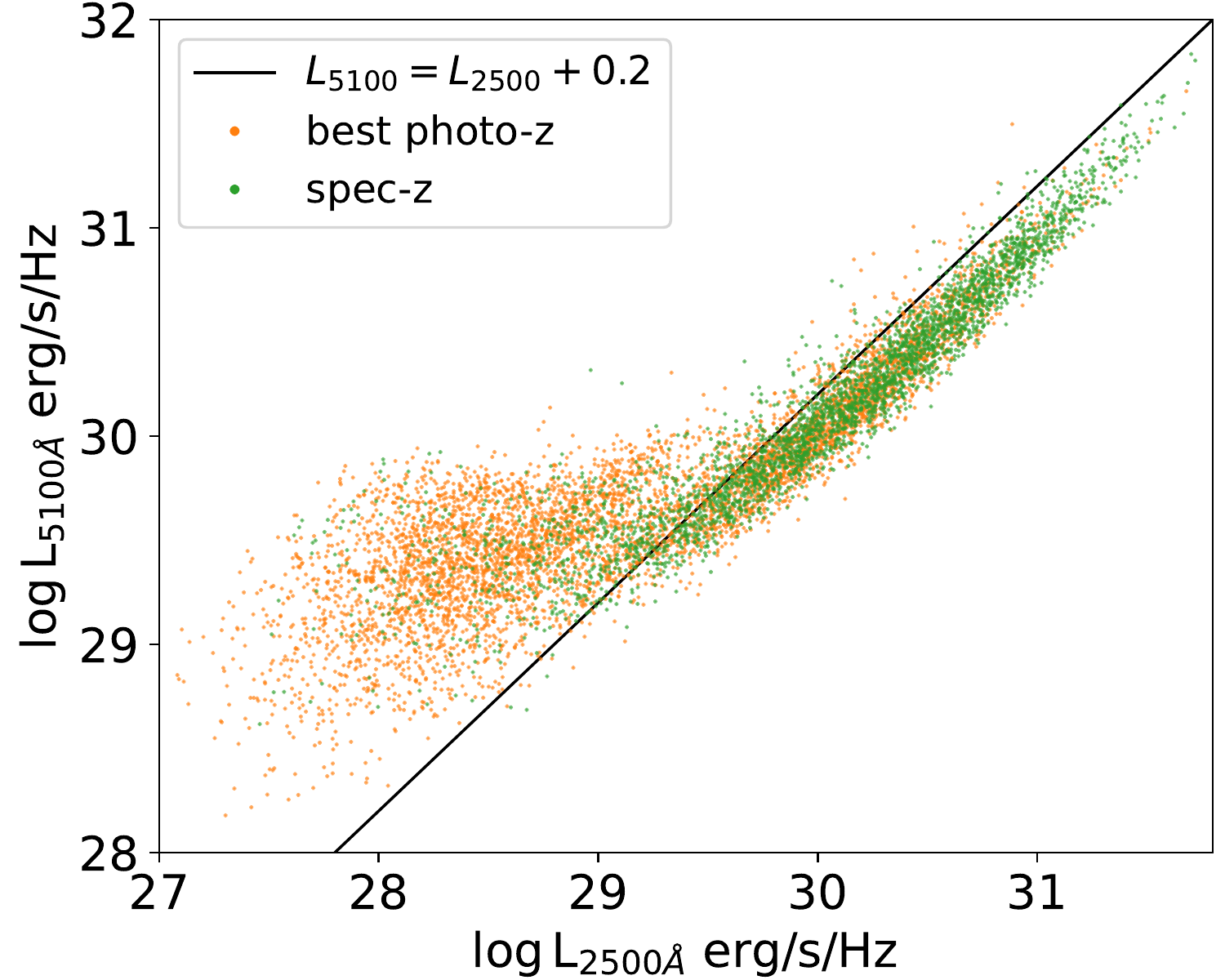}
\caption{The left and middle panels display the $L_{2500}$--$z$ and $L_{5100}$--$z$ distributions. The right panel display the $L_{2500}$--$L_{5100}$ correlation of the sources with both good $L_{2500}$ and $L_{5100}$ measurements. Only the sources with spec-z (green) or with the highest-quality photo-z ($zG=$4; orange) are included.}
\label{fig:L2500_L5100}
\end{center}
\end{figure*}

Paper II has measured photo-z of all the eFEDS sources through SED fitting.
Based on the multi-band photometry data and the best-fit SED model, we measure the UV 2500\A{A} and optical 5100\A{A} rest-frame monochromatic luminosities $L_{2500}$ and $L_{5100}$.
In the catalog, spec-z are adopted when available, which could be different from the photo-z measurements.
For those sources with spec-z, we rerun the SED fitting with the redshift fixed at the spec-z.
If with the fixed spec-z the data cannot be well fitted with any SED model, we abandon the SED data ($16$ sources).
We use the photometry data between rest-frame 1500\A{A} and 3500\A{A} to measure $L_{2500}$ and the data between rest-frame 4100\A{A} and 6100\A{A} to measure $L_{5100}$.
In these two bands, we normalize the best-fit SED component of AGN plus host galaxy to the data separately, multiplying the model by the mean data to model ratio $f_{2500}$ and $f_{5100}$ to calculate $L_{2500}$ and $L_{5100}$, respectively.
We consider the $L_{2500}$ and $L_{5100}$ as good measurements only if the data quality satisfies the following criteria.
First, the source must have either spec-z or the highest-quality photo-z ($zG=$4).
We require at least three photometry data points in the 1500\A{A}--3500\A{A} band, thus excluding the low-redshift sources.
For $L_{5100}$, which is in the optical band where most sources have relatively better photometry, we require at least one data point in the 4100\A{A}--6100\A{A} band.
We require the mean data to model ratio in these two bands $f_{2500}$ and $f_{5100}$ in the range of $0.7$--$1.4$ ($0.15$ dex). When the data to model deviation exceeds this range, we consider the SED model as less reliable.
These criteria result in $10346$ $L_{2500}$ measurements and $12274$ $L_{5100}$ measurements.

\begin{figure}[h]
\begin{center}
     \includegraphics[width=0.9\columnwidth]{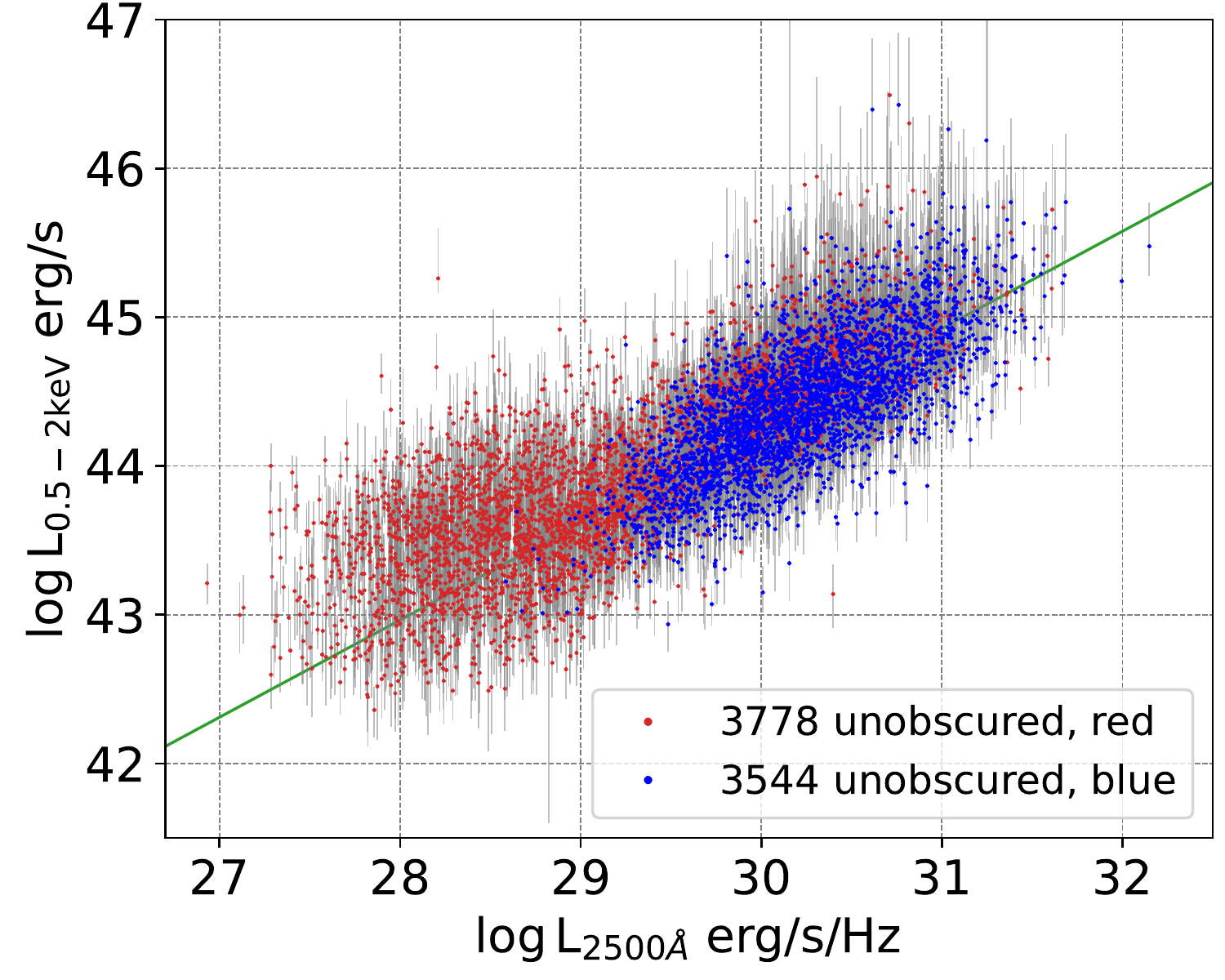}
     \includegraphics[width=0.9\columnwidth]{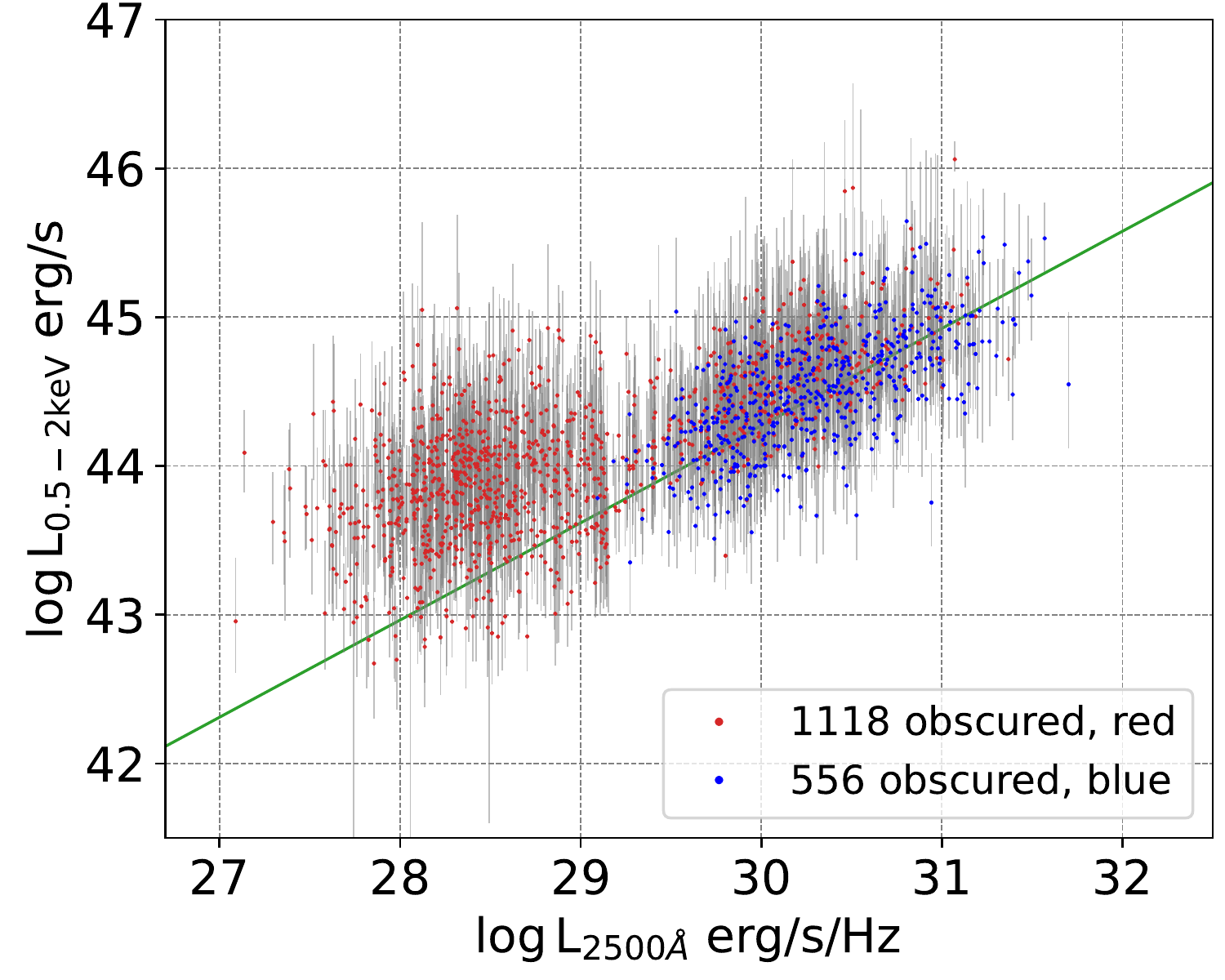}
\caption{
  The X-ray 0.5-2 keV luminosity vs UV 2500\A{A} luminosity scatter plots for the X-ray unobscured AGN (upper panel) and X-ray obscured AGN (lower), respectively.
  UV-strong AGN and UV-weak AGN (with $L_{5100}-L_{2500}$ below and above $0.2$) are plotted in blue and red, respectively.
   The green lines are obtained by linear regression among the unobscured, UV-strong AGN (blue points in the upper panel).
}
\label{fig:Luv_Lx}
\end{center}
\end{figure}

The distributions of $L_{2500}$ and $L_{5100}$ are displayed in Fig.~\ref{fig:L2500_L5100}.
Discontinuity and bimodality can be seen in the luminosities--redshift distributions, which reflect various sample selection effects in the hybrid multi-band redshift catalog and various SED models adopted in the photo-z measurements.
At high $L_{2500}$, the sources show a strong correlation between $L_{2500}$ and $L_{5100}$ , indicating they are typical type-I AGN with a blue UV-optical continuum.
At low $L_{2500}$, the $L_{5100}$ becomes higher than $L_{2500}$, suggesting strong UV extinction in type-II AGN and/or strong contamination from host galaxies.
We use a criterion of $L_{5100}-L_{2500}<0.2$ to select blue AGN. The others are called red AGN.

To analyze the correlation between the X-ray and UV emission, we select only the sources with spectral measurements of $L_X$, excluding the counts measurements based on model 5 (``shape-fixed-powerlaw'').
We select X-ray obscured sources as the \texttt{well-measured} or \texttt{mildly-measured} sources with median $\log$\N{H}$\geqslant$21.5 and select X-ray unobscured sources as the ones with median $\log$\N{H}$<$21.5 excluding any sources classified as \texttt{uninformative}).
Fig.~\ref{fig:Luv_Lx} displays the correlation between the 0.5-2 keV X-ray luminosity $L_X$ and the UV luminosity $L_{2500}$ for the X-ray unobscured and X-ray obscured sources separately.
Since $L_X$ is corrected for absorption, both the X-ray unobscured and obscured blue AGN show a strong $L_X$--$L_{2500}$ correlation.
For red AGN, the $L_{2500}$ is relatively lower at low luminosities, indicating UV extinction in such sources.
This trend of lower $L_{2500}$ to $L_X$ ratio is more significant X-ray obscured red AGN, indicating a larger fraction of type-II AGN among such sources.

\begin{figure}[h]
\begin{center}
  \includegraphics[width=0.9\columnwidth]{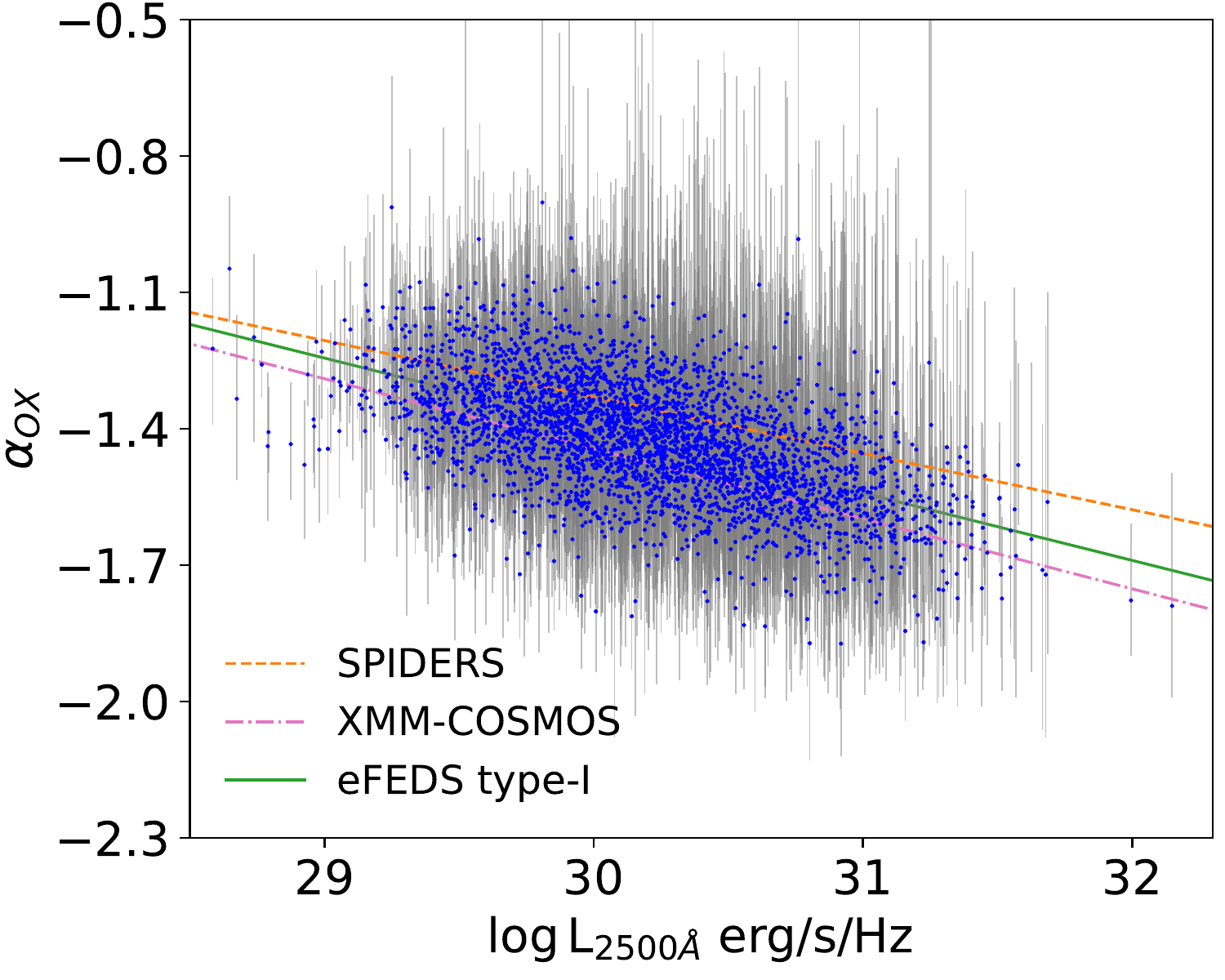}
     \includegraphics[width=0.9\columnwidth]{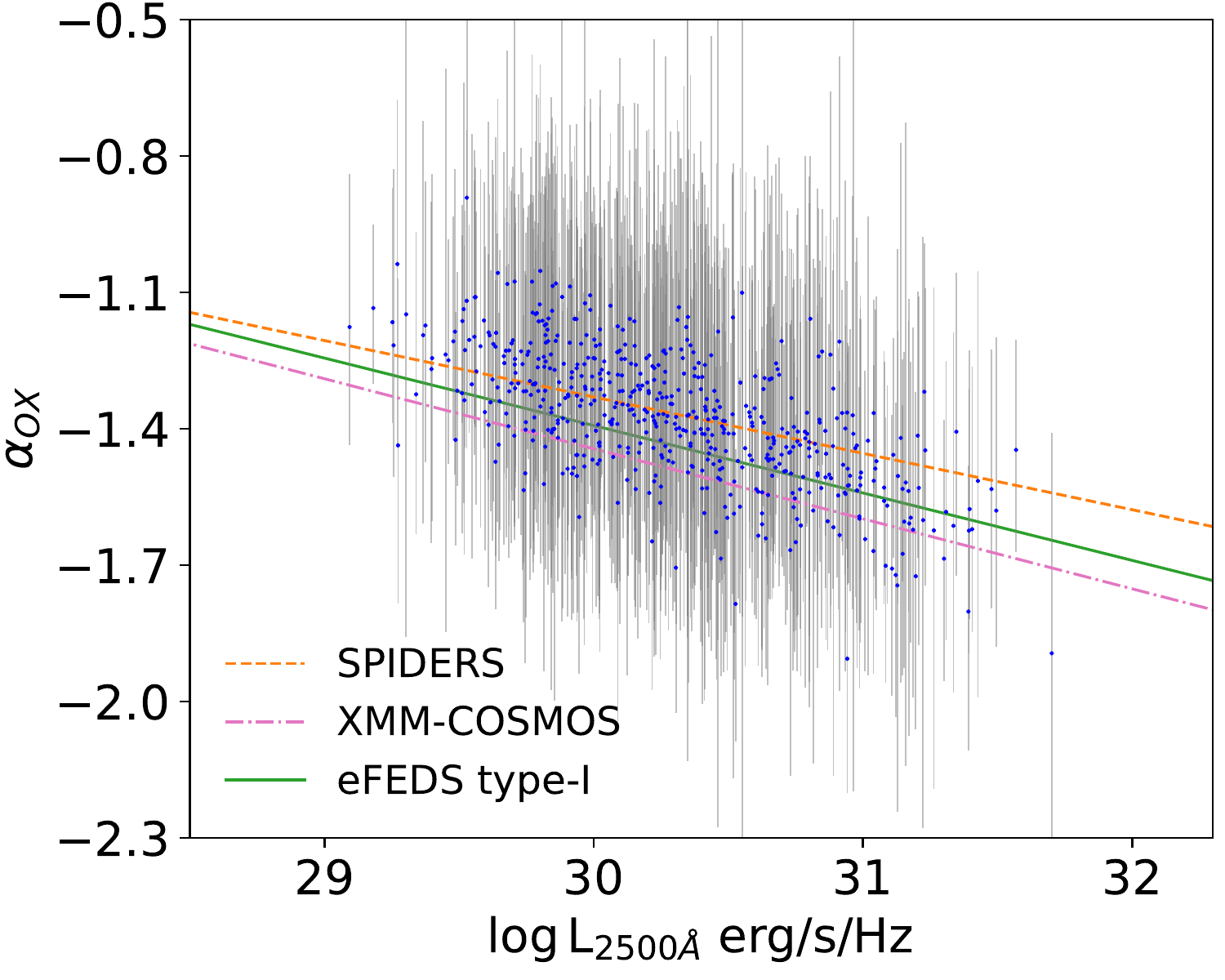}
\caption{
  The $\alpha_{OX}$--$L_{2500}$ scatter plots of the X-ray unobscured (upper panel) and obscured (lower) blue AGN.
  The green solid line ($\alpha_{OX} = 2.990-0.146\log L_{2500}$) is based on the eFEDS type-I subsample (upper panel); the orange dashed line ($\alpha_{OX} = 2.39-0.124\log L_{2500}$) is measured from the SPIDERS AGN catalog \citep{Coffey2019}; the magenta dash-dotted line ($\alpha_{OX} = 3.176-0.154\log L_{2500}$) is from the XMM-COSMOS AGN catalog \citep{Lusso2010}.
}
\label{fig:alphaox}
\end{center}
\end{figure}

Fig.~\ref{fig:alphaox} displays the $\alpha_{OX}$ of the blue AGN, which is defined as
$$\alpha_{OX} = \frac{\log(L_\textrm{2keV}/L_{2500})}{\log(\nu_\textrm{2keV}/\nu_{2500})},$$
where $L_\textrm{2keV}$ , $L_{2500}$, $\nu_\textrm{2keV}$ , and $\nu_{2500}$ are the monochromatic luminosities and frequencies at 2 keV and 2500\A{A}, respectively.
We consider the X-ray unobscured blue AGN as type-I AGN. By linear regression on them, we find the following relation (upper panel of Fig.~\ref{fig:alphaox}):
$$\alpha_{OX} = 2.990\pm0.108 - (0.146\pm0.004) \log L_{2500}.$$
The X-ray obscured blue AGN show slightly larger $\alpha_{OX}$ than the X-ray unobscured blue AGN, possibly because of selection bias against low-$L_X$ obscured AGN.
For comparison, we also plot the $\alpha_{OX}$--$L_{2500}$ relations measured from the ROSAT and \XMM{} detected SDSS-IV/SPIDERS (SPectroscopic IDentification of eROSITA Sources) type-I AGN catalog \citep{Coffey2019} and the XMM-COSMOS type-I AGN catalog \citep{Lusso2010}.
The differences between these relations might be caused by differences in the intrinsic properties of the AGN, e.g., the black hole mass and accretion rate.

\subsection{Particular AGN}
In this section, we introduce some particular AGN in our catalog, including the most luminous AGN and a candidate of Compton-thick type-I AGN.

\begin{figure}[hptb]
\centering
\includegraphics[width=0.9\columnwidth]{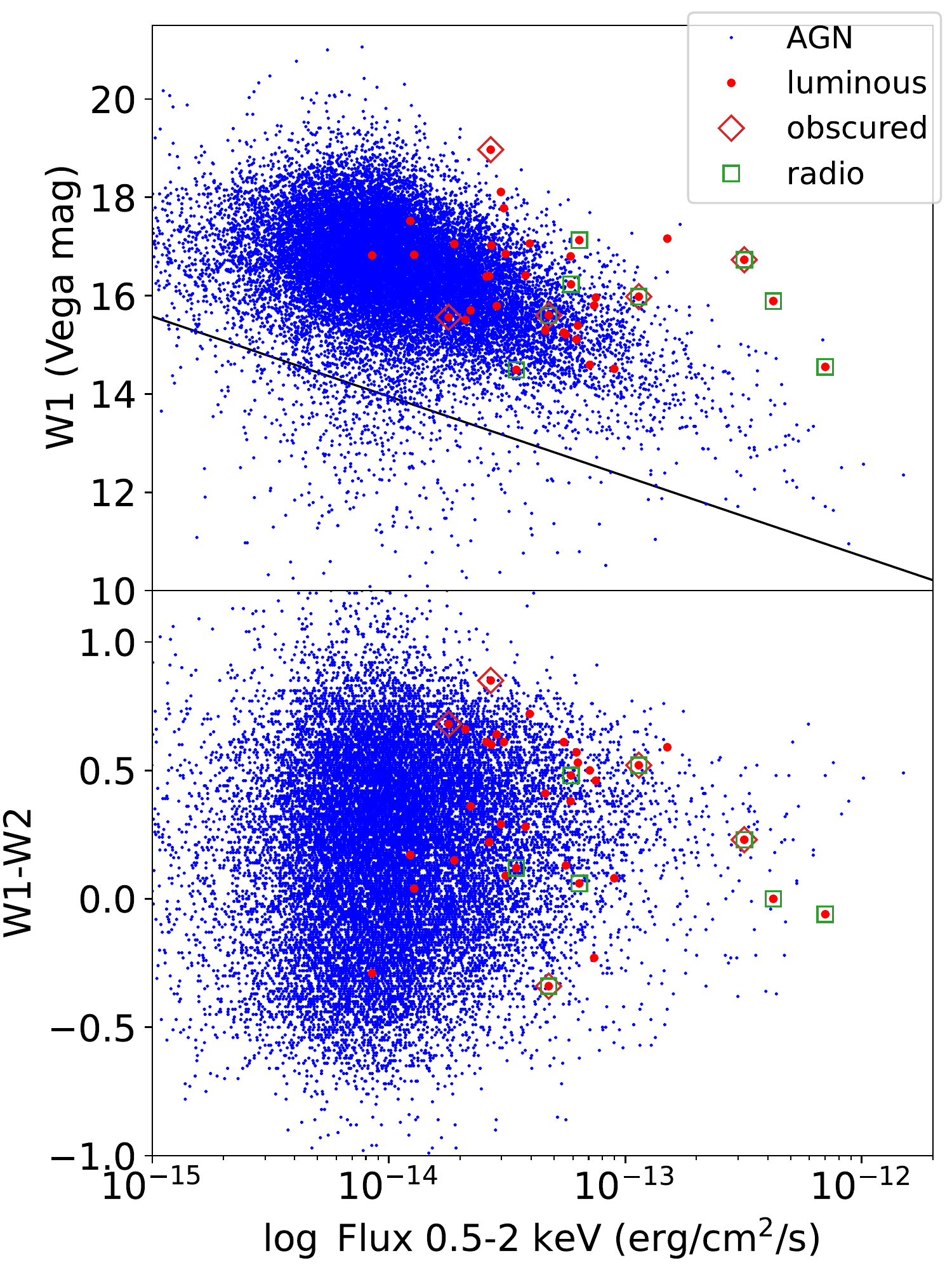}
\caption{Distributions of the eFEDS AGN in the space of WISE W1 magnitude (upper panel) and W1-W2 color (lower) versus observed 0.5--2 keV flux. The $36$ luminous AGN are plotted in red points. The red diamonds and green squares indicate the five obscured luminous AGN and the eight radio-detected ones, respectively. The black line corresponds to $W1=-1.625\log \mathrm{Flux}_\mathrm{0.5-2keV}-8.8$ \citep{Salvato2018}.}
\label{fig:Flux_W1W2}
\end{figure}

There are $91$ AGN in our catalog with 0.5--2 keV luminosities (\texttt{LumiIntr\_Med\_s}) above 10$^{46}$ erg/s, out of which we select $36$ luminous AGN with good redshift measurements ($zG\geqslant$3) and have spectral measurements of luminosities (\texttt{Model}$<$5).
Among them, only two have spec-z (ID 86, 830).
There are five obscured luminous AGN with $\log$\N{H} above 21.5.
Four of these five sources (ID 505,2267,4944,and 21377) have high photo-z ($z>$4) with $zG=$3 (not the best quality), and only one of them (ID$=$86) has a spec-z of 3.277.
We look for radio counterparts of these luminous AGN in the 1.4~GHz FIRST survey catalog \citep{White1997} and the 3~GHz VLASS Epoch 1 catalog \citep{Gordon2020}, adopting a searching radius of 2\arcsec.
Eight sources (ID 19, 86, 830, 988, 2157, 2267, 505, 4005) are detected in the VLASS catalog, and the first six of them are also detected in the FIRST catalog.
In Fig.~\ref{fig:Flux_W1W2}, we compare these luminous AGN with others in the space of infrared magnitude/color and X-ray flux.
A few of them have abnormally large X-ray to infrared flux ratios, especially the radio-detected ones, indicating the existence of powerful or even beamed jets.
Among the five obscured luminous AGN, three are radio-detected.
The X-ray obscuration of them might be due to intergalactic medium instead of the AGN themselves \citep{Arcodia2018}.

We noticed that a type-I AGN 2MASX J09325962+0405062 (ID$=$352, $z=$0.0592) has an extremely flat power-law slope of 1.25 ($1.07\sim1.42$). Its flat spectral shape might be due to Compton-thick absorption or due to warm absorber. To test the type-I Compton-thick scenario, we performed simultaneous optical and X-ray spectroscopic observations. We observed this source with Chandra on Jan. 26 and 28, with a total exposure time of 50 ks. We performed optical photometry on Jan. 10, 11, 13, 16, 18, 26, 27, 28, and 29 with the GROND instrument on the 2.2m ESO/MPG telescope and optical spectroscopy on Jan. 24, 26, 27, and 29 at Asiago. The optical observations confirmed the type-I AGN nature of this source at the time of the Chandra observation. However, the Chandra spectrum can be well described by a powerlaw with photon index of $1.63\pm0.09$, and does not require a Compton-thick absorption model or a warm absorber. Therefore, this source is more of a typical type-I AGN. The flat slope measured by eROSITA might be caused by spectral variability or by additional uncertainties in the extraction of source and background spectra for this particular source.
\section{Conclusion}

We present the AGN catalog ($21952$ sources) selected from the eFEDS main X-ray catalog ($27910$ sources).
To investigate the X-ray spectral properties of AGN, we extract and analyze the spectra of all the eFEDS sources, assuming all the them are point sources.
As the first systematical analysis of eROSITA AGN spectra, we describe in details the point-source spectra extraction methodology of eROSITA in the scanning mode.
Using a Bayesian method, we fit the spectra of all the sources, with a second aim of exploring the lower limit of spectral constraining capability of eROSITA.
We adopt a single-temperature plasma model for stars and a few power-law based models for AGN.
As simplified versions of the power-law model, we fix the power-law slope parameter at $2.0$, and even fix the \N{H} at 0.
As enriched versions, we add additional soft power-law or soft blackbody component to the primary power-law in order to model potential soft excess component.
For the sake of information completeness, the spectral fitting results of all the models are presented together with the AGN catalog, as summarized in Table.~\ref{table:tables}.

We use the ``single-powerlaw'' model (model 1) to measure the AGN obscuring \N{H} and use the ``double-powerlaw'' model (model 2) to measure the slopes of the primary power-law.
The output of Bayesian spectral analysis cannot always be adopted directly without considering the impact of the chosen prior on the posterior.
We introduce a method of quantifying this impact based on Kullback-Leibler divergence.
Using this method, we classify the \N{H} measurements as \texttt{uninformative}, \texttt{unobscured}, \texttt{mildly-measured}, and \texttt{well-measured}, and select obscured AGN as the sources with \texttt{well-measured} or \texttt{mildly-measured} \N{H} and having a median $\log$\N{H}$>$21.5.
To cope with the large parameter uncertainty of each source, we use the HBM method to estimate the parameter distribution of samples.
Using the HBM method, we recover the intrinsic \N{H} distribution of the sample, which is largely dominated by unobscured sources and has a 10\% fraction of obscured sources with $\log$\N{H} above 21.5.
Through HBM analysis with a Gaussian model on the AGN with at least $10$ counts, we find a mean $\Gamma$ of $1.94\pm0.01$ and a standard deviation of $0.22\pm0.01$.
A number of $82$ AGN have soft excess detected, which can be described by a power-law with a slope between $4$ and $8$.
Among the $50$ brightest AGN with $>500$ photon counts, nine sources have soft excess detected.
There are also a small fraction of sources with extremely flat slopes ($<1.4$), which might be due to warm absorber. 

According to spectral counts and parameter constraining capability, we choose the most appropriate model to measure the X-ray luminosity $L_X$ for each AGN.
For all the eFEDS sources, we measure the 0.5-2 keV and 2.3-5 keV observed fluxes from the spectra.
In addition to the broad-band spectral fitting, we also run narrow, soft- and hard-band fitting in order to improve the flux measurement accuracy.
Our flux measurement assumes all the sources are point sources and is thus invalid for $\sim$500 galaxy clusters in eFEDS (Liu A. et al. submitted).

We also present the rest-frame 2500\A{A} and 5100\A{A} luminosities, when multi-band photometry is available within a $\pm 1000$\A{A} wavelength range around rest-frame 2500\A{A} and 5100\A{A}.
The eFEDS AGN can be divided into two types by applying a threshold to the $L_{2500}$ to $L_{5100}$ ratio, i.e., blue AGN that are likely type-I and red AGN that are likely type-II or have strong contamination from host galaxies.
The blue AGN show a strong correlation between X-ray and UV emission.
The $\alpha_{OX}$ of the blue AGN is anti-correlated with $L_{2500}$.
 
Since eROSITA is much more sensitive in the soft band ($<2.3$ keV) than in the hard band, the eFEDS AGN catalog is more of a soft-X-ray selected catalog and thus is biased for unobscured, steep-slope sources. A detailed study of the hard-band selected eFEDS AGN is presented in Nandra et al. (in prep.).
 The eFEDS field has been observed by both SDSS-IV and SDSS-V, a more detailed investigation of the physical properties of the eFEDS AGN based on the optical properties will be presented in a following paper.


\begin{acknowledgements}
This work is based on data from eROSITA, the soft X-ray instrument aboard SRG, a joint Russian-German science mission supported by the Russian Space Agency (Roskosmos), in the interests of the Russian Academy of Sciences represented by its Space Research Institute (IKI), and the Deutsches Zentrum für Luft- und Raumfahrt (DLR). The SRG spacecraft was built by Lavochkin Association (NPOL) and its subcontractors, and is operated by NPOL with support from the Max Planck Institute for Extraterrestrial Physics (MPE).

The development and construction of the eROSITA X-ray instrument was led by MPE, with contributions from the Dr. Karl Remeis Observatory Bamberg \& ECAP (FAU Erlangen-Nuernberg), the University of Hamburg Observatory, the Leibniz Institute for Astrophysics Potsdam (AIP), and the Institute for Astronomy and Astrophysics of the University of Tübingen, with the support of DLR and the Max Planck Society. The Argelander Institute for Astronomy of the University of Bonn and the Ludwig Maximilians Universität Munich also participated in the science preparation for eROSITA.

The eROSITA data shown here were processed using the eSASS/NRTA software system developed by the German eROSITA consortium.

Funding for the Sloan Digital Sky Survey IV has been provided by the Alfred P. Sloan Foundation, the U.S. Department of Energy Office of Science, and the Participating Institutions. SDSS acknowledges support and resources from the Center for High-Performance Computing at the University of Utah. The SDSS web site is www.sdss.org.

SDSS is managed by the Astrophysical Research Consortium for the Participating Institutions of the SDSS Collaboration including the Brazilian Participation Group, the Carnegie Institution for Science, Carnegie Mellon University, Center for Astrophysics | Harvard \& Smithsonian (CfA), the Chilean Participation Group, the French Participation Group, Instituto de Astrofísica de Canarias, The Johns Hopkins University, Kavli Institute for the Physics and Mathematics of the Universe (IPMU) / University of Tokyo, the Korean Participation Group, Lawrence Berkeley National Laboratory, Leibniz Institut für Astrophysik Potsdam (AIP), Max-Planck-Institut für Astronomie (MPIA Heidelberg), Max-Planck-Institut für Astrophysik (MPA Garching), Max-Planck-Institut für Extraterrestrische Physik (MPE), National Astronomical Observatories of China, New Mexico State University, New York University, University of Notre Dame, Observatório Nacional / MCTI, The Ohio State University, Pennsylvania State University, Shanghai Astronomical Observatory, United Kingdom Participation Group, Universidad Nacional Autónoma de México, University of Arizona, University of Colorado Boulder, University of Oxford, University of Portsmouth, University of Utah, University of Virginia, University of Washington, University of Wisconsin, Vanderbilt University, and Yale University.

The Hyper Suprime-Cam (HSC) collaboration includes the astronomical communities of Japan and Taiwan, and Princeton University. The HSC instrumentation and software were developed by the National Astronomical Observatory of Japan (NAOJ), the Kavli Institute for the Physics and Mathematics of the Universe (Kavli IPMU), the University of Tokyo, the High Energy Accelerator Research Organization (KEK), the Academia Sinica Institute for Astronomy and Astrophysics in Taiwan (ASIAA), and Princeton University. Funding was contributed by the FIRST program from Japanese Cabinet Office, the Ministry of Education, Culture, Sports, Science and Technology (MEXT), the Japan Society for the Promotion of Science (JSPS), Japan Science and Technology Agency (JST), the Toray Science Foundation, NAOJ, Kavli IPMU, KEK, ASIAA, and Princeton University. 

This paper makes use of software developed for the Large Synoptic Survey Telescope. We thank the LSST Project for making their code available as free software at  http://dm.lsst.org

The Pan-STARRS1 Surveys (PS1) have been made possible through contributions of the Institute for Astronomy, the University of Hawaii, the Pan-STARRS Project Office, the Max-Planck Society and its participating institutes, the Max Planck Institute for Astronomy, Heidelberg and the Max Planck Institute for Extraterrestrial Physics, Garching, The Johns Hopkins University, Durham University, the University of Edinburgh, Queen’s University Belfast, the Harvard-Smithsonian Center for Astrophysics, the Las Cumbres Observatory Global Telescope Network Incorporated, the National Central University of Taiwan, the Space Telescope Science Institute, the National Aeronautics and Space Administration under Grant No. NNX08AR22G issued through the Planetary Science Division of the NASA Science Mission Directorate, the National Science Foundation under Grant No. AST-1238877, the University of Maryland, and Eotvos Lorand University (ELTE) and the Los Alamos National Laboratory.

Based [in part] on data collected at the Subaru Telescope and retrieved from the HSC data archive system, which is operated by Subaru Telescope and Astronomy Data Center at National Astronomical Observatory of Japan.

M.K. acknowledges support by DFG grant KR 3338/4-1.
\end{acknowledgements}

\begin{appendix}
\section{Spectral vs imaging flux}
\begin{figure}[h]
\begin{center}
\includegraphics[width=0.9\columnwidth]{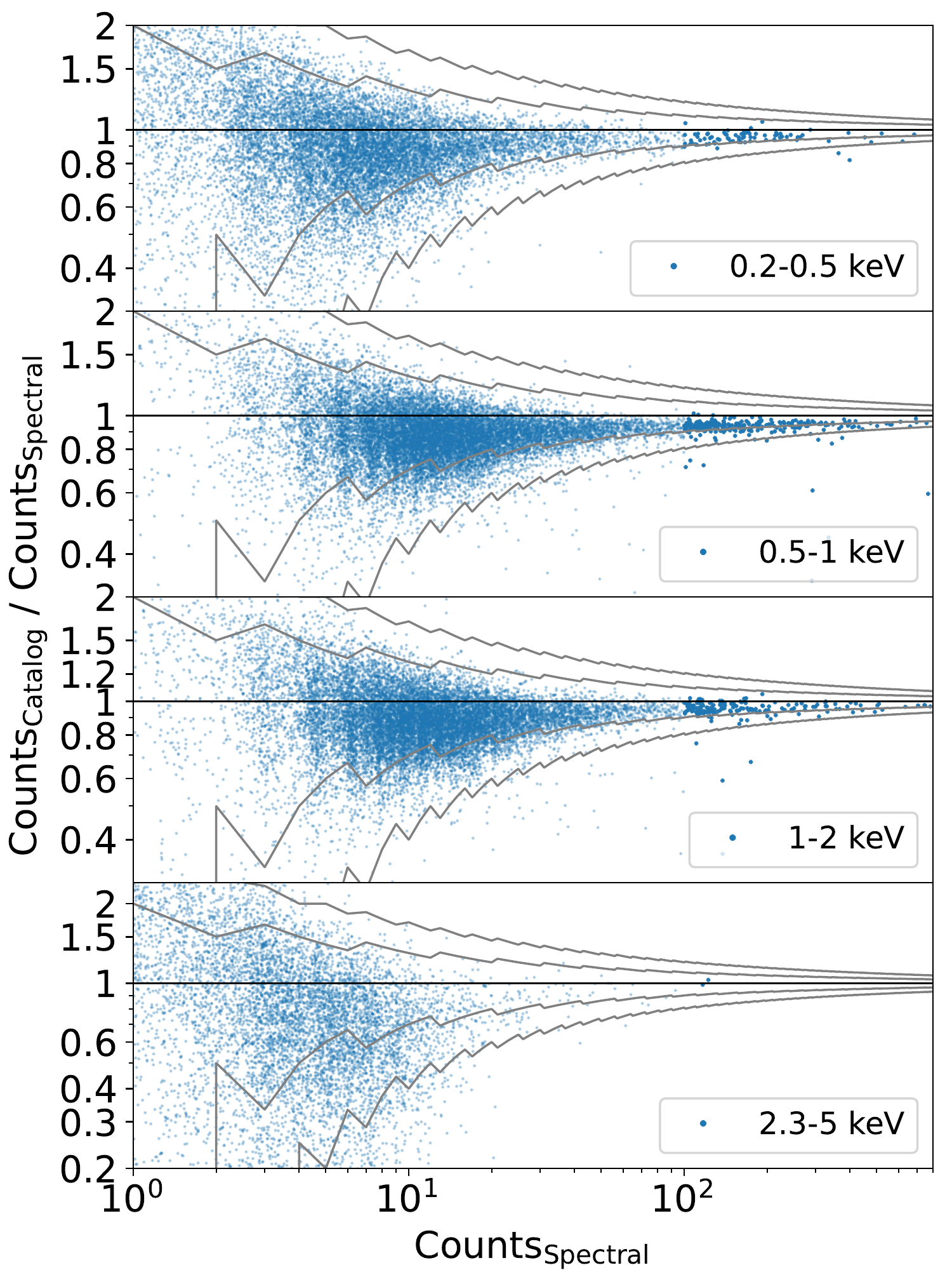}
\caption{A comparison between the catalog source counts (\texttt{ML\_CTS}) and the net counts measured from the spectra, both of which are corrected for PSF-loss outside the extraction region but not corrected for vignetting. The grey lines indicate the 1$\sigma$ and 2$\sigma$ interval expected by the Poisson distribution.}
\label{fig:cts_cts}
\end{center}
\end{figure}
With the source and background spectra, a net source count rate can be calculated by subtracting the scaled background signal form the source signal.
We correct it for the PSF-loss using the average correction values in the corresponding energy range stored in the ARF. Fig.~\ref{fig:cts_cts} compares the source counts measured from the spectra and from the catalog in a few bands. The source counts in the catalog are relatively lower by a few percent. Such a lower counts measurements in the catalog was also found by simulation with respect to the input counts (Liu et al. submitted).
This deviation is possibly caused by different PSF models adopted by different eSASS tasks.
Both the simulation in Liu et al. (submitted) and the spectra extraction in this work use the 2D PSF model \citep{Dennerl2020}; and only the count-rate measurement in the catalog uses the shapelet PSF model (Paper I). 

\begin{figure}[h]
\begin{center}
\includegraphics[width=\columnwidth]{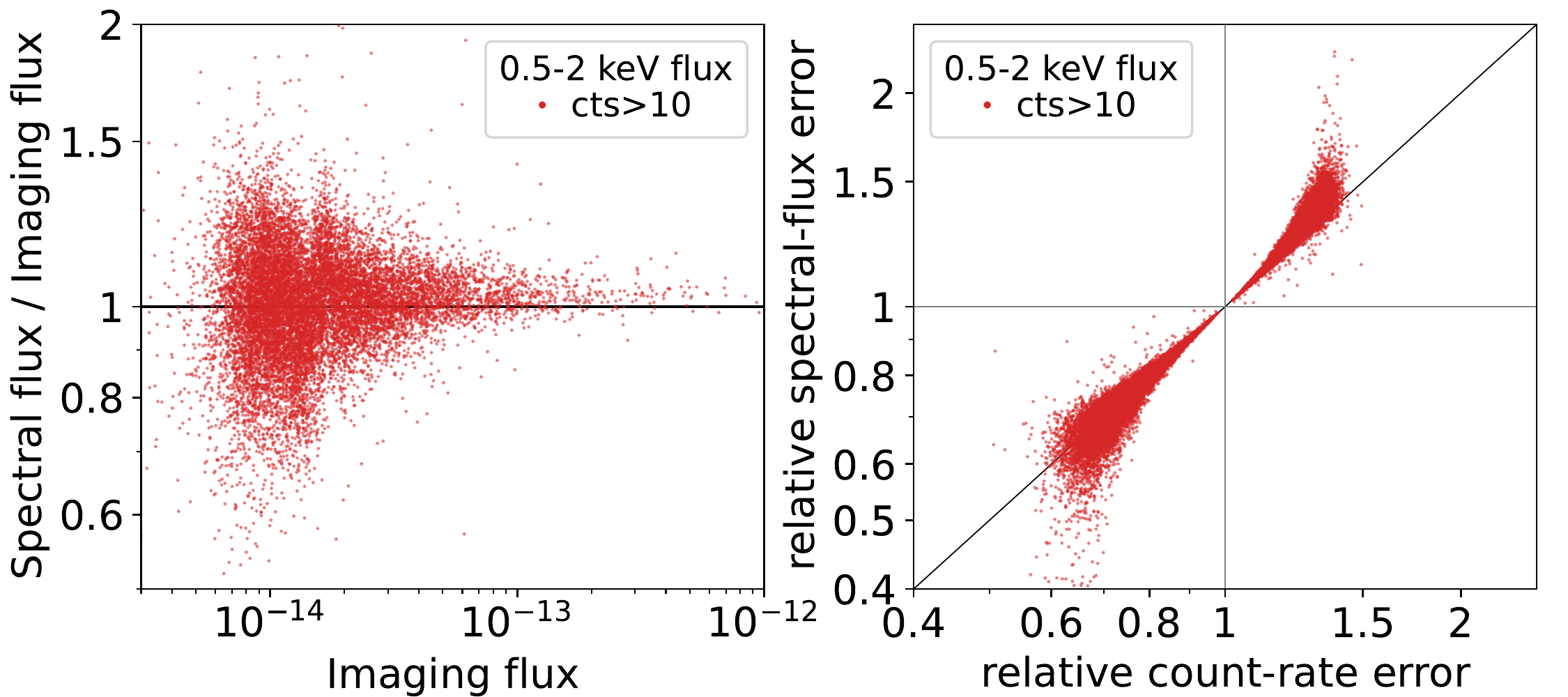}
\includegraphics[width=\columnwidth]{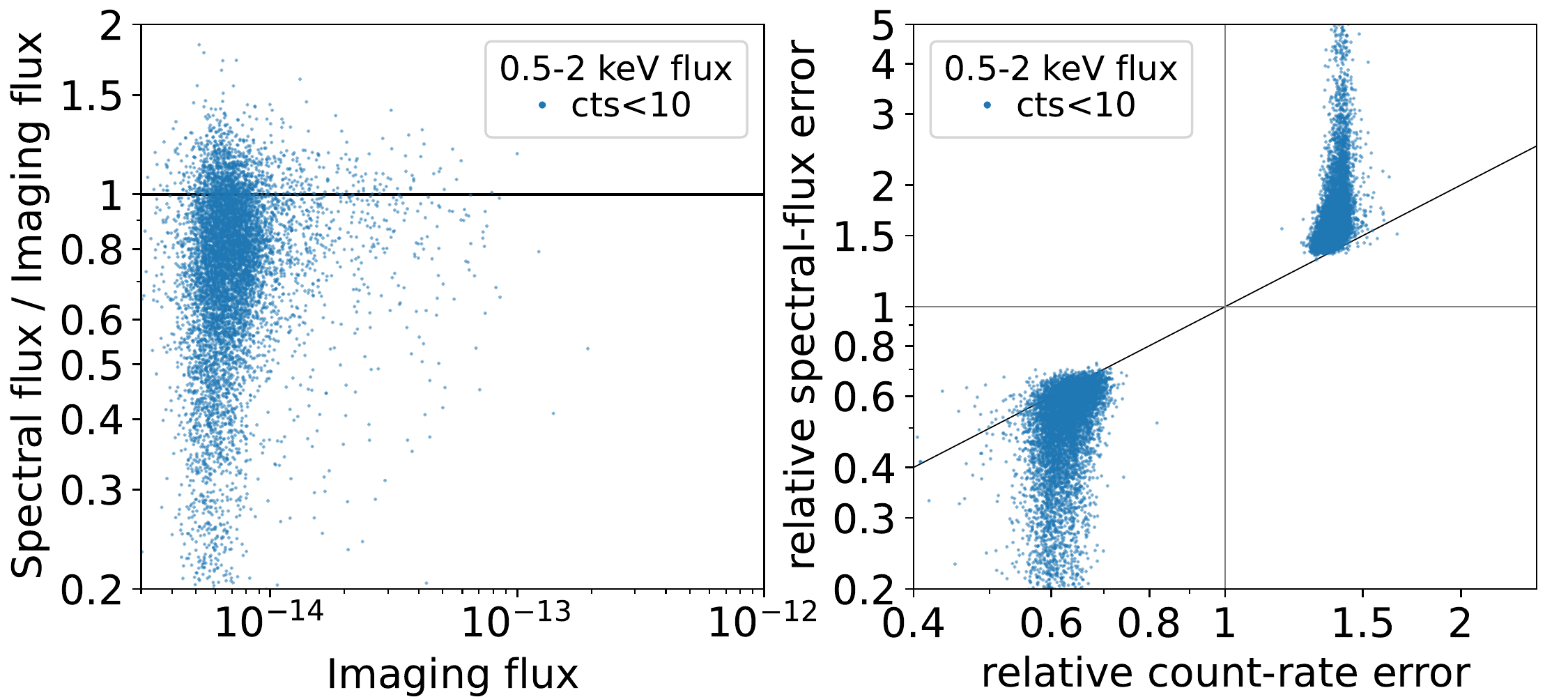}
\caption{A comparison between 0.5-2 keV fluxes measured from the 0.2--2.3 keV spectral fitting and from the 0.2--2.3 keV count rate (imaging), displaying the sources with 0.2--5 keV counts above and below 10 separately in the upper (in red color) and lower panels (in blue).
  The relative errors are the ratio between the 1-$\sigma$ upper/lower limits and the median value, thus the upper and lower limits are displayed in the ranges $>1$ and $<1$, respectively.
  }
\label{fig:flux_flux}
\end{center}
\end{figure}

\begin{figure}[h]
\begin{center}
\includegraphics[width=0.7\columnwidth]{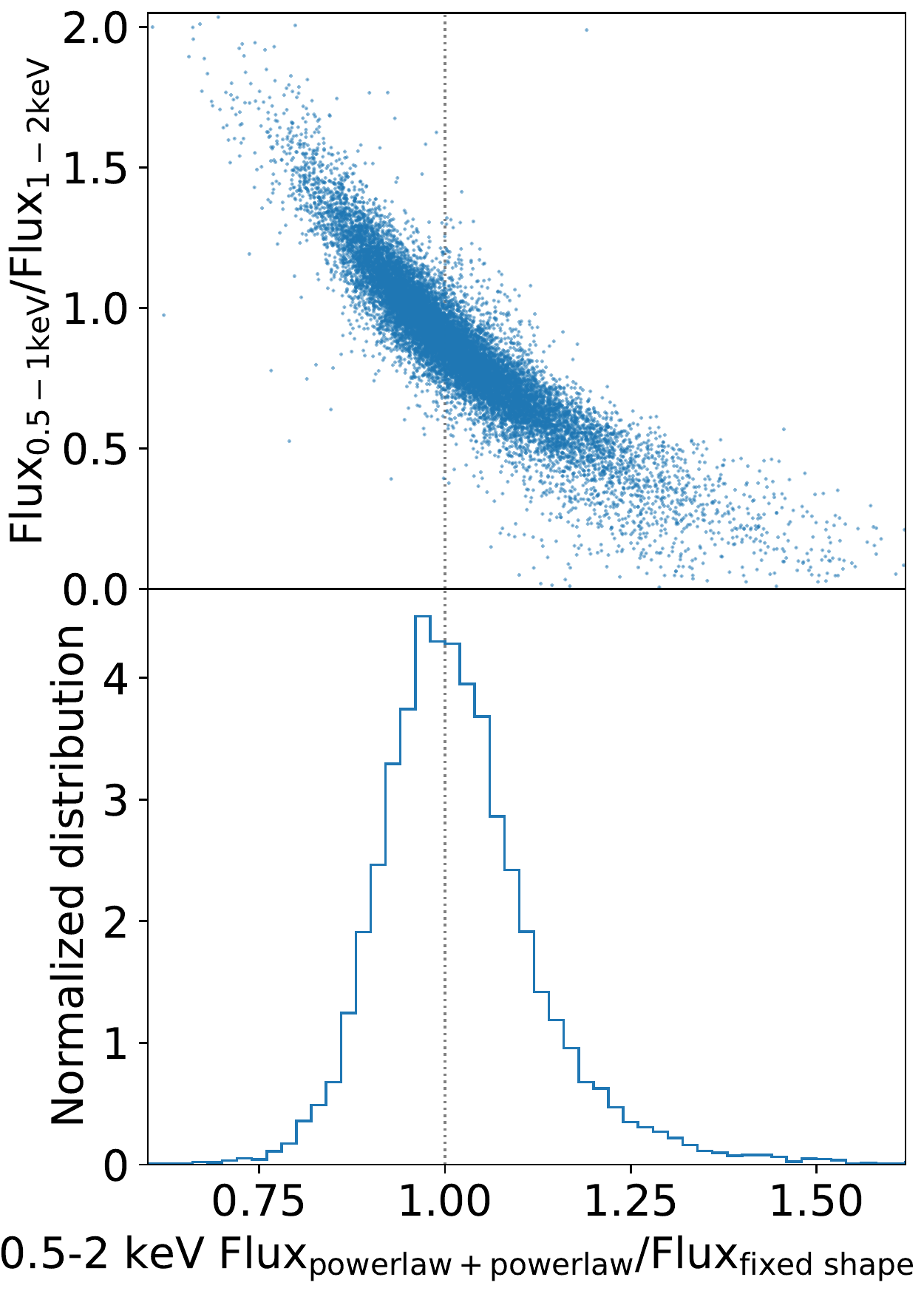}
\caption{Ratio between the 0.5-2 keV fluxes measured using the ``double-powerlaw'' model and using the ``shape-fixed-powerlaw'' model (lower panel), and its correlation with the spectral softness, i.e., the ratio between the 0.5-1 keV and 1-2 keV fluxes.}
\label{fig:softflux_power2_powfix}
\end{center}
\end{figure}

Assuming a power-law with a slope of $2.0$ and with Galactic \N{H}=$3\times10^{20}$ cm$^{-2}$, the energy conversion factor (ECF) from the 0.2--2.3 keV count rate to the 0.5-2 keV observed flux is $1.557\times 10^{12}$ cm$^{2}$/erg.
With this ECF, we convert the 0.2--2.3 keV count rate from the eFEDS catalog into 0.5-2 keV flux.
Fig.~\ref{fig:flux_flux} compares the count-rate flux and the spectral flux measured using the ``shape-fixed-powerlaw'' model, displaying the sources with 0.2--5 keV counts above and below 10 separately, just for representation.
Note that the two fluxes are both measured in the 0.2--2.3 keV band, assuming the same spectral model ($\Gamma=2$).
At high fluxes, the two measurements are broadly consistent, but the spectral flux is slightly higher (by a few percent) than the count-rate flux, likely because of the deviation shown in Fig.~\ref{fig:cts_cts}.
In the low-counts regime, the count rate measured by maximum-likelihood PSF-fitting tend to truncate near the sensitivity limit; their uncertainties are also truncated accordingly.
This is caused by Eddington bias in the space of counts.
Through Bayesian spectral fitting, we can avoid this bias and measure flux and its uncertainty more accurately.

For high-quality spectra, a detailed spectral fitting that models all the spectral features is required to measure the observed fluxes accurately.
As displayed in Fig.~\ref{fig:softflux_power2_powfix}, the ratio between the 0.5-2 keV fluxes measured using the ``double-powerlaw'' model and that using the ``shape-fixed-powerlaw'' has a median of $1.01$ and a standard deviation of $0.12$, indicating that $\Gamma=2$ is a good assumption in ECF calculation.
However, this ratio shows a strong correlation with the spectral softness, i.e., the ratio between the 0.5-1 keV and 1-2 keV fluxes measured using the ``double-powerlaw'' model.
This correlation indicate an additional flux uncertainty caused by spectral shape variety, which is not considered when calculating fluxes from count rate with a single ECF value.

  \section{Catalog Content description}
\label{sec:tabledescription}

\begin{table*}[hbtp]
  \centering
  \begin{tabular}{p{0.24\textwidth} p{0.75\textwidth}}
           \hline
    Column name & Description\\
          \hline
ID\_SRC                  	&ID of the sources in the eFEDS main X-ray catalog (Paper I) \\
RA\_CORR                   	&X-ray right ascension (J2000), astrometric corrected (Paper I) \\
DEC\_CORR                  	&X-ray declination (J2000), astrometric corrected (Paper I) \\
DET\_LIKE                  	&0.2-2.3 keV source detection likelihood (Paper I) \\
inArea90                  	&Whether located inside the inner 90\%-area region of eFEDS (Paper I) \\
CTP\_LS8\_UNIQUE\_OBJID      	&ID of the best LS8 counterpart (Paper II) \\
CTP\_LS8\_RA                	&Right ascension (J2000) of the best LS8 counterpart (Paper II) \\
CTP\_LS8\_Dec               	&Declination (J2000) of the best LS8 counterpart (Paper II) \\
CTP\_quality               	&Counterpart quality (Paper II). A threshold >=2 is adopted for the AGN catalog. \\
CTP\_CLASS                 	&Classification of the optical counterpart (Paper II). For AGN it can be 2: likely extraGalactic or 3: secure extraGalactic.  \\
CTP\_REDSHIFT              	&Redshift of the optical counterpart (Paper II) \\
CTP\_REDSHIFT\_GRADE        	&Redshift Grade (Paper II). A threshold >=3 is adopted for the AGN catalog. The highest value 5 indicates spec-z. \\
in\_KiDS\_flag              	&Whether located inside the region of the KiDS survey (Paper II) \\
Model                     	&Index of selected model for X-ray luminosity measurement. 1: single-powerlaw; 2: double-powerlaw; 4: powerlaw with Gamma fixed at 2.0; 5: shape-fixed-powerlaw. The ones with model 1,2,4 are called spectral measurements. \\
FSclass                   	&Class of soft-band (0.5-2 keV) flux measurement (\S~\ref{sec:lum_flux}). 0: model 5 (shape-fixed-powerlaw; for the faintest sources); 1: model 2 (double-powerlaw); 2: model 6 (0.4-2.2keV fit); 3: model 0 (APEC; for stars) \\
FHclass                   	&Class of hard-band (2.3-5 keV) flux measurement (\S~\ref{sec:lum_flux}). 0: model 5 (shape-fixed-powerlaw; for the faintest sources); 1: model 7 (powerlaw fitting in 2.3-6 keV) \\
NHclass                     &Class of measurement of AGN \N{H} (\S~\ref{sec:NH_Gm_KL}) based on model 1, which can be 1: \texttt{uninformative}, 2: \texttt{unobscured}, 3: \texttt{mildly-measured}, and 4: \texttt{well-measured}.\\
galNH                     	&Galactic absorption column density from HI4PI (cm$^{-2}$) \\
    FluxObsv\_{\sl\small suffix}\_{\sl\small band}&Observed energy flux (erg/cm$^2$/s) in an observed-frame energy band {\sl\small s}  or {\sl\small t} .\\
    FluxCorr\_{\sl\small suffix}\_{\sl\small band}&Absorption corrected energy flux (erg/cm$^2$/s) in an observed-frame energy band {\sl\small s}  or {\sl\small t} .\\
    LumiIntr\_{\sl\small suffix}\_{\sl\small band}&Intrinsic (absorption corrected) luminosity (erg/s) in a rest-frame energy band {\sl\small s} , {\sl\small h} , or {\sl\small 2keV} .\\
    FluxIntr\_{\sl\small suffix}\_{\sl\small band}&Absorption corrected energy flux (erg/cm$^2$/s) in a rest-frame energy band {\sl\small s} , {\sl\small h} , or {\sl\small 2keV} .\\
lognH\_{\sl\small suffix}\_m1&log AGN absorption column density (cm$^{-2}$) in model 1. The {\sl\small suffix} include ``\_KL'' (KL divergence), ``\_Med'' (posterior median), ``\_HLo'' (1-$\sigma$ HDI lower limit), and ``\_HUp'' (1-$\sigma$ HDI upper limit).\\
Gamma\_{\sl\small suffix}\_m2&Slope of the primary power-law in model 2. The {\sl\small suffix} include ``\_KL'' (KL divergence), ``\_Med'' (posterior median), ``\_HLo'' (1-$\sigma$ HDI lower limit), and ``\_HUp'' (1-$\sigma$ HDI upper limit).\\
logZ\_m$i$                   &log Bayesian evidence with model $i$, where $i$ is the model index 0,1,2,3,4,5.\\
L2500                     	&The rest-frame 2500\A{A} luminosity in erg/s/Hz \\
L5100                     	&The rest-frame 5100\A{A} luminosity in erg/s/Hz \\
W1                        	&LS8-WISE W1 AB magnitude (Paper II) \\
W1\_ERR                    	&LS8-WISE W1 magnitude error (Paper II) \\
W2                        	&LS8-WISE W2 AB magnitude (Paper II) \\
W2\_ERR                    	&LS8-WISE W2 magnitude error (Paper II) \\
    \hline
    \end{tabular}
    \caption{Columns of the eFEDS AGN catalog. 
      For the flux-related columns, i.e., ``FluxObsv'', ``FluxCorr'', ``FluxIntr'', and ``LumiIntr'', the {\sl\small suffix} can be ``\_Med'' (posterior median), ``\_Lo1'' (1-$\sigma$ percentile lower limit), ``\_Up1'' (1-$\sigma$ percentile upper limit), ``\_Lo2'' (2-$\sigma$ percentile lower limit), ``\_Up2'' (2-$\sigma$ percentile upper limit), or ``\_BF'' (best-fit model), and the energy bands include {\sl\small s} for 0.5--2 keV, {\sl\small t} for 2.3--5 keV, {\sl\small h} for 2--10 keV, and {\sl\small 2keV} for 1.999--2.001 keV. \\
    }
    \label{table:AGN}
\end{table*}

\begin{table*}[h]
  \centering
  \begin{tabular}{p{0.18\textwidth} p{0.81\textwidth}}
    \hline
    Column name & Description\\
           \hline
ID\_SRC                          &ID of the sources in the eFEDS main X-ray catalog (Paper I)\\
RA\_CORR                         &X-ray right ascension (J2000), astrometric corrected (Paper I)\\
DEC\_CORR                        &X-ray declination (J2000), astrometric corrected (Paper I)\\
DET\_LIKE                        &0.2-2.3 keV source detection likelihood (Paper I)\\
inArea90                        &Whether located inside the inner 90\%-area region of eFEDS (Paper I)\\
galNH                           &Galactic absorption column density from HI4PI (cm$^{-2}$)\\
Exposure                        &Spectra exposure time (s)\\
SrcCts                          &Source net counts in the 0.2-5 keV band measured from the spectra\\
RA                              &X-ray right ascension (J2000) before astrometric correction (Paper I), used in spectra extraction\\
DEC                             &X-ray declination (J2000) before astrometric correction (Paper I), used in spectra extraction\\
Backscal\_s                 & Source BACKSCAL (deg$^2$)\\
Backscal\_b                 & Background BACKSCAL (deg$^2$)\\
Rad                        & Source extraction radius (arcsec)\\
Ann1                       & Inner radius of background extraction region (arcsec)\\
Ann2                       & Outer radius of background extraction region (arcsec)\\
REGAREA\_s                  & Geometry area of source extraction region (deg$^2$)\\
Nempty                     & Number of empty channels between channel 20 and 900\\
Rate\_{\sl\small band}      & Net count rate in 0.2-2.3, 0.2-0.5, 0.5-1, 1-2, 2-4.5, 2.3-5, and 5-8 keV (with {\sl\small band} suffixes of d2\_2d3, d2\_d5, d5\_1, 1\_2, 2\_4d5, 2d3\_5, and 5\_8) \\
RateErr\_{\sl\small band}   & Net count rate error in the corresponding energy band\\
BkgCts\_{\sl\small band}    & Background counts in 0.2-0.6, 0.6-2.3, 2.3-5, and 5-8 keV (with {\sl\small band} suffixes of d2\_d6, d6\_2d3, 2d3\_5, and 5\_8)\\
BkgCtsErr\_{\sl\small band} & Background counts error in the corresponding energy band\\
PSFCor\_{\sl\small band}    & ARF CORRPSF averaged in 0.2-0.5, 0.5-1, 1-2, and 2.3-5 keV (with {\sl\small band} suffixes of d2\_d5, d5\_1, 1\_2, and 2d3\_5)\\
FluxObsv\_{\sl\small suffix}\_{\sl\small band}&Observed energy flux (erg/cm$^2$/s) in an observed-frame energy band {\sl\small s} (0.5--2 keV)  or {\sl\small t} (2.3--5 keV). The {\sl\small suffix} include ``\_Med'' (posterior median), ``\_Lo1'' (1-$\sigma$ percentile lower limit), and ``\_Up1'' (1-$\sigma$ percentile upper limit).\\
FSclass                   	&Class of soft-band (0.5-2 keV) flux measurement (\S~\ref{sec:lum_flux}). 0: model 5 (shape-fixed-powerlaw; for the faintest sources); 1: model 2 (double-powerlaw); 2: model 6 (0.4-2.2keV fit); 3: model 0 (APEC; for stars) \\
FHclass                   	&Class of hard-band (2.3-5 keV) flux measurement (\S~\ref{sec:lum_flux}). 0: model 5 (shape-fixed-powerlaw; for the faintest sources); 1: model 7 (powerlaw fitting in 2.3-6 keV) \\
           \hline
    \end{tabular}
    \caption{The columns of the basic spectral property table (table 2 ``Spec'' in Table.~\ref{table:tables}).
    }
    \label{table:spec}
\end{table*}

\begin{table*}[h]
  \centering
  \begin{tabular}{p{0.18\textwidth} p{0.81\textwidth}}
           \hline
    Column name & Description\\
    \hline
    \multicolumn{2}{l}{Spectral model parameters}\\
           \hline
Gamma\_{\sl\small suffix}&                      Powerlaw slope\\
lognH\_{\sl\small suffix}&                       AGN absorption column density (cm$^{-2}$)\\
logBkgNorm\_{\sl\small suffix}&                  Background normalization\\
logPowNorm\_{\sl\small suffix}&                  Power-law normalization\\
logApecNorm\_{\sl\small suffix}&                 APEC normalization\\
logBBNorm\_{\sl\small suffix}&                   Blackbody normalization\\
logkT\_{\sl\small suffix}&                       Temperature (keV) of blackbody or APEC\\
dGm\_{\sl\small suffix}&                        Slope of the additional soft power-law minus slope of the primary power-law\\
logFrac\_{\sl\small suffix}&                     Constant factor multiplied to the additional power-law\\
           \hline
    \multicolumn{2}{l}{Fluxes and luminosities}\\
           \hline
FluxObsv\_{\sl\small suffix}\_{\sl\small band}&            Observed energy flux (erg/cm$^2$/s) in an observed-frame energy band \\
FluxCorr\_{\sl\small suffix}\_{\sl\small band}&            Absorption corrected energy flux (erg/cm$^2$/s) in an observed-frame energy band \\
FluxIntr\_{\sl\small suffix}\_{\sl\small band}&            Absorption corrected energy flux (erg/cm$^2$/s) in a rest-frame energy band \\
LumiIntr\_{\sl\small suffix}\_{\sl\small band}&            Intrinsic (absorption corrected) luminosity (erg/s) in a rest-frame energy band \\
          \hline
    \multicolumn{2}{l}{Other columns of spectral fitting results}\\
           \hline
ID\_SRC                          &ID of the sources in the eFEDS main X-ray catalog (Paper I)\\
Redshift& The redshift adopted in the spectral model\\
logZ&                       Logarithmic Bayesian evidence\\
logZerr&                    Uncertainty of logZ\\
statistic& {\sl C} statistic of the best-fit model\\
chi25& the $\chi^2$ of the best-fit model against the rebinned data with at least 25 counts in each bin\\
dof25& the DOF of the best-fit model against the rebinned data with at least 25 counts in each bin\\
NHclass                     &Class of measurement of AGN \N{H} (\S~\ref{sec:NH_Gm_KL}) based on model 1, which can be 1: \texttt{uninformative}, 2: \texttt{unobscured}, 3: \texttt{mildly-measured}, and 4: \texttt{well-measured}; only for model 1.\\
    \hline
    \end{tabular}
    \caption{The columns of spectral fitting results (table 3$\sim$10 in Table.~\ref{table:tables}).
      A prefix ``log'' in parameter name indicates the value is in logarithm.
      The suffixes of ``\_Med'', ``\_Mean'', and ``\_Std'' in the column names indicate the median, mean, and standard deviation values of the posterior distribution.
      The suffixes of ``\_Lo1'' (or ``\_Lo'') and ``\_Up1'' (or ``\_Up'') indicate 1-$\sigma$ (68\%) percentile confidence interval around the median; ``\_Lo2'' and ``\_Up2'' correspond to 2-$\sigma$.
      The parameters corresponding to the best-fit model are also presented with a ``\_BF'' suffix in the column name. 
      For spectral shape parameters, we also measure the 1-$\sigma$ HDI lower and upper limits (with suffixes of ``\_HLo'' and ``\_HUp'' respectively) and the KL divergence between the prior and posterior distributions (with a suffix of ``\_KL'').
    }
    \label{table:pars}
\end{table*}


As listed in Table.~\ref{table:tables}, we present the eFEDS AGN catalog and the X-ray spectral properties of all the eFEDS sources.
The eFEDS AGN catalog provides the AGN \N{H} measured with the ``single-powerlaw'' model, the power-law slopes measured with the ``double-powerlaw'' model, and the fluxes/luminosities based on the most appropriate models selected in \S~\ref{sec:modelselection}.
The columns of this catalog are described in Table.~\ref{table:AGN}.

In the second table as listed in Table.~\ref{table:tables}, we present the basic properties of the X-ray spectra for all the eFEDS sources, including the spectra extraction information and the source count rates and observed fluxes measured from the spectra. The columns of this table are described in the first section of Table.~\ref{table:spec}.

We also present eight sets of spectral fitting results with different spectral models and settings, i.e., the tables 3$\sim$10 listed in Table.~\ref{table:tables}.
The spectral fitting results in these tables are named in a uniform way as described in Table.~\ref{table:pars}, although some columns only apply to particular models.
For any parameter of the spectral models or any value (e.g., flux and luminosity) derived from the spectral models, we can measure a few statistical quantities of its posterior distribution, including the median, the mean, the standard deviation, the 1-$\sigma$/2-$\sigma$ percentile confidence intervals, and the HDI intervals.
The values corresponding to the best-fit model are also presented, although we recommend to use the above quantities obtained through Bayesian inference.
For spectral shape parameters, which are of more interests than normalization parameters, the KL divergence between the prior and posterior distributions is also provided.
In the observed-frame 0.5--2 and 2.3--5 keV bands, we provide measurements of both observed and absorption corrected fluxes.
In the rest-frame 0.5--2, 2--10, and 1.999--2.001 keV bands, we measure the absorption corrected fluxes and luminosities.

\end{appendix}
\bibliography{spec}
\end{document}